\documentclass{svmult}
\usepackage{mathptmx}
\usepackage{helvet}
\usepackage{courier}
\usepackage{makeidx}
\usepackage{multicol}
\usepackage{footmisc}
\usepackage[utf8x]{inputenc}
\usepackage{graphicx}
\usepackage{color}
\usepackage{amsmath}
\usepackage{subfig}
\usepackage{float}
\usepackage{multirow}
\begin{document}
\title*{Timing interactions in social simulations: The voter model}
\author{Juan Fern\'andez-Gracia, V\'ictor M. Egu\'iluz and Maxi San Miguel}
\institute{IFISC, Instituto de F\'isica interdisciplinar y Sistemas Complejos (CSIC-UIB), Campus Universitat Illes Balears, E-07122 Palma de Mallorca, Spain}

\titlerunning{Timing interactions in social simulations}
\authorrunning{ J. Fern\'andez-Gracia et al. }

\maketitle

\abstract{
The recent availability of huge high resolution datasets on human activities has revealed the heavy-tailed nature of the interevent time distributions. In social simulations of interacting agents the standard approach has been to use Poisson processes to update the state of the agents, which gives rise to very homogeneous activity patterns with a well defined characteristic interevent time. As a paradigmatic opinion model we investigate the voter model and review the standard update rules and propose two new update rules which are able to account for heterogeneous activity patterns. For the new update rules each node gets updated with a probability that depends on the time since the last event of the node, where an event can be an update attempt (exogenous update) or a change of state (endogenous update). We find that both update rules can give rise to power law interevent time distributions, although the endogenous one more robustly. Apart from that for the exogenous update rule and the standard update rules the voter model does not reach consensus in the infinite size limit, while for the endogenous update there exist a coarsening process that drives the system toward consensus configurations.
}

\section{Introduction}

Individual based models of collective social behavior include traditionally two basic ingredients:
the mechanism of interaction and the network of interactions \cite{castellano_stat_phys}. The idea of choosing a mechanism of interaction, such as random imitation \cite{clifford_sudbury,holley_liggett,opinion} or threshold behavior under social pressure \cite{Granovetter,Watts,DC}, is to isolate this mechanism and to determine its consequences at the collective level of emergent properties. The network of interactions determines  who interacts with whom. The topology of the network incorporates among other things the heterogeneity of ties among individuals. In addition ties can be non persistent, so that the network structure changes with time. In particular, the network and the state of the individuals can evolve in similar time scales (co-evolution). Such entangled process of \textit{dynamics of the network} and the \textit{dynamics on the network} can describe how to go from  \textit{interacting with neighbors} to \textit{choosing neighbors}  \cite{coevolution1,coevolution2,coevolution3,interface,pre_fede_jc}. A third ingredient of individual based models, which was not considered in detail in the past, is the timing of interactions: When do individuals interact? The usual assumption in simulation models was that of a constant rate of interaction. In this paper we revise this assumption addressing the consequences of the heterogeneity in the timing of interactions.

Addressing this question is timely due to the availability of massive and high resolution data on human activity patterns. Information and knowledge extracted from this data needs to be included in a realistic modeling of collective social behavior. Indeed, many interevent time distributions measured recently in empirical studies about human activities such as e-mail communication, surface mail, timing of financial trades, visits to public places, long-range travels, online games, response time of cybernauts, printing processes and phone calls, among others \cite{malmgrem_univ_corr, Dar_Eins_BAr, Eckmann_orig, moro, vazquez_non_poiss, small_but_slow, vazquez_spread, malmgrem_poiss, origin_bursts, model_bursts}, show heavy tails for large times. Motivated by these findings there are two current lines of research:
\begin{itemize}
 \item Origin of these heavy-tailed distributions
 \begin{itemize}
  \item Explain these tails based on circadian cycle and seasonality, via a non-homogeneous Poisson process with a cascading mechanism \cite{malmgrem_univ_corr,malmgrem_poiss}.
  \item Root these heavy tails in the way individuals organize and prioritize their tasks modeling it via priority queuing models \cite{Dar_Eins_BAr,vazquez_non_poiss,origin_bursts,model_bursts}. Cite \textbf{Byungjoon Min, K.-I. Goh, and I.-M. Kim. Burstiness: Measures, models, and dynamic consequences}
 \end{itemize}
 \item Effects of this interaction timing heterogeneity on certain dynamics: independently of the origin of this feature it has been noticed that a non-homogeneous interaction in time can give rise to non-trivial behavior. An example considered so far in some detail is spreading and infection dynamics: SI-type spreading dynamics have been investigated, showing that this peculiar timing gives rise to a slowing down of the dynamics that cannot be explained just by a change of time scale but it changes the functional form of the prevalence of a disease \cite{moro,vazquez_non_poiss,small_but_slow,vazquez_spread}. Cite \textbf{Alexei Vazquez. Spreading dynamics following bursty activity patterns}
\end{itemize}
Our work \cite{PREJuan} goes along the second of these research lines. It considers the implementation of human activity patterns in simulation models of interacting individuals, and the consequences of the timing of interactions.  As an illustrative model we explore this general question in the context of the voter model \cite{clifford_sudbury,holley_liggett}. The voter model is a very stylized model that serves as a null model for the competition of two equivalent states under a dynamics of random imitation. A difference with previous work in spreading and infection dynamics is that in the voter model each individual can be in two equivalent states. Then the question is  when the system reaches consensus in either of these two states or when there is asymptotic dynamical coexistence of the two states. We will see that the answer to this question depends crucially on the timing of interactions.  Related work on the voter model, discussed later, include the papers by Stark \textit{et al.} \cite{tessone}, Baxter \cite{baxter} and Takaguchi and Masuda \cite{masuda}.

In Sect. 2 we revise the definition of the voter model and the different quantities used to monitor its macroscopic dynamics. Sect. 3 considers the voter model dynamics with different standard update rules, i.e. update rules that incorporate a constant rate of interaction. In Sect.4 we introduce new update rules to account for heterogeneous activity patterns. We consider two update rules: \emph{endogenous update}, coupled to the dynamics of the states of the agents; and \emph{exogenous update} which is independent of the states of the agents. Section 5 includes a discussion of our results and related work.

\section{The voter model}

\subsection{Definition of the voter model}

The voter model is a microscopic model first considered in Ref.~\cite{clifford_sudbury} in 1973 as a model for the competition of species for their habitats, and named \textit{voter model} in Ref.~\cite{holley_liggett} in 1975 as the natural interpretation of its rules in terms of opinion dynamics \cite{castellano_stat_phys,opinion}. However it has been investigated not just in the context of social dynamics but also in fields such as probability theory \cite{holley_liggett} and population dynamics \cite{clifford_sudbury}.

The voter model consists of a set of $N$ agents placed on the nodes of an interacting network. The links of the network are the connections among agents. Two nodes are first neighbors if they are directly connected by a link in the network. The agents have a binary variable (opinion, state...) which can take the values $+1$ or $-1$. The behavior of the agents is characterized by an imitation process, because, whenever they interact, they just copy the state of a randomly chosen first neighbor.

The model has two absorbing configurations, \textit{i.e.},  configurations in which the dynamics stop, which consist either of all agents in state $+1$ or in state $-1$. These absorbing configurations are also typically called \textit{consensus}, as the whole population has agreed in the same state.

This model has been studied by computer simulations using what we later define as \textit{random asynchronous update} for node dynamics. In this case the basic steps in the dynamics are:
\begin{enumerate}
 \item Randomly choose an agent $i$ with opinion $x_i$.
 \item Randomly choose one of $i$'s neighbors, $j$, with opinion $x_j$. Agent $i$ adopts $j$'s opinion; $x_i(t+1/N)=x_j(t)$.
 \item Resume at 1.
\end{enumerate}
The alternative link dynamics is considered in Ref.~\cite{conserv_sucheki}.

Usually the time is measured in units of $N$ basic steps, \textit{i.e.}, a Monte Carlo step, following the idea that every agent gets updated on average once per unit time.

\subsubsection{Macroscopic description}

A basic question is under which conditions consensus will be reached and how. In order to answer this question we have to define some macroscopic quantities which will describe the state of the system and its dynamical behavior.
\begin{description}
 \item {\textit{Magnetization} $m(t)$: }{ It is the average state of the population and is defined as $$m(t)=\frac{1}{N}\sum_{i=1}^N x_i.$$}
 \item {\textit{Density of interfaces} $\rho(t)$: }{It is the fraction of links connecting agents with different states. It is defined as $$\rho(t)=\frac{\text{\# of links between $-1$ and $+1$}}{\text{\# of links in the network}}=\frac{1}{\sum_{i=1}^Nk_i}\left(\sum_{\langle ij \rangle}\frac{1-x_ix_j}{2}\right),$$ where $\langle ij \rangle$ stands for summing over neighboring nodes.}
\end{description}
In numerical simulations finite size effects come into play. In finite size systems consensus will be reached, but we have to differentiate if consensus is reached due to the inherent dynamics or to a finite size fluctuation. We use averages over many realizations to extract the mean behavior. This is what is called an ensemble average and will be denoted by $\langle \cdot \rangle$. When doing the ensemble averages some conservation laws can be found. For the case of regular networks, where every node has the same number of neighbors (same degree), the ensemble average of the magnetization $\langle m(t) \rangle$ is conserved under node dynamics \cite{conserv_sucheki,sci_rep_klemm}. For this reason the magnetization is not a good order parameter and we have to define the density of interfaces $\rho$, which in general is not conserved. This is a proper order parameter as it measures the degree of order in the system. It is nonzero while the system is not in one of the absorbing states and is zero otherwise. A decrease of $\rho(t)$ describes the coarsening process with growth of domains with agents in the same state. If the network is heterogeneous, \textit{i.e.},  the degrees of the nodes are not all the same, the conservation law for $\langle m(t)\rangle$ breaks down unless we use link dynamics.

In order to gain more insight into the dynamics for finite size systems we also introduce two other quantities to characterize the dynamics. These quantities are:
\begin{description}
 \item {\textit{Survival probability $S(t)$: }}{It is the probability that a realization of the system has not reached one of the absorbing states at time $t$. The mean time $\langle T\rangle$ to reach consensus is then given by\footnote{$S(t)$ is the probability of being in an active configuration at simulation time $t$. Then the probability of reaching an absorbing state at time $t$ is $\frac{d}{dt}(1-S(t))=-\frac{d}{dt} S(t)$. The average time to reach consensus is then $\langle T\rangle=-\int_0^{\infty}(t \frac{d}{dt}S(t))dt$ and, integrating by parts one finds that $\langle T\rangle=\int_0^{\infty}S(t)$.} $$\langle T\rangle=\int_0^{\infty}S(t)dt.$$}
 \item {\textit{Density of interfaces averaged over surviving runs} $\langle\rho^*(t)\rangle$: }{This quantity is basically the same as the density of interfaces, but disregarding the realizations that have already reached an absorbing state when doing the ensemble average. It tells us the degree of order in the system for the realizations that are still in an active state. This quantity is related to the density of interfaces averaged over all realizations by $$\langle \rho(t) \rangle=S(t)\langle\rho^*(t)\rangle.\label{rel_rho_rho*}$$}                                                                                                                                                                                                                                                                                                                                                                                                                                                                                                                                                          
\end{description}

A novel quantity in the study of the voter model has to be introduced in order to characterize the temporal activity patterns. This quantity is:

\begin{description}
 \item {\textit{Interevent time (IET) distribution $M(\tau)$: }}{It is the probability that, given two consecutive changes of state of a node, the time interval between them equals $\tau$. We will also use the complementary cumulative distribution\footnote{In the remainder we will refer to the complementary cumulative distribution just as cumulative distribution.} of this, $C(\tau)=1-\int_0^{\tau}M(t)dt$.}
\end{description}

\section{Standard update rules}

In this section we review standard update rules used in simulations of agent based models (ABM's). We also investigate the behavior of the voter model for these different rules. In ABM's agents are placed on the nodes of a network. The state of the agents is characterized by a variable that can take one of various values. The specific dynamics tells how the states of the nodes are updated. But in addition to the dynamical rules, simulation incorporate rules that determine when an agent is given the opportunity to update her state. Standard update rules implement a homogeneous pattern of updates in time.

The simulations all over this chapter where done with random initial conditions, \textit{i.e.} every agent has the same probability in the beginning to have one state or the other, and setting their initial time since the last change of state equal to zero.

\subsection{Definitions of standard update rules}

Typically the update rules implemented are
\begin{description}
 \item{\textbullet\textbf{Asynchronous update:}}{ At each simulation step only one of the agents is updated. The unit of time is typically defined as $N$ simulation steps (a Monte Carlo step), where $N$ is the number of agents in the system.
\begin{description}
 \item{\textbf{Random asynchronous update (RAU):}}{ the agents are updated in a random order.}
 \item{\textbf{Sequential asynchronous update (SAU):}}{ the agents are always updated in the same order.}
\end{description}
}
 \item{\textbullet\textbf{Synchronous update (SU):}}{ All the agents are updated at the same time. The time is measured in units of simulation steps.}
\end{description}
The most commonly used update for the voter model has been the RAU. Most of the results have been derived for that update.
As we can see from the definitions of these standard update rules, there exists a well defined characteristic time between two consecutive updates of the same node. In the case of SAU and SU every agent is updated exactly once per unit time, while for RAU this only happens on average.

\subsection{Voter model with standard update rules}\label{voter_standard}

\begin{figure}
  \centering
  \subfloat{\label{usual_rho_meanf_suchecki}\includegraphics[draft=false,height=0.31\textwidth,angle=-90]{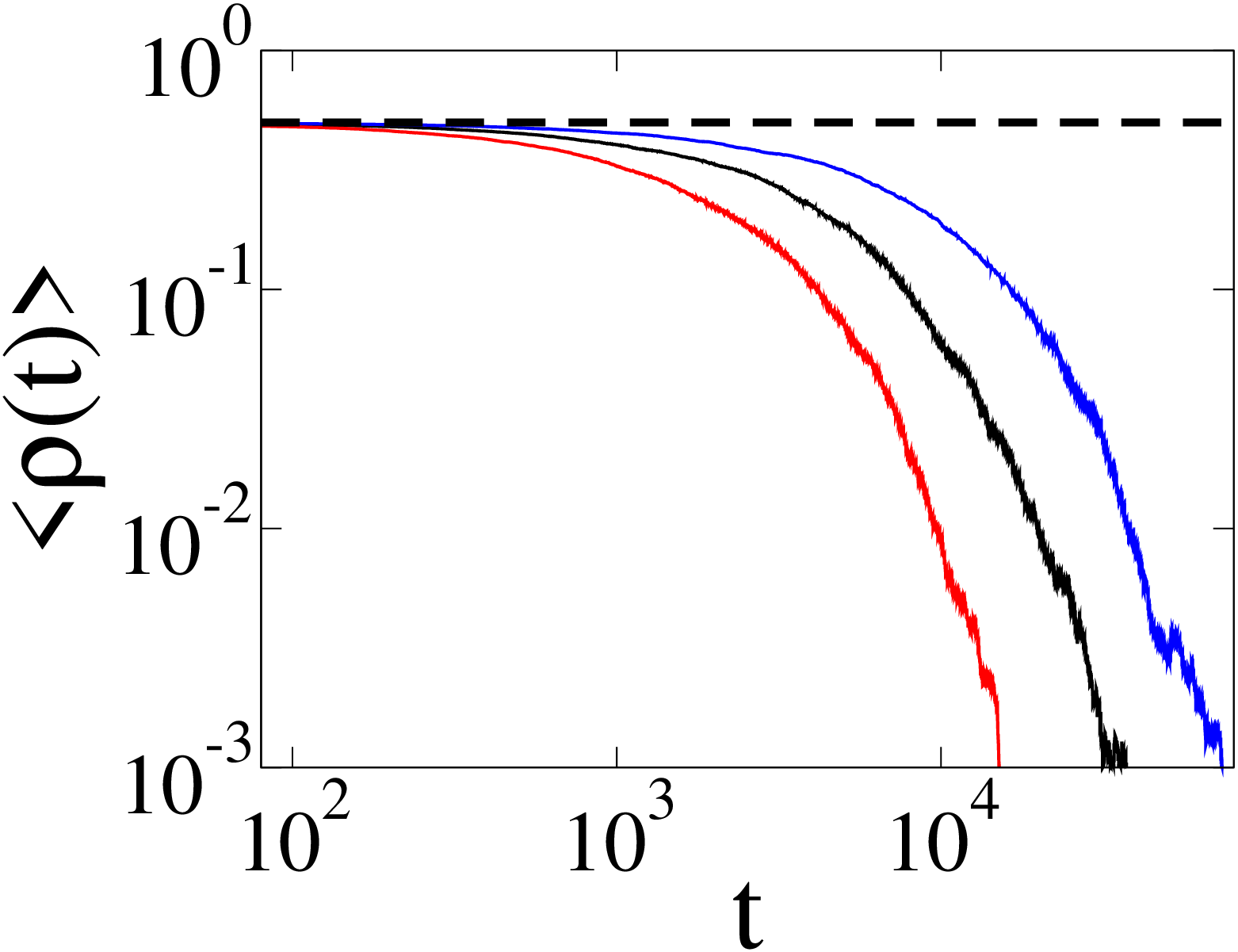}}\ 
  \subfloat{\label{usual_rho_randNm3_suchecki}\includegraphics[draft=false,height=0.31\textwidth,angle=-90]{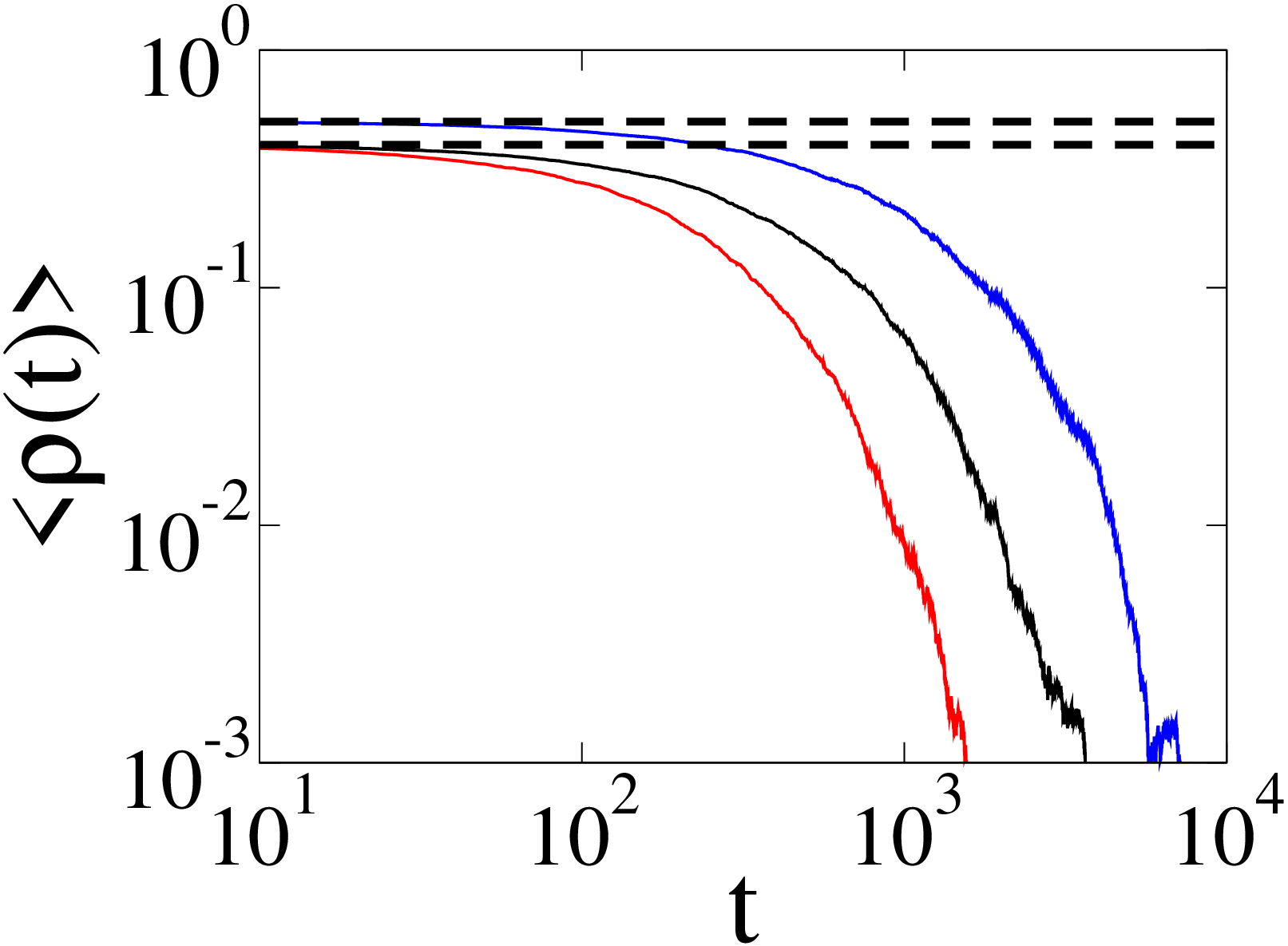}}\ 
  \subfloat{\label{usual_rho_BAm3_suchecki}\includegraphics[draft=false,height=0.31\textwidth,angle=-90]{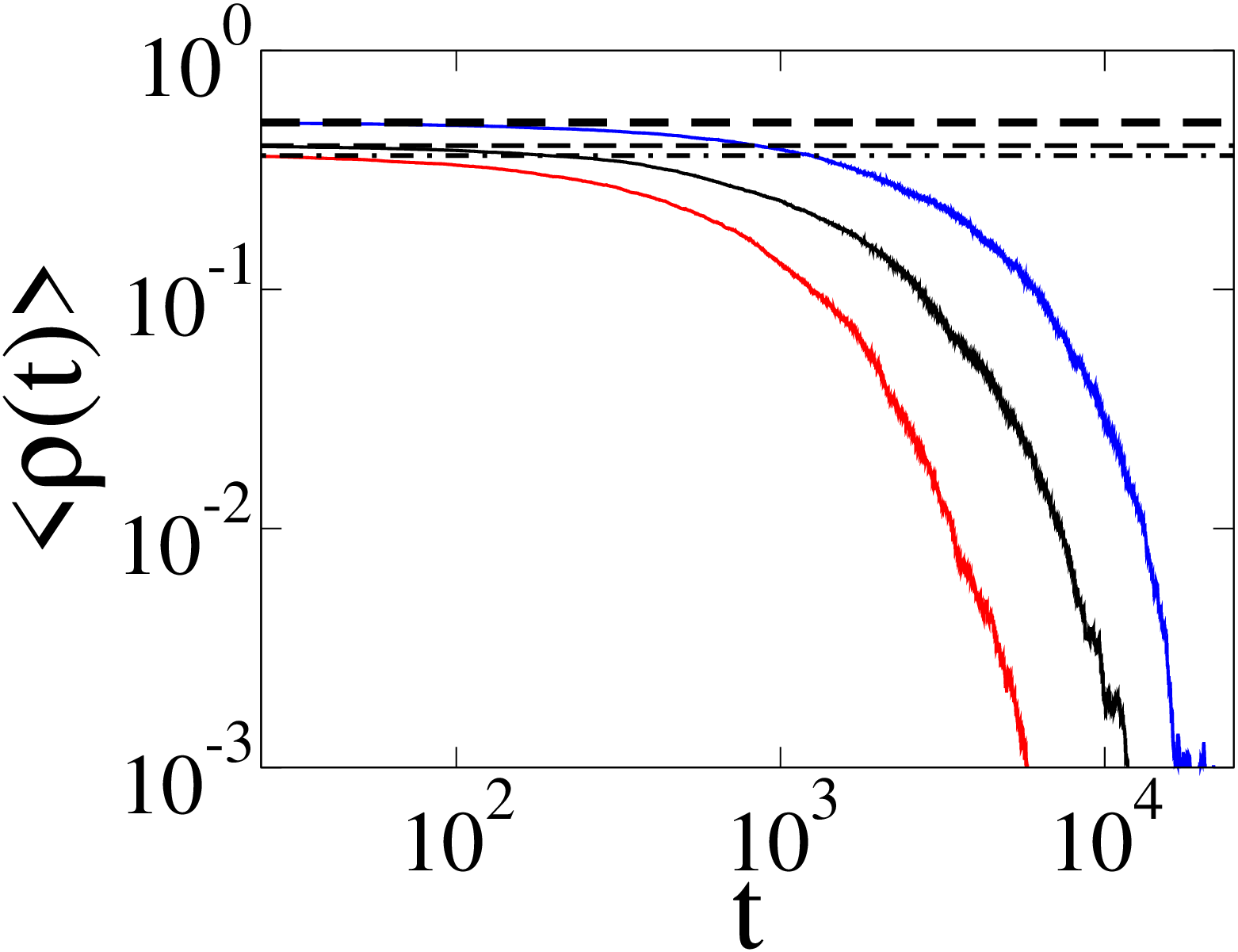}}\\
  \subfloat{\label{usual_rhostar_meanf}\includegraphics[draft=false,height=0.31\textwidth,angle=-90]{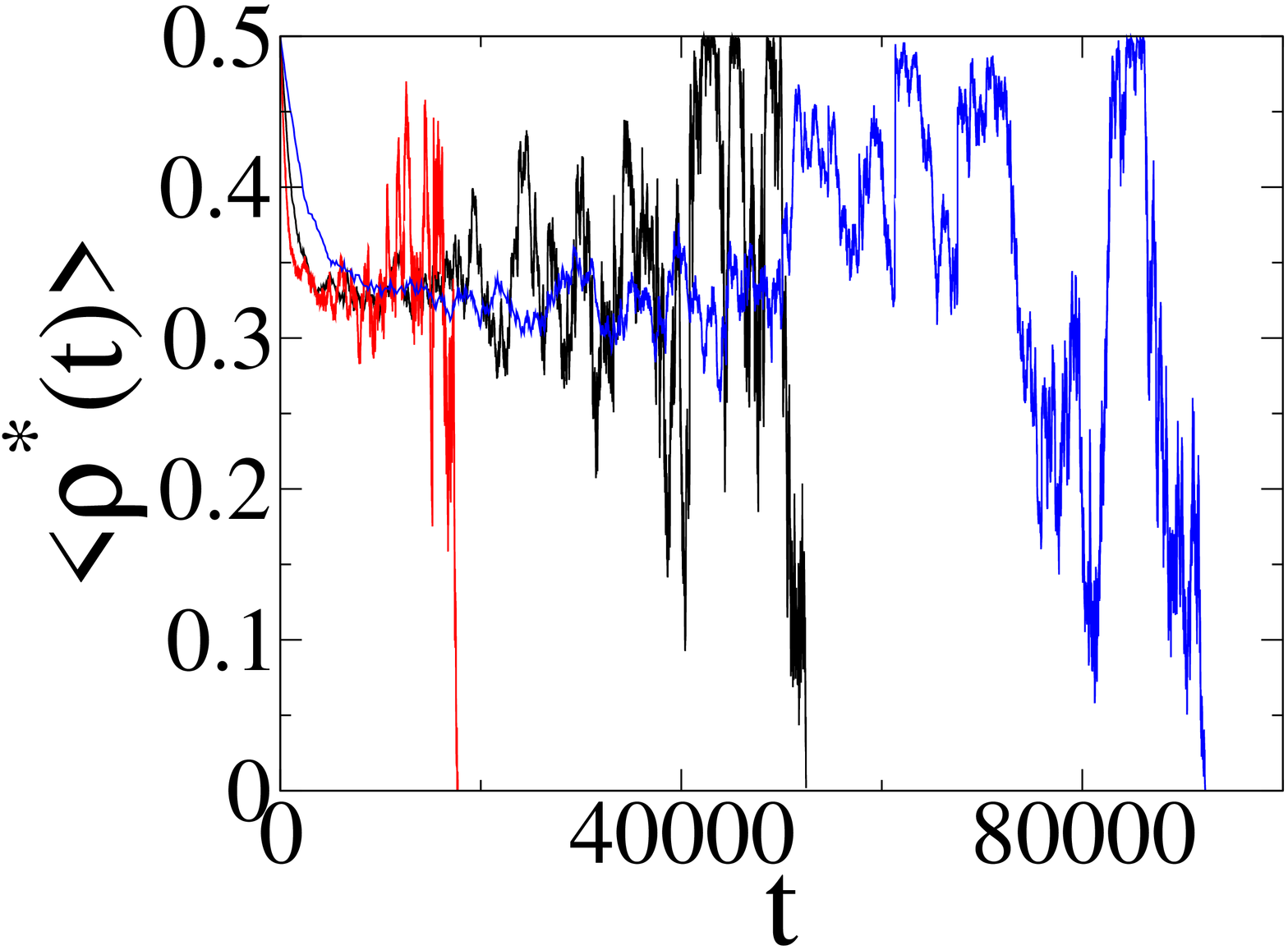}}\ 
  \subfloat{\label{usual_rhostar_randNm3}\includegraphics[draft=false,height=0.31\textwidth,angle=-90]{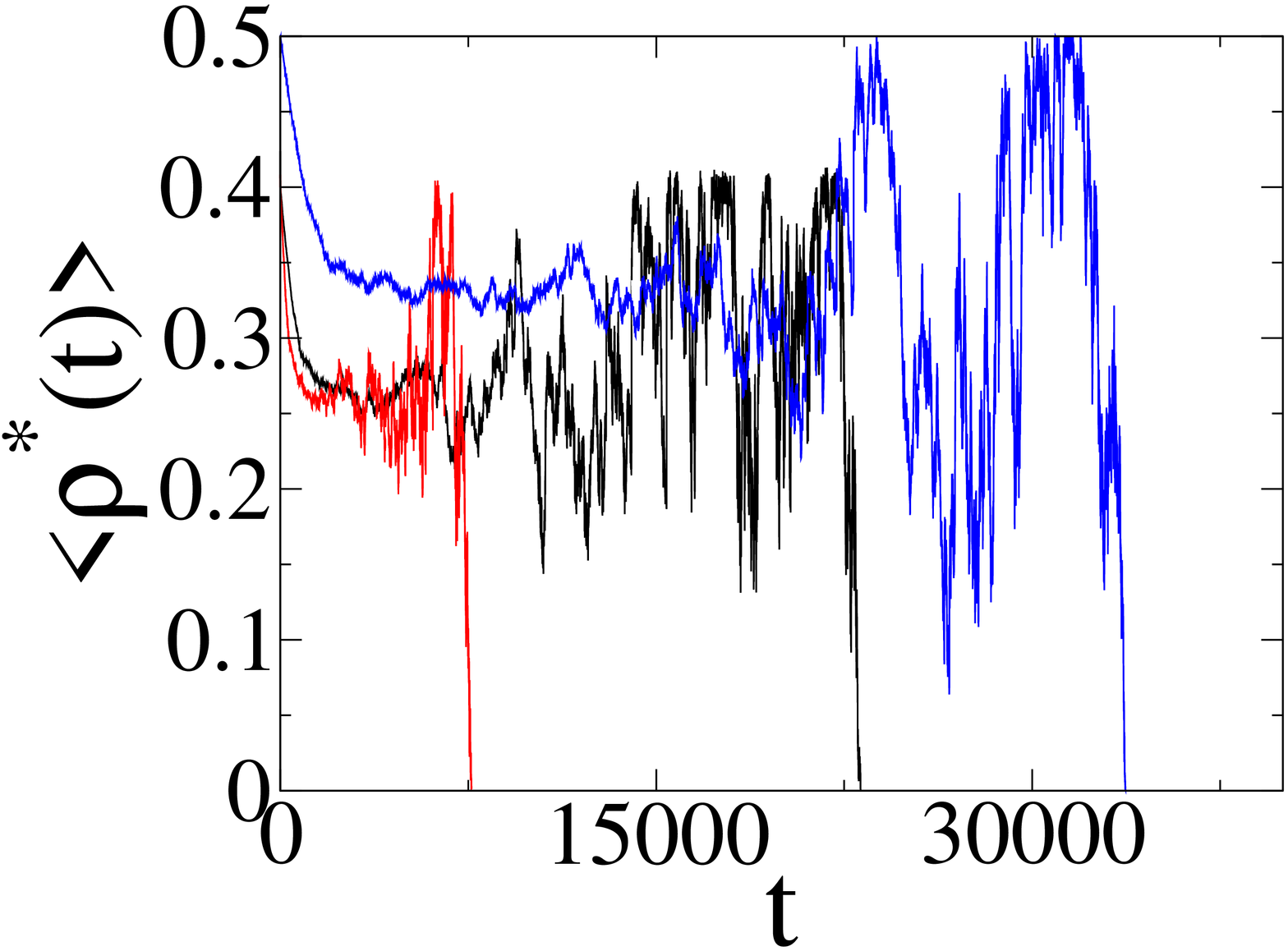}}\ 
  \subfloat{\label{usual_rhostar_BAm3}\includegraphics[draft=false,height=0.31\textwidth,angle=-90]{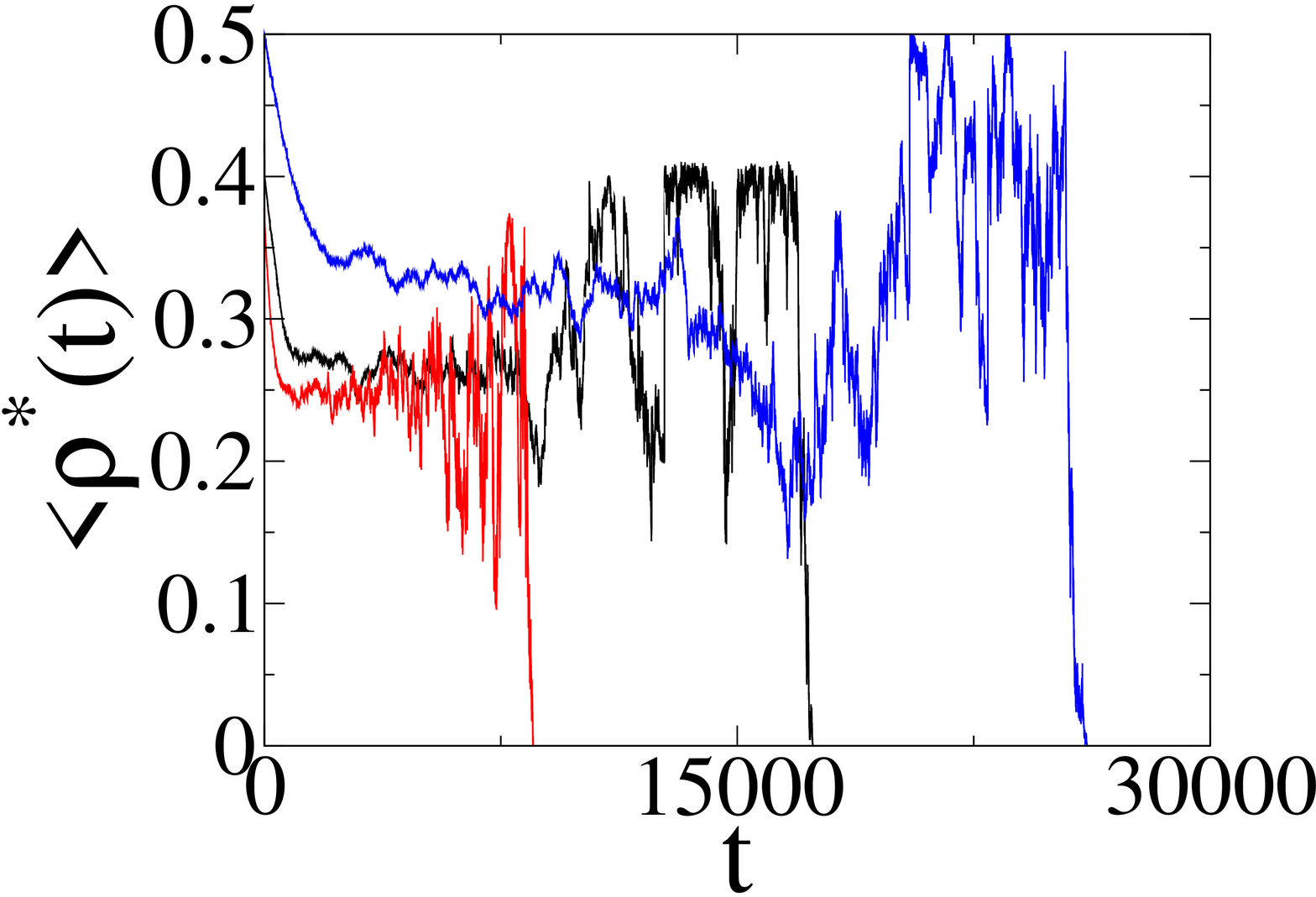}}\\
  \subfloat{\label{usual_rhosingle_meanf}\includegraphics[draft=false,height=0.31\textwidth,angle=-90]{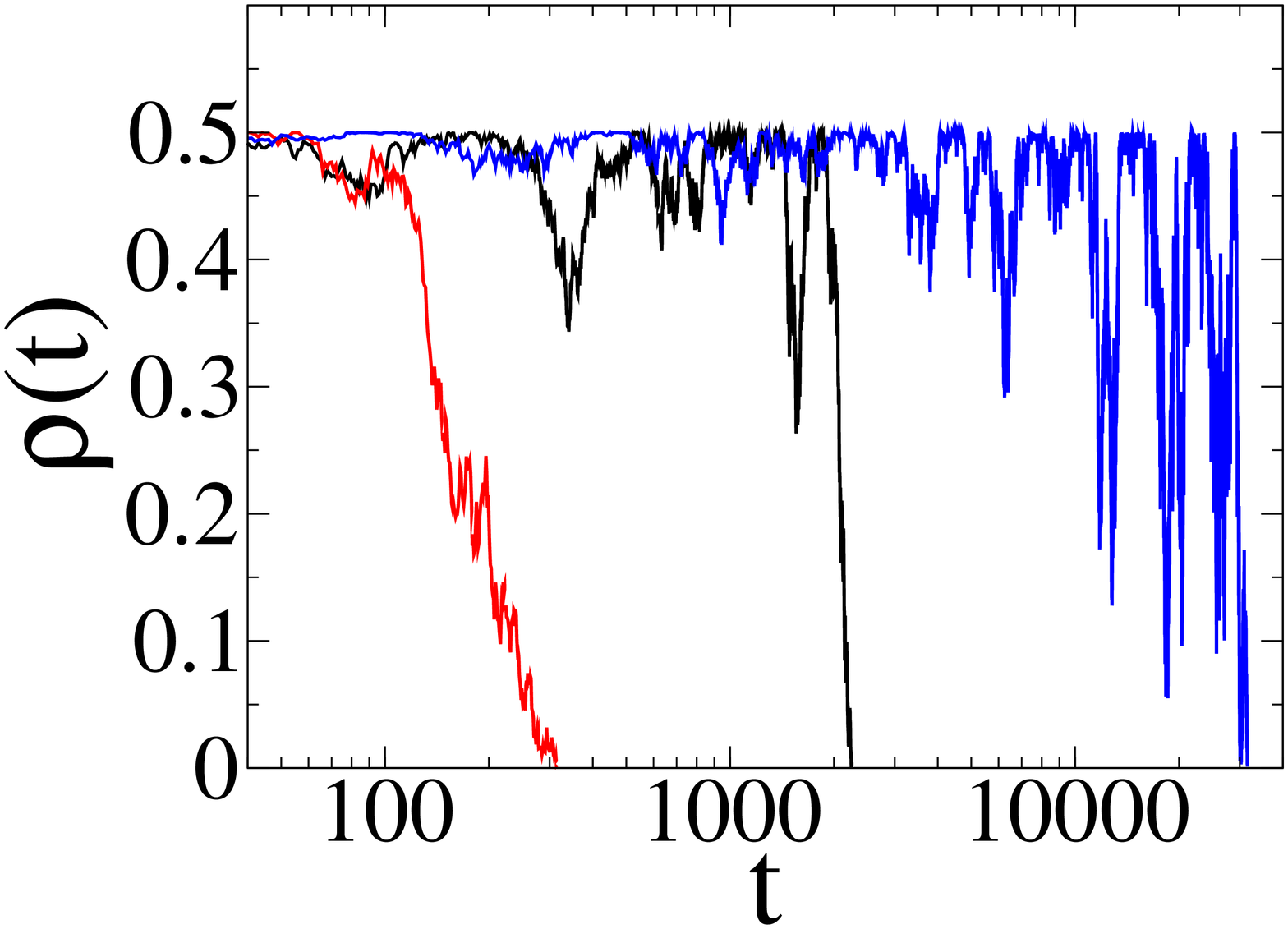}}\ 
  \subfloat{\label{usual_rhosingle_randNm3}\includegraphics[draft=false,height=0.31\textwidth,angle=-90]{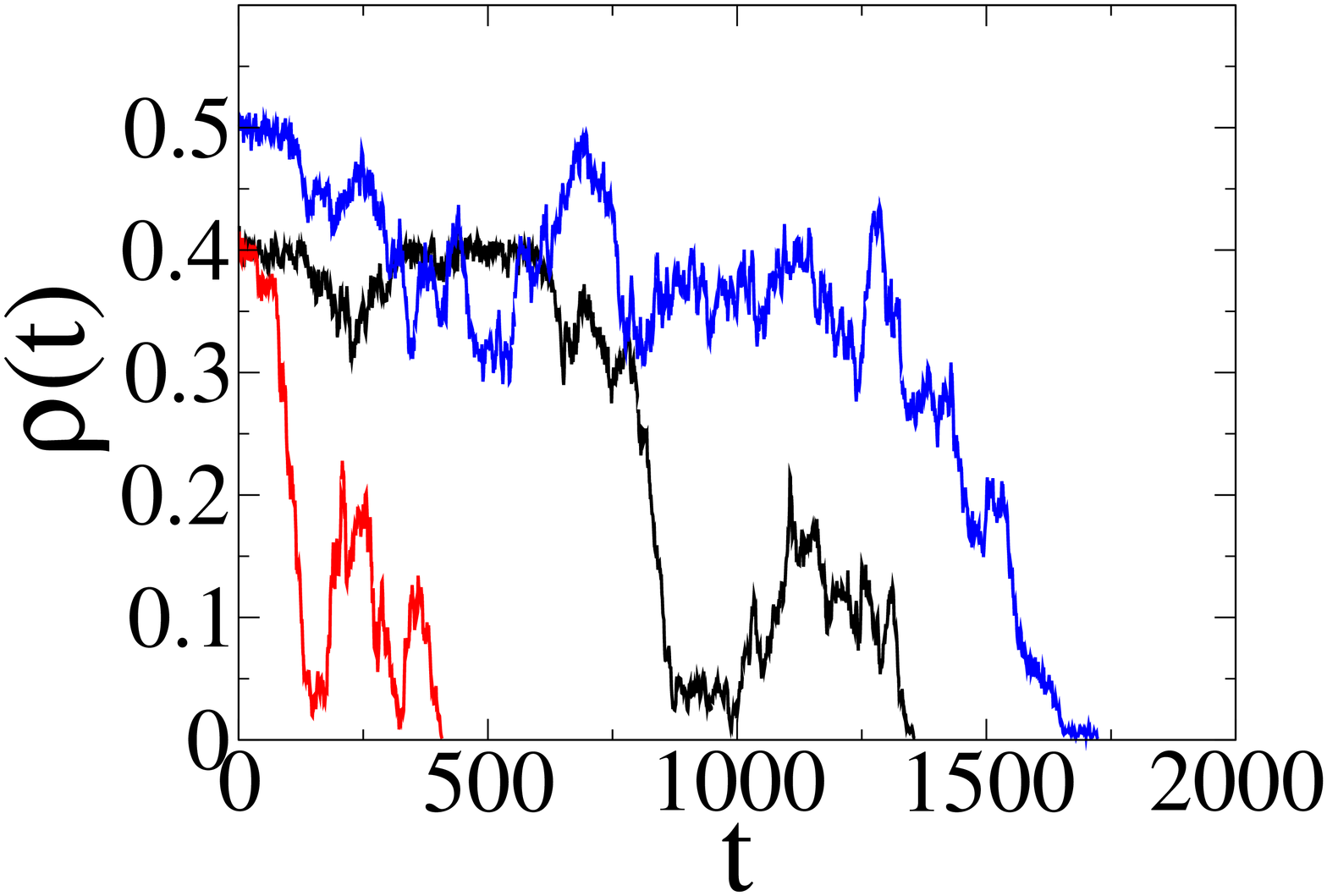}}\ 
  \subfloat{\label{usual_rhosingle_BAm3}\includegraphics[draft=false,height=0.31\textwidth,angle=-90]{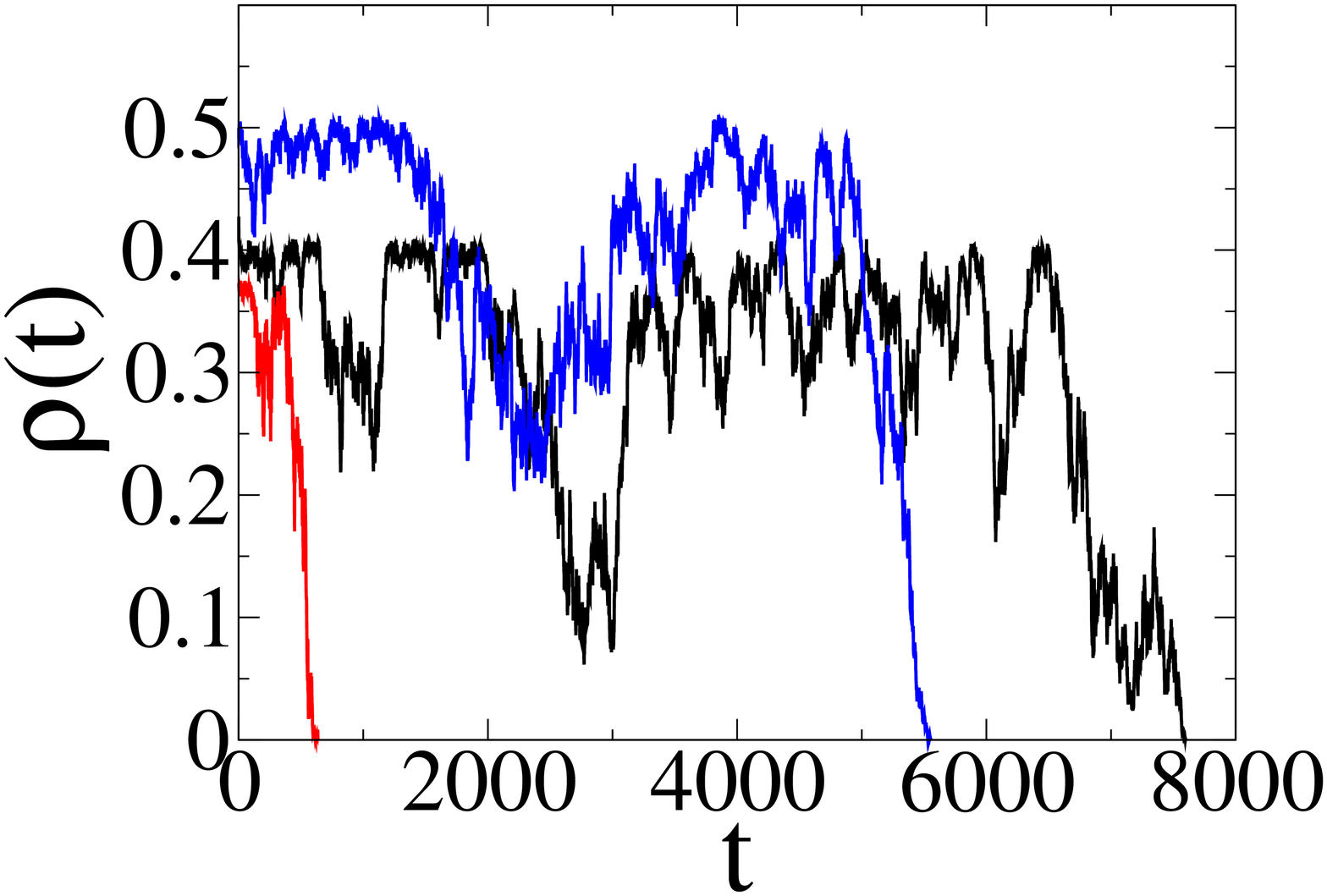}}\\
  \subfloat{\label{usual_surv_meanf}\includegraphics[draft=false,height=0.31\textwidth,angle=-90]{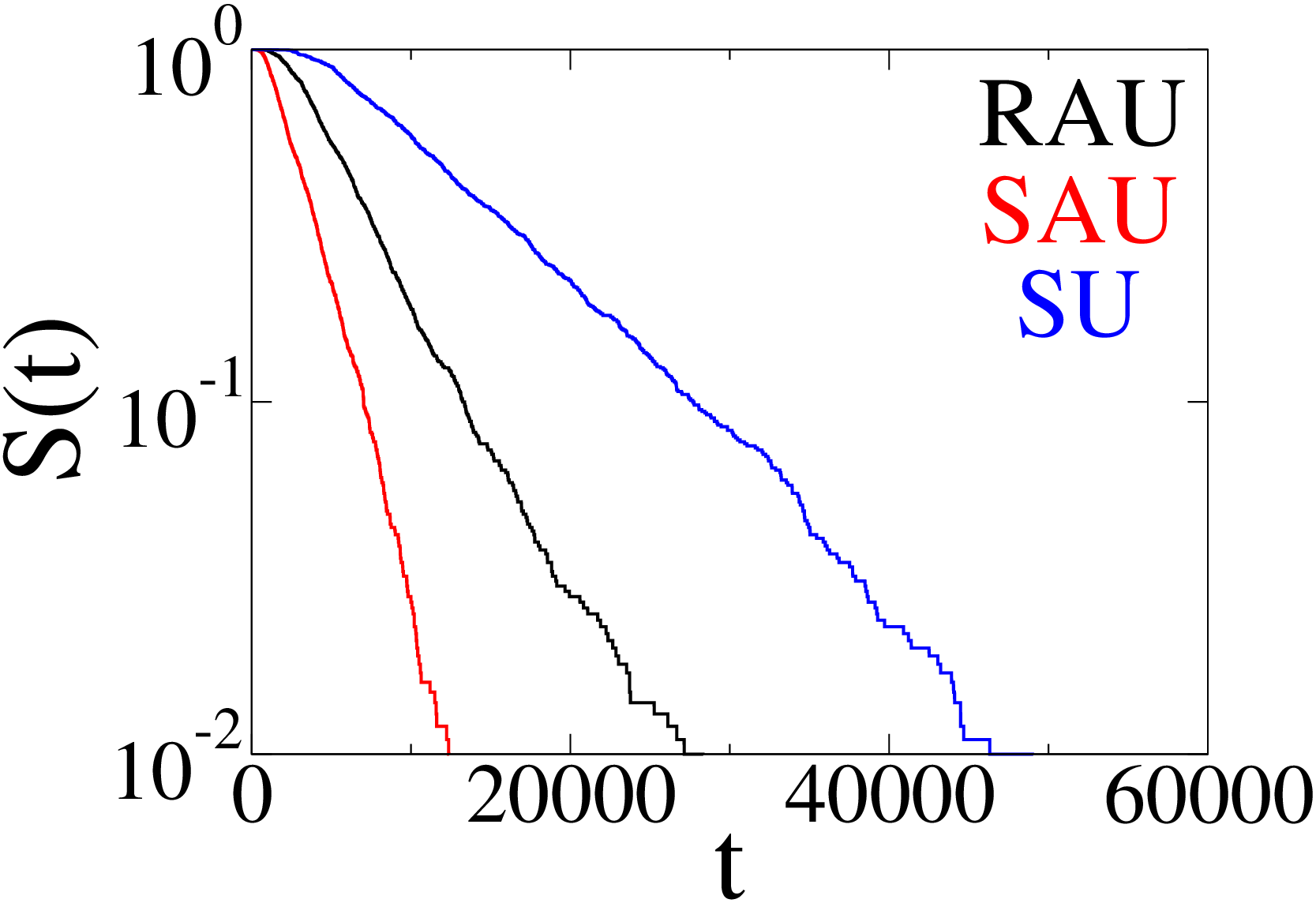}}\
  \subfloat{\label{usual_surv_randNm3}\includegraphics[draft=false,height=0.31\textwidth,angle=-90]{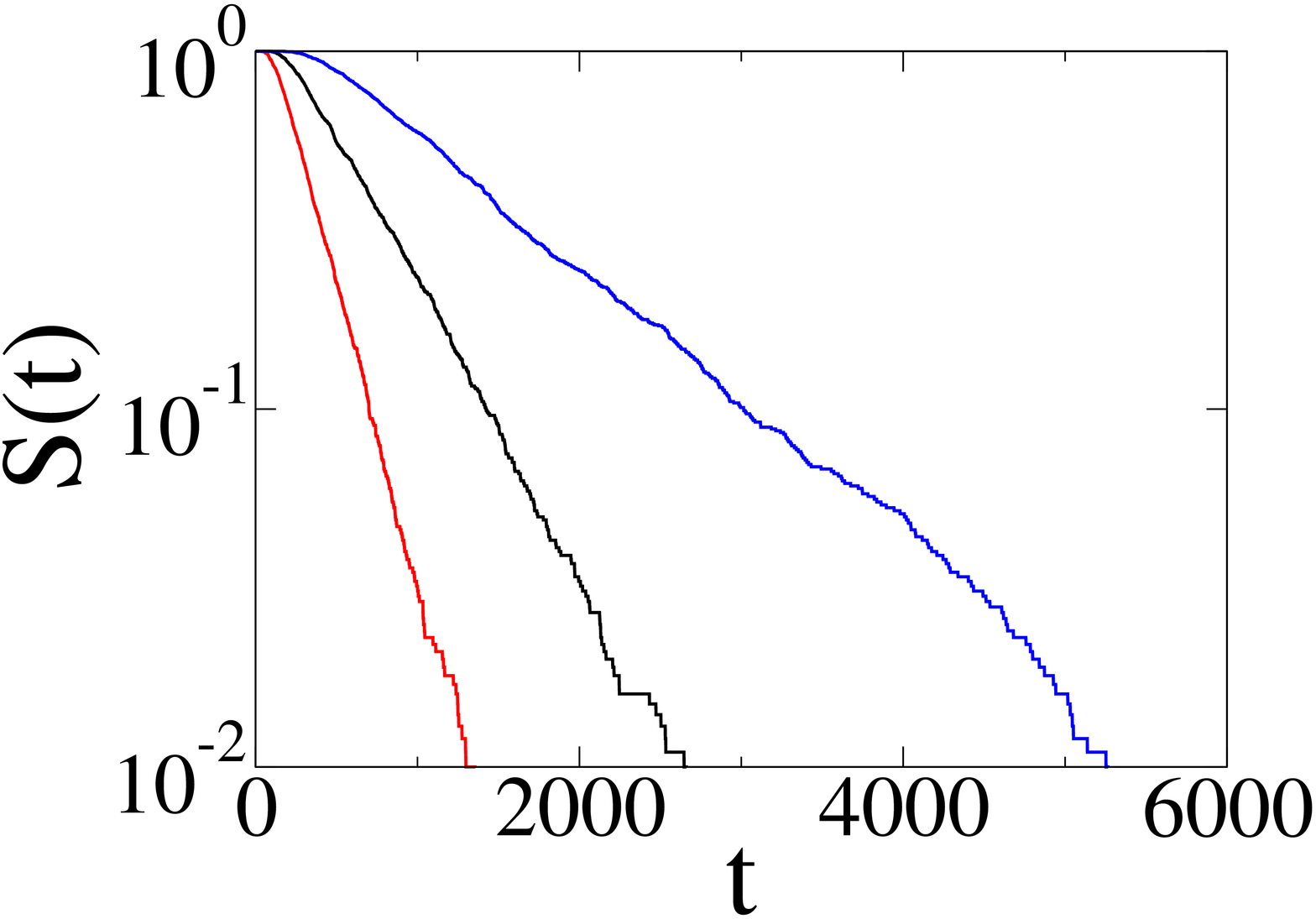}}\
  \subfloat{\label{usual_surv_BAm3}\includegraphics[draft=false,height=0.31\textwidth,angle=-90]{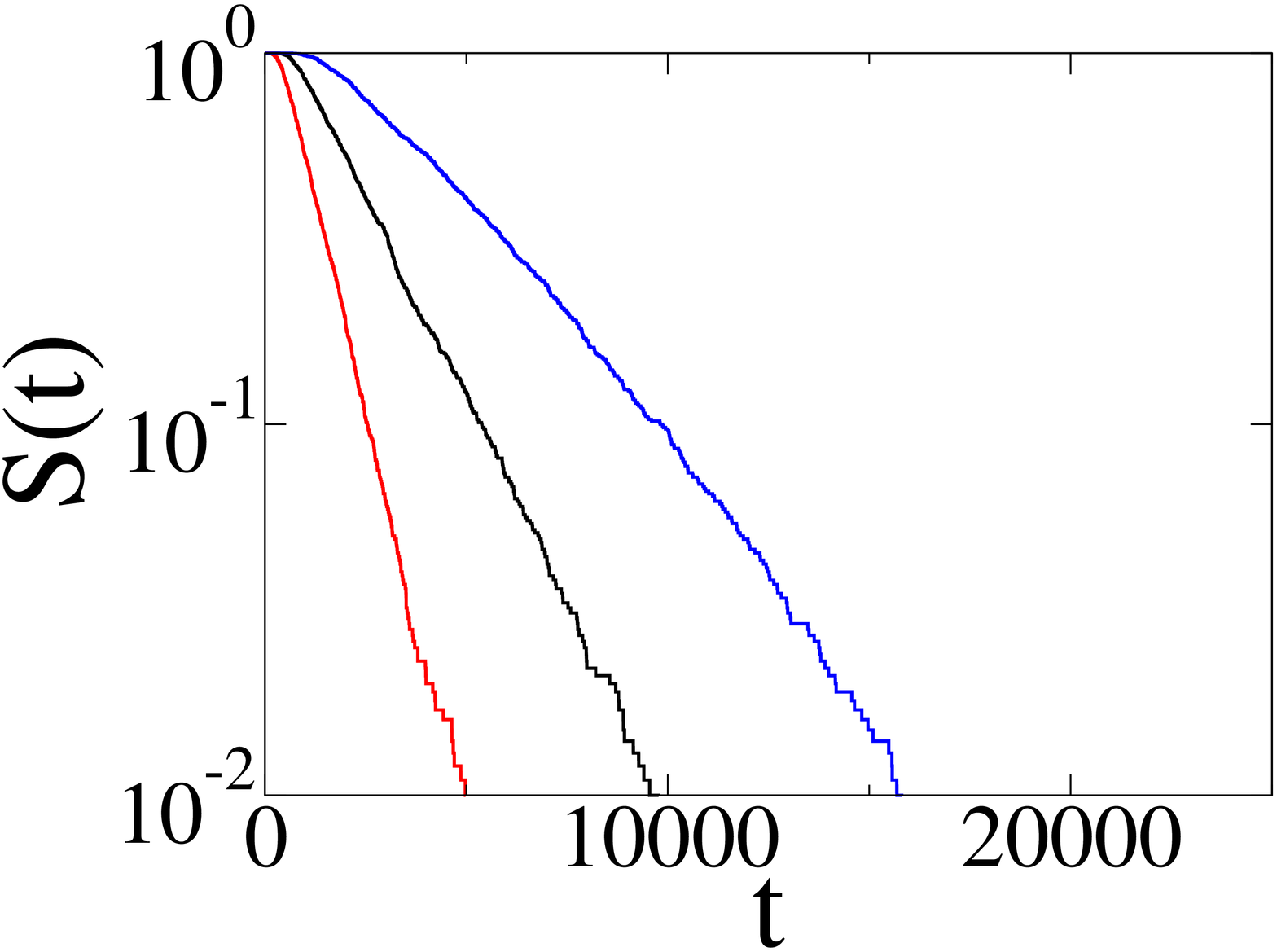}} 
  \caption{The voter model under the usual update rules (RAU in black, SAU in red and SU in blue) on different networks. All the averages where done over $1000$ realizations. The left column is for a complete graph, middle column for a random graph with average degree $\langle k \rangle=6$ and right column a scale-free graph with average degree $\langle k \rangle=6$. Top row contains plots for the average density of interfaces $\langle \rho \rangle$ with dashed lines at the value of the plateau that will only exist in the thermodynamic limit, second row shows the density of interfaces averaged only over surviving runs $\langle \rho^* \rangle$, third row shows the density of interfaces for single realizations and the bottom row contains the survival probability. System size is $N=1000$.}
  \label{results_usual}
\end{figure}

\begin{figure}
  \centering
  \subfloat{\label{usual_intert_meanf}\includegraphics[draft=false,height=0.31\textwidth,angle=-90]{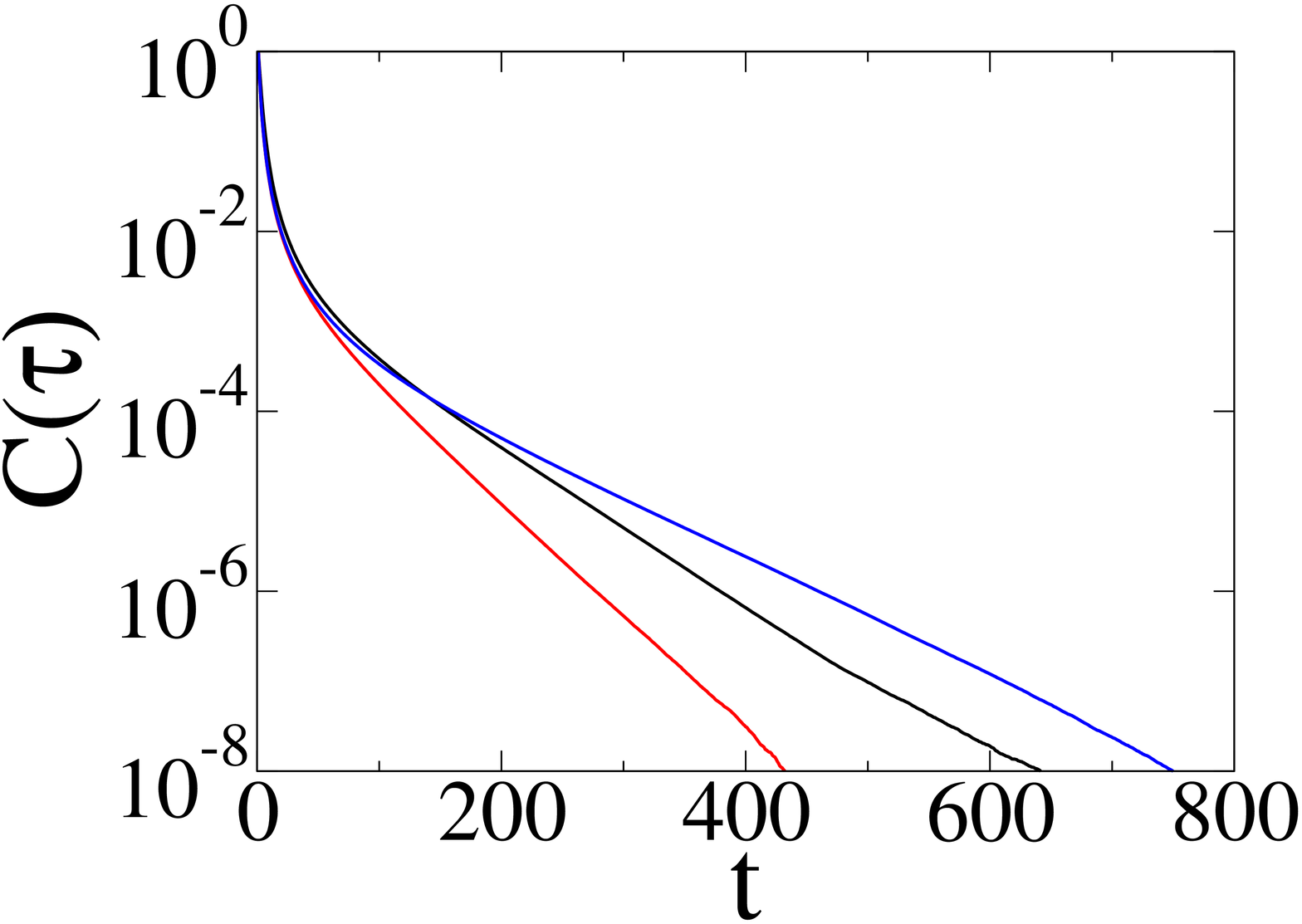}}\ 
  \subfloat{\label{usual_intert_randNm3}\includegraphics[draft=false,height=0.31\textwidth,angle=-90]{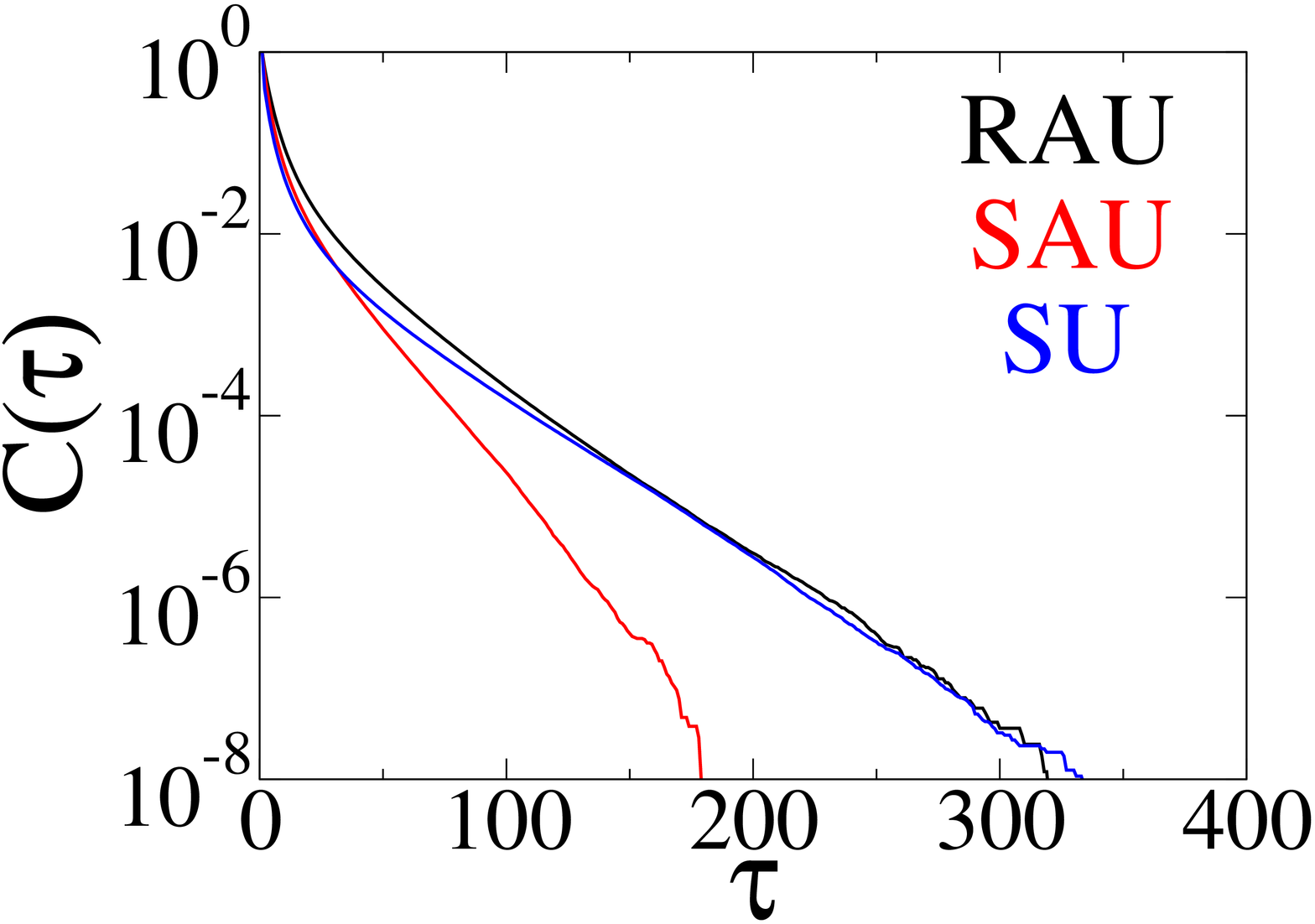}}\ 
  \subfloat{\label{usual_intert_BAm3}\includegraphics[draft=false,height=0.31\textwidth,angle=-90]{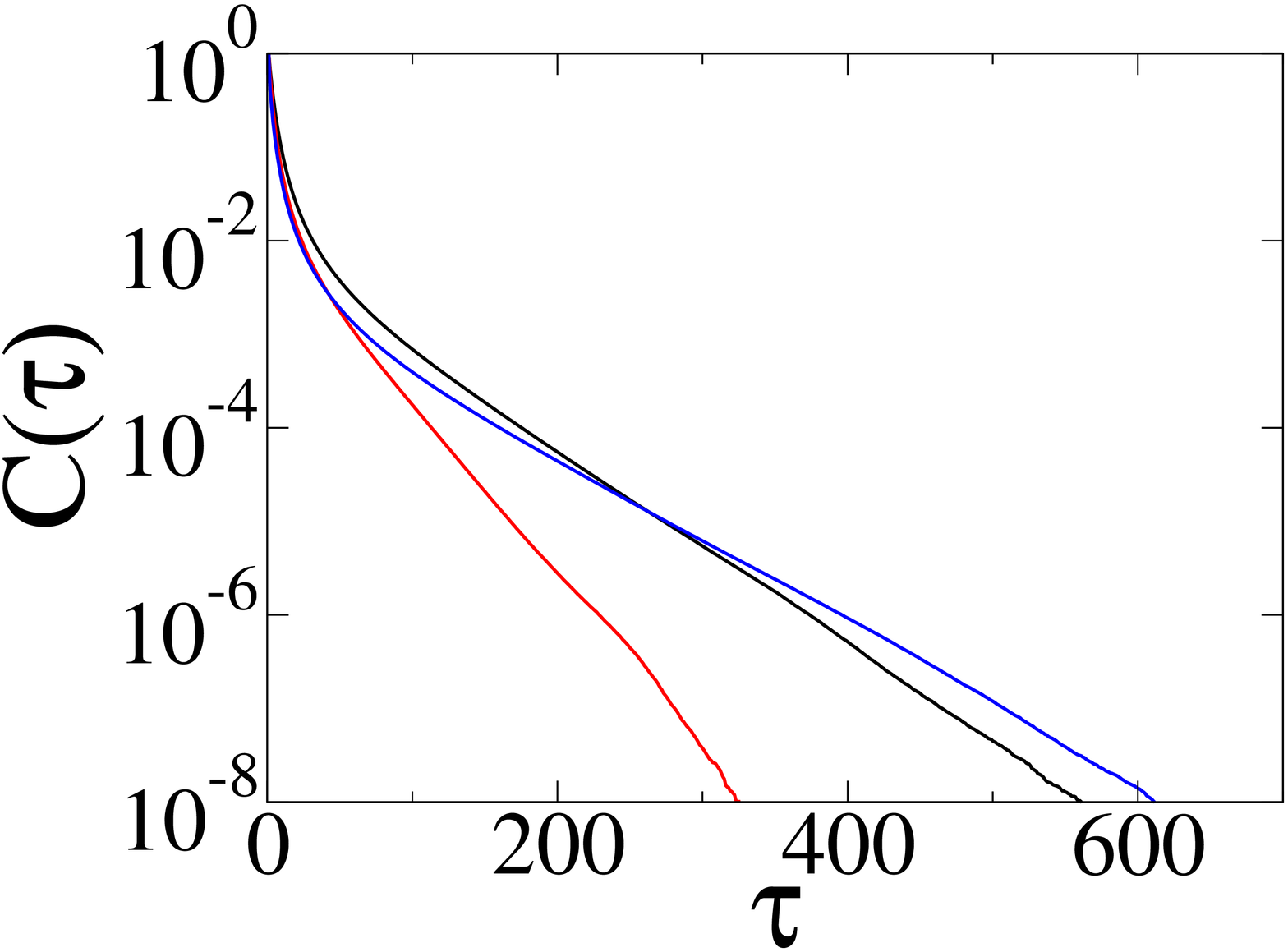}}
  \caption{Cumulative IET distributions for the voter model under the usual update rules (RAU in black, SAU in red and SU in blue) on different networks. All the averages where done over $1000$ realizations. Left plot is for a complete graph, middle plot for a random graph with average degree $\langle k \rangle=6$ and right plot for a scale-free graph with average degree $\langle k \rangle=6$. System size is $N=1000$.}
  \label{results_usual_cdet}
\end{figure}

In Fig.~\ref{results_usual} we can see the outcome of the simulations on a complete graph, a random graph of average degree $\langle k\rangle=6$  and on a scale-free graph of average degree $\langle k\rangle=6$. These figures include plots of the averaged density of active links $\langle \rho(t)\rangle$, the evolution of $\rho$ in a single realization and the survival probability $S(t)$. The cumulative IET distribution $C(\tau)$ is plotted for the three updates and the three different networks in Fig.\ref{results_usual_cdet}. The question of interest is whether $C(\tau)$ is Poisson-like or a more heterogeneous distribution.

Results for RAU, SAU and SU are plotted together for comparison purposes. We observe that the averaged density of active links $\langle \rho(t)\rangle$, the survival probability $S(t)$ and the tail of the cumulative IET distribution $C(\tau)$ display an exponential decay $\exp(-t/\tau(N))$, with a characteristic time that depends on the system size. These characteristic times have been extracted by fitting the data for many system sizes and computing the scaling behavior of $\tau(N)$. The results of this analysis are summarized in Table \ref{table_taus} for the different update rules and networks. Both the average density of interfaces $\langle\rho(t)\rangle$ and the survival probability $S(t)$ display the same characteristic time. This feature gives rise to the appearance of a plateau in the density of interfaces averages over surviving runs, as  $\langle\rho^*(t)\rangle=\langle\rho(t)\rangle/S(t)$, which is a signature of the system not being ordered by the dynamics.

\begin{table}
\centering
\small
\begin{tabular}{c|c|c|c|c|}
   & & RAU $\tau(N)$ & SAU $\tau(N)$ & SU $\tau(N)$ \\
  \hline
  \hline
 \multirow{2}{*}{CG} &  $\langle\rho\rangle|S(t)$ & $N/2$ & $0.23(4)N^{1.01(2)}$ & $0.9(1)N^{1.01(2)}$\\
  \cline{2-5}
  & $C$ & $0.63(7) N^{0.47(2)}$ & $0.33(4)N^{0.50(1)}$ & $0.6(1) N^{0.51(2)}$\\
  \hline
  \hline
 \multirow{2}{*}{RG $\langle k\rangle =6$} & $\langle\rho\rangle|S(t)$ & $0.57(7) N^{0.99(2)}$&$0.34(6) N^{0.97(2)}$ & $1.0(1) N^{1.01(2)}$\\
  \cline{2-5}
  & $C$ & $1.0(2) N^{0.47(2)}$&$0.38(6) N^{0.51(2)}$ & $0.74(9)N^{0.51(2)}$\\
  \hline
  \hline
 \multirow{2}{*}{SFG $\langle k\rangle =6$} & $\langle\rho\rangle|S(t)$ & $0.25(5)N^{0.88(2)}$& $0.19(3)N^{0.92(5)}$& $1.6(4) N^{0.84(3)}$\\
  \cline{2-5}
  & $C$ &$0.35(7) N^{0.52(2)}$ &$0.18(7)N^{0.53(4)}$ &$1.0(3) N^{0.43(3)}$ \\
  \hline
\end{tabular}
\caption{System size dependence of the characteristic times in the density of active links, $\langle\rho(t)\rangle$ and in the cumulative distribution of interevent times, $C(\tau)$, for different network topologies and node update rules. CG stands for complete graph, RG for random graph and S-FG for scale-free graph.}
\label{table_taus}
\end{table}
Our results indicate that the voter model has the same qualitative dynamical behavior under RAU, SAU and SU node update rules. These results can be summarized as follows:
\begin{description}
 \item {\textit{Density of active links: }}{\\$\langle \rho(t) \rangle$: For the ensemble average over all realizations we find an exponential decay in $$\langle \rho(t) \rangle \propto e^{-t/\tau(N)}$$ with a characteristic time that scales as $\tau(N)\propto N$ for a complete graph and random graphs. For the case of Barab\'asi-Albert scale-free graphs the scaling is compatible with the analytical result $\tau(N)\propto N/\log(N)$ \cite{conserv_sucheki,vazquez_eguiluz,voter_redner}. We can see that the characteristic time diverges with the system size, so that $\langle \rho(t) \rangle$ remains constant in the infinite size limit for any of these networks. This is telling us already that the system is not reaching an ordered state in the thermodynamic limit.\\$\langle\rho^*(t)\rangle$: Decays exponentially until it reaches a plateau. The plateau height is independent of the system size, meaning that, on average, the realizations that have not yet reached an absorbing state stay at a disordered state with a
finite and large fraction of active links.}
 \item{\textit{Survival probability: }}{\\$S(t)$: The survival probability decays exponentially, $$S(t)\propto e^{-t/\tau(N)},$$ with the same characteristic time as $\langle \rho(t) \rangle$. Thus when combining $\langle \rho(t) \rangle/S(t)=\langle\rho^*(t)\rangle$ we find a constant value for $\langle\rho^*(t)\rangle$. The mean times to reach consensus for finite systems are well defined. In the infinite size limit, as $\tau(N)$ diverges with the system size, we can conclude again that the system does not reach an ordered state and the survival probability is just equal to one for all times in the thermodynamic limit.}
 \item{\textit{Cumulative IET distribution: }}{\\$C(\tau)$: This distribution shows an exponential tail, $$S(t)\propto e^{-t/\tau'(N)}$$ indicating that there is a well defined average IET. The characteristic time in the exponential tail scales approximately $$\tau'(N)\propto\sqrt{N}.$$}
\end{description}
These are the features shared by all standard node update rules. There are also differences, since the precise characteristic times and the plateau heights of $\langle\rho(t)\rangle$ and $\langle\rho^*(t)\rangle$ depend on the update rule. See left column in Fig.~\ref{results_usual} where the plateaus for the different update rules are plotted with a dashed black line. It is clear that the difference between RAU and SAU update rule lies in correlations that will be present in SAU and not in RAU. For the case of SU, the differences come from the fact that for this update rule the dynamics is purely discrete. Still the main result is that the qualitative behavior is the same: for these three update rules the system remains, in the thermodynamic limit, in an active disordered configuration for the voter model dynamics in a complete graph and in complex networks of infinite effective dimensionality such as Erd\"os-Renyi and Barab\'asi-Albert networks. Also the activity patterns
are very homogeneous, with a well defined IET.

\section{Update rules for heterogeneous activity patterns}\label{coev_update_rule}

A set of $N$ agents are placed on the nodes of a network of interaction, as was explained generally for agent based models in Sect. 3. Each agent $i$ is characterized by its state $x_i$ and an internal variable that we will call \textit{persistence time} $\tau_i$. For any given interaction model (Ising, voter, contact process, ...), the dynamics is as follows: at each time step,
\begin{enumerate}
 \item with probability $p(\tau_i)$ each agent $i$ becomes active, otherwise it stays inactive;
 \item active agents update their state according to the dynamical rules of the particular interaction model;
 \item all agents increase their persistence time $\tau_i$ in one unit
\end{enumerate}
The persistence time measures the time since the last event for each agent. Typically an event is an interaction (\textit{exogenous update}: active agents reset $\tau =0$ after step (ii)) or a change of state (\textit{endogenous update}: only active agents that change their state in step (ii) reset $\tau =0$).

There are two interesting limiting cases of this update when $p(\tau)$ is independent of $\tau$: when $p(\tau)=1$, all agents are updated synchronously; when $p(\tau)=1/N$, every agent will be updated on average once per $N$ unit time steps. The latter corresponds to the usual random asynchronous update (RAU). We are interested in non-Poissonian activation processes, with probabilities $p(\tau)$ that decay with $\tau$, that is, the longer an agent stays inactive, the harder is to activate. To be precise, we will later consider that
\begin{equation}
p(\tau) = \frac{b}{\tau}~,\label{pdetau}
\end{equation}
where $b$ is a parameter that controls the decay with $\tau$. This mechanism is similar to the reinforcement dynamics explained in Ch.\textbf{Kun Zhao, Márton Karsai and Ginestra Bianconi. Models, entropy and information of temporal social networks}. An alternative mechanism, also based on a process with memory is the Hawkes process discussed in Ch. \textbf{Naoki Masuda, Taro Takaguchi, Nobuo Sato and Kazuo Yano. Self-exciting point process modeling of conversation event sequences}

We expect the IET distribution $M(t)$ to be related to the activation probability $p(\tau)$. Neglecting the actual dynamics and assuming that at each update event, the agent changes state we can find an approximate relation between $M(t)$ and $p(\tau)$. Recall that $M(t)$ is the probability that an agent changes state (updating and changing state coincide in this approximation) $t$ timesteps after her last change of state. Therefore the probability that an agent has not changed state in $t-1$ timesteps is $1-\sum_{j=1}^{t-1}M(j)$ and the probability of changing state having persistence time $t$ is $p(t)$. Therefore we can write for $t$ larger than one:
\begin{equation}
\left(1-\sum_{j=1}^{t-1}M(j)\right)p(t)=M(t),
\end{equation}
with $p(1)=M(1)$. Taking the continuous limit and expressing this equation in terms of the cumulative IET distribution we obtain
\begin{equation}
d\ln(C(t))=-p(t)dt.\label{cdetau}
\end{equation}
Setting $p(\tau)=b/\tau$ the cumulative IET distribution decays as a power law $C(t)\sim t^{-\beta}$ with $\beta=b$. Numerical simulations show that this approximation holds for the voter model on a fully connected network for endogenous updates and for a small range of $b$-values in the exogenous update for any topology of the ones considered in this study.

The modification of the model is investigated more exhaustively for the case in which the cumulative IET distribution is set to a power law $C(\tau)\propto\tau^{-\beta}$, but any distribution $C(\tau)$ can be plugged into the definition of $p(\tau)$ (Eq.~(\ref{cdetau})). In fact the case $\beta=1$ will be studied in more detail.

\begin{figure}
 \centering
 \includegraphics[draft=false,width=\textwidth]{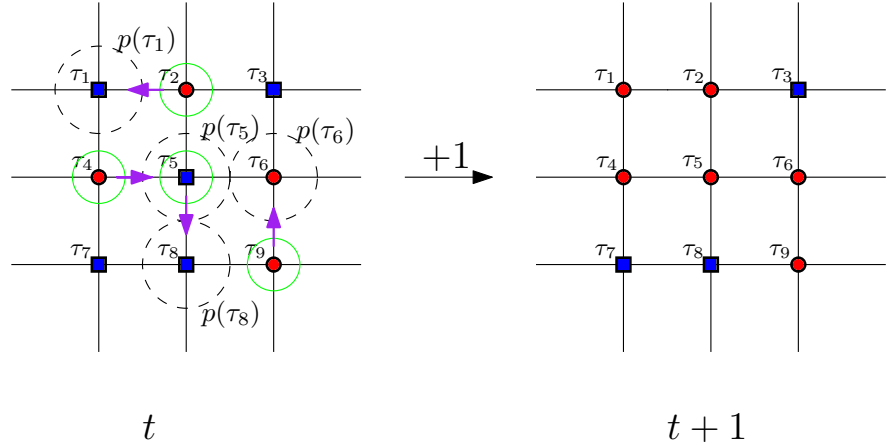}
 \caption{Example of the new update rule. Every agent gets updated with her own probability $p(\tau_i)$, being $\tau_i$ her persistence time. The two possible states of the nodes are represented by blue squares and red circles. The node or nodes inside a black dashed circle are the ones that are updated. The nodes inside a green circle are the randomly chosen neighbors for the interaction and the purple arrow tells in which direction the state will be copied.}
 \label{memU}
\end{figure}

When applied to the voter model the new update rule we changes the transition rates for node-dependent rates that are function of the persistence time of each node.

\subsection{Application to the voter model}

First of all, and to have a better idea of the kind of dynamics that arise from the new update rules, we exemplify them in Tables (\ref{tab:rau})-(\ref{tab:endo}). In those Tables we show snapshots of the evolution of the voter model under the different update rules on a square lattice. In particular we show the configuration of nodes states, times since the last change of state and time since last update at different points in time: for RAU (Table \ref{tab:rau}), for exogenous update (Table \ref{tab:exo})  and for endogenous update (Table \ref{tab:endo}).

\begin{table}
\centering
\begin{tabular}{|cccc|}
\hline
Time&Configuration&Time since state change&Time since update\\
\hline \hline
$t=1$&\includegraphics[width=0.23\textwidth]{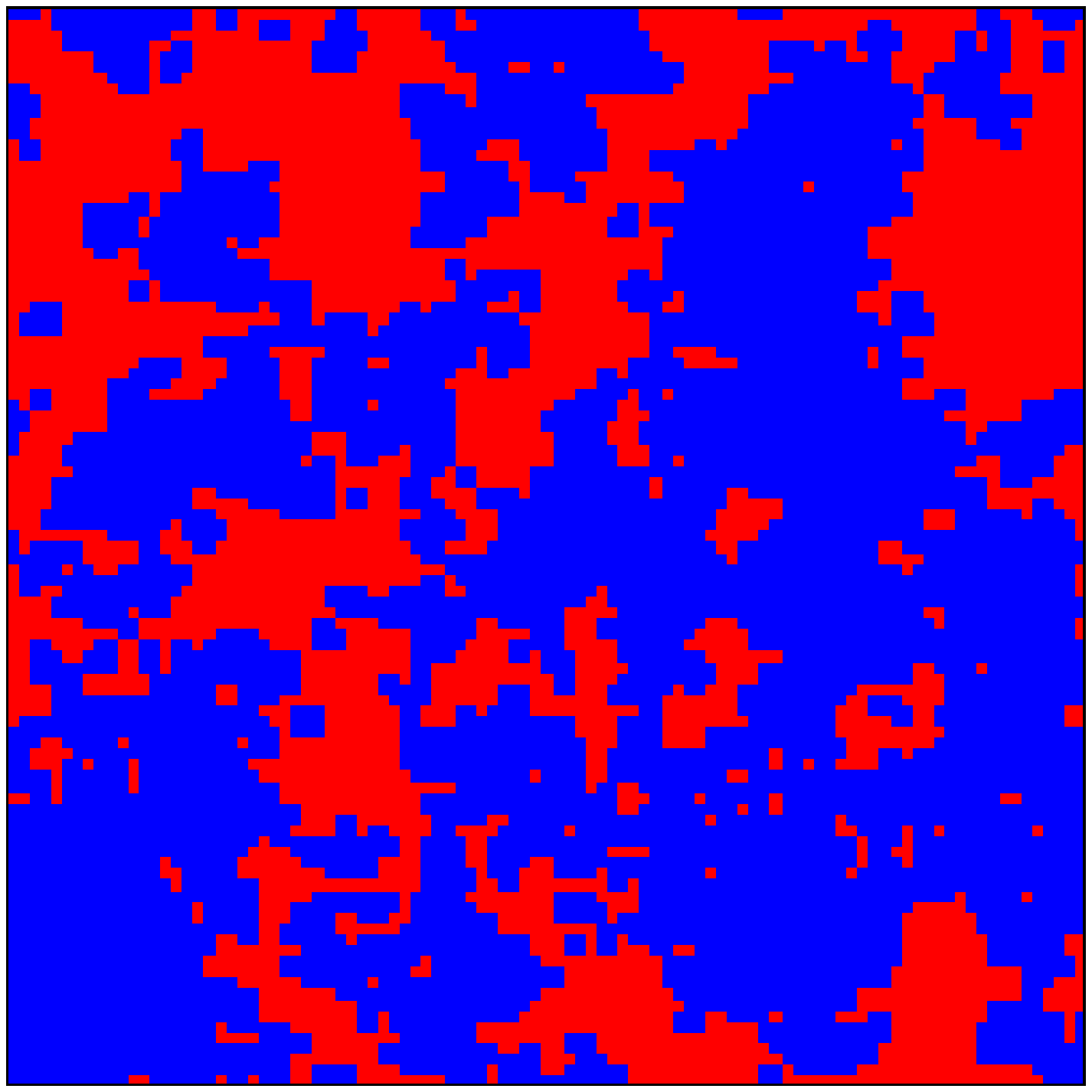}&\includegraphics[width=0.23\textwidth]{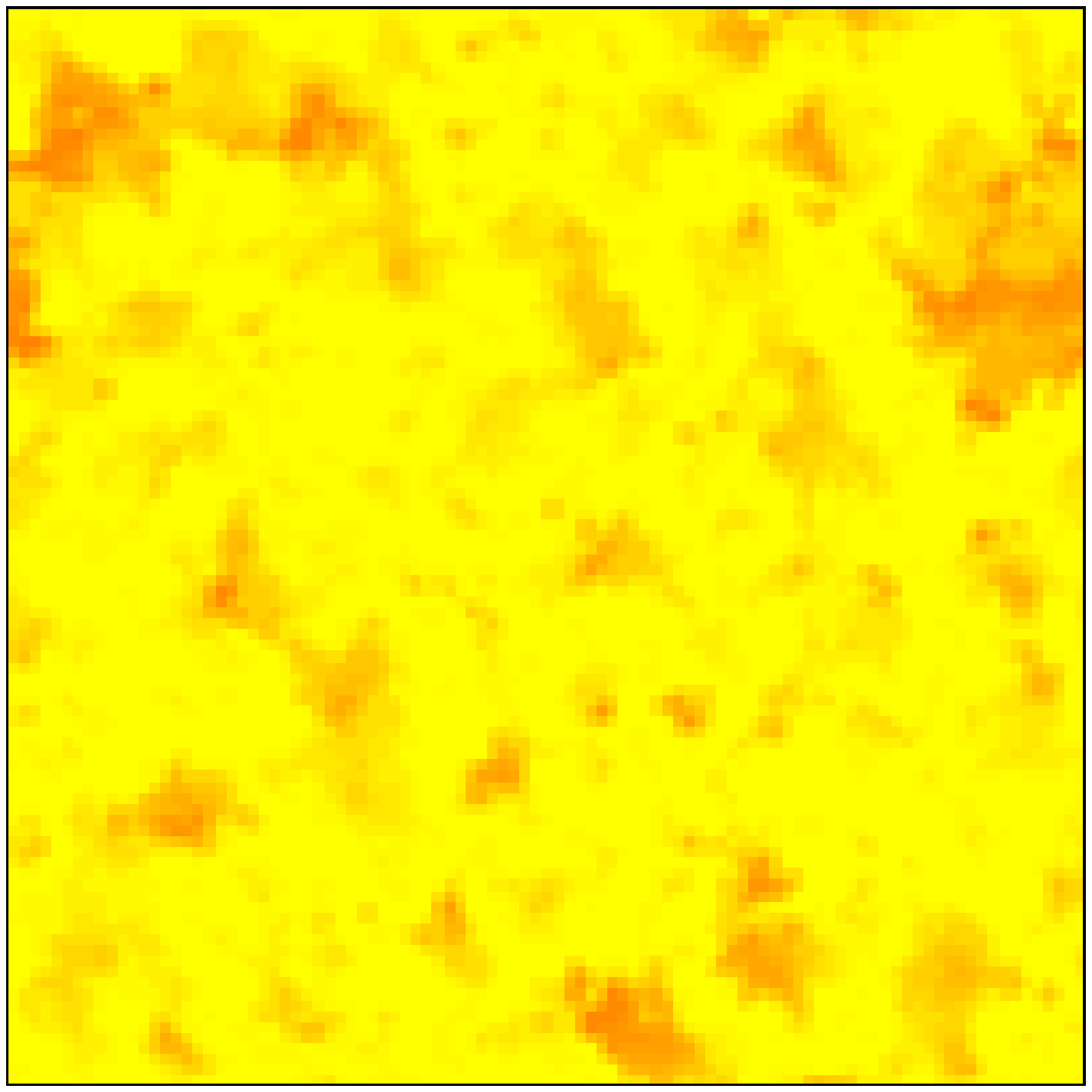}&\includegraphics[width=0.23\textwidth]{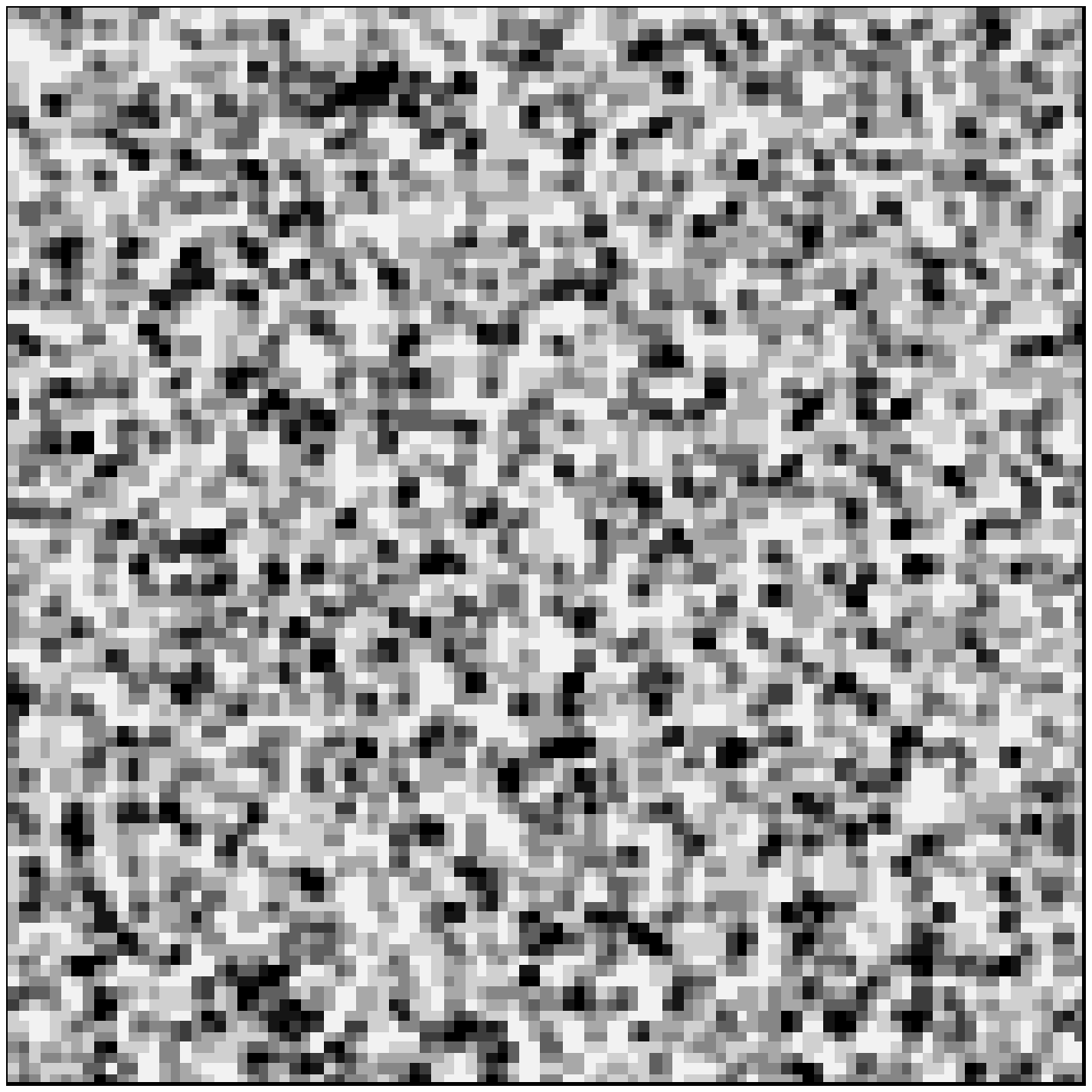}\\
\hline
$t=10$&\includegraphics[width=0.23\textwidth]{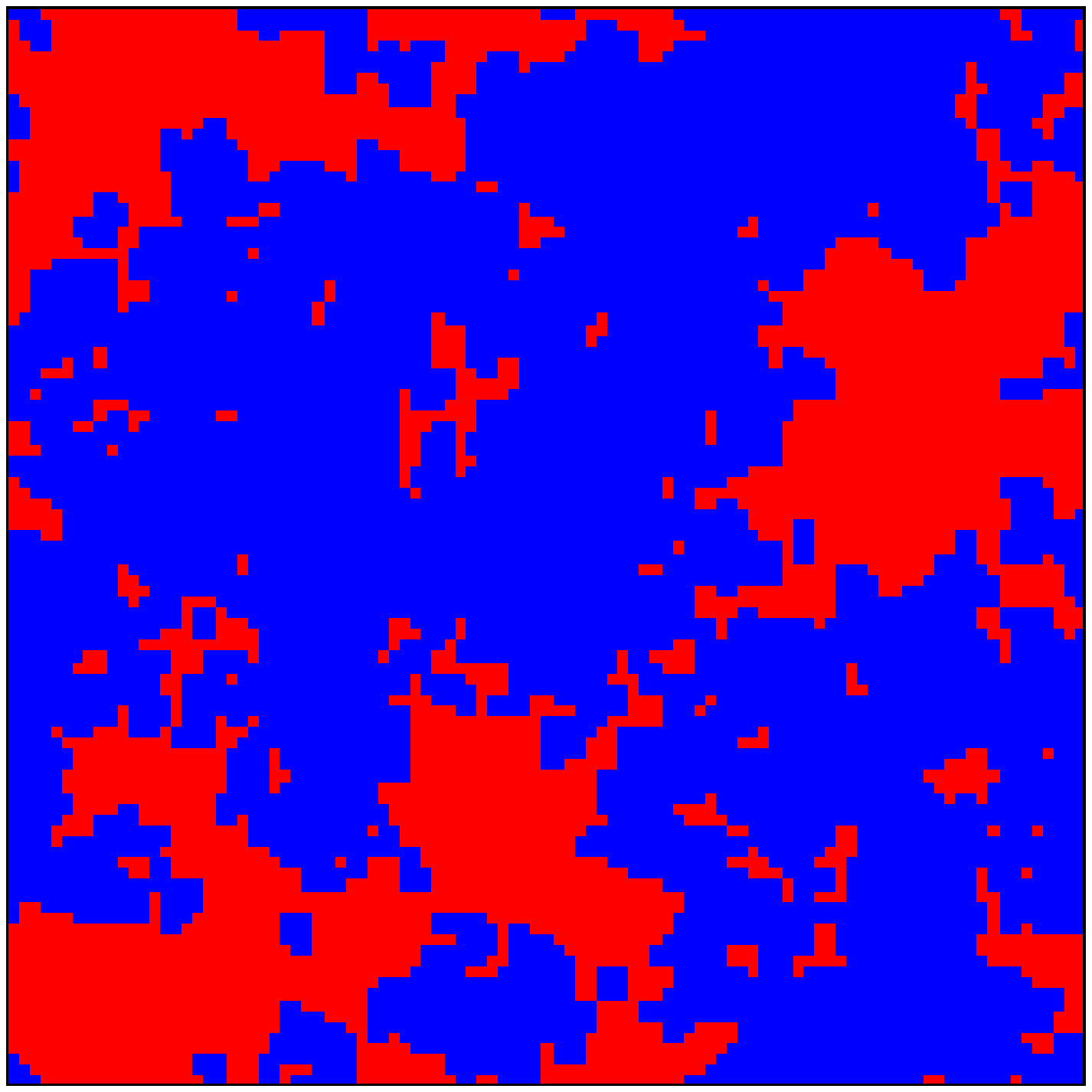}&\includegraphics[width=0.23\textwidth]{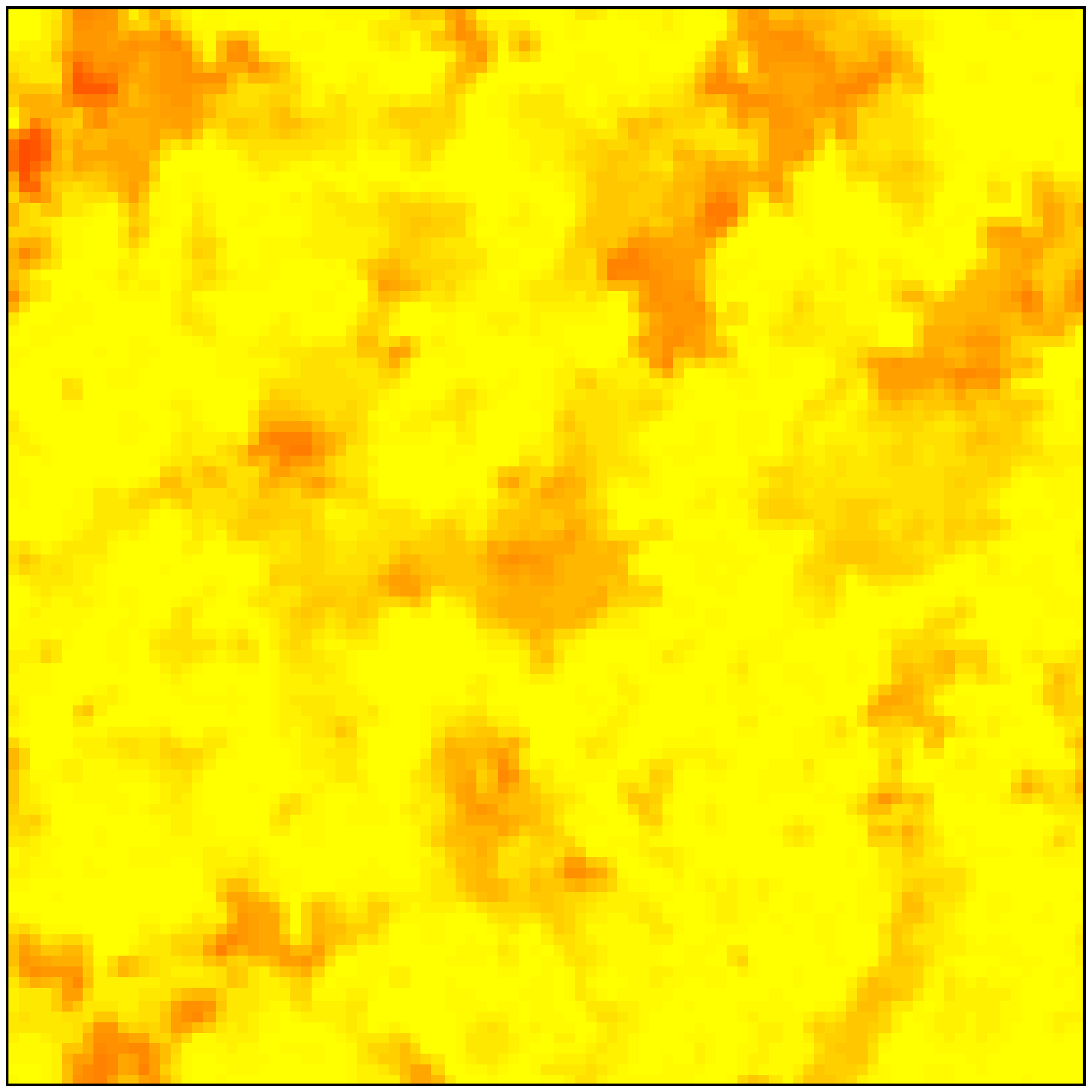}&\includegraphics[width=0.23\textwidth]{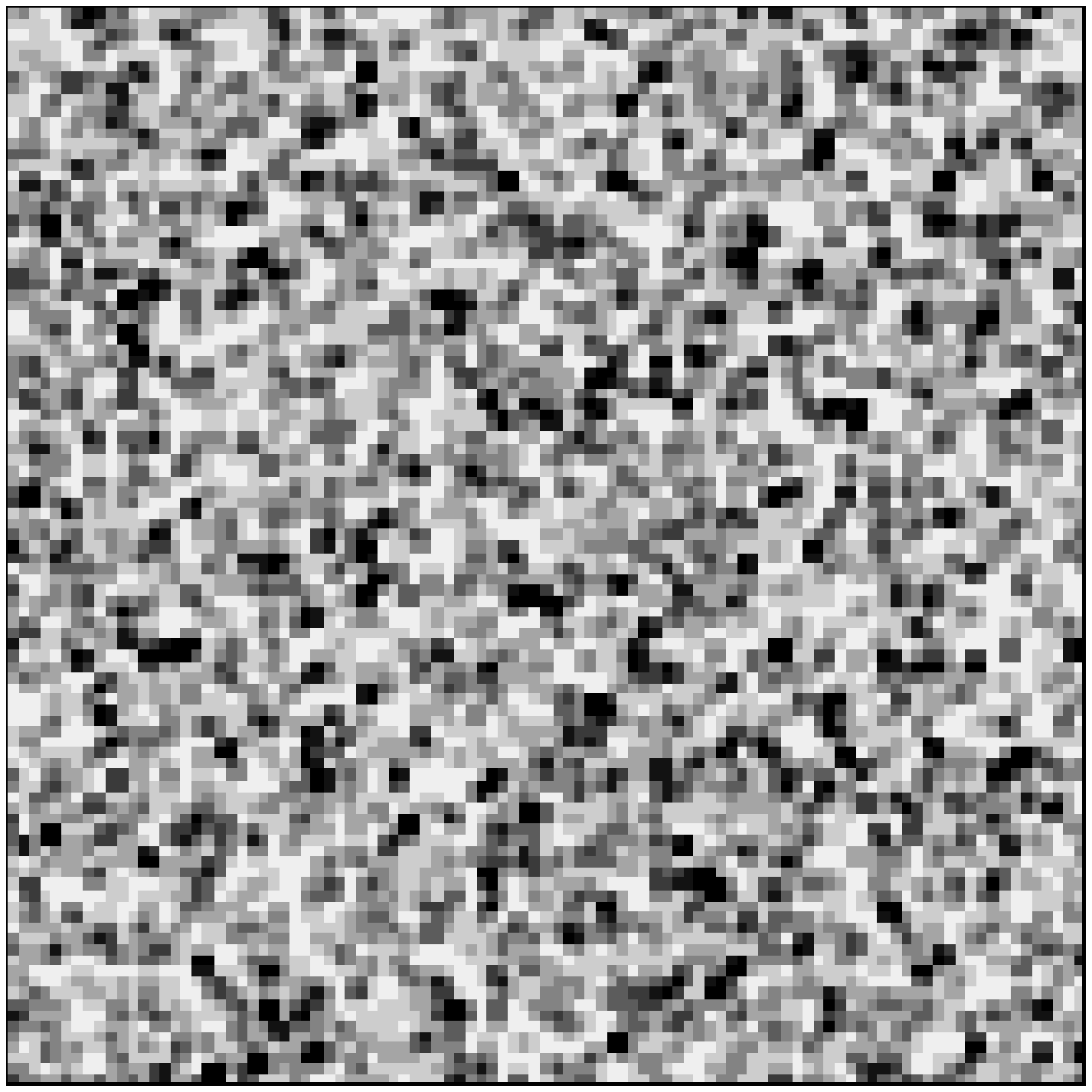}\\
\hline
$t=100$&\includegraphics[width=0.23\textwidth]{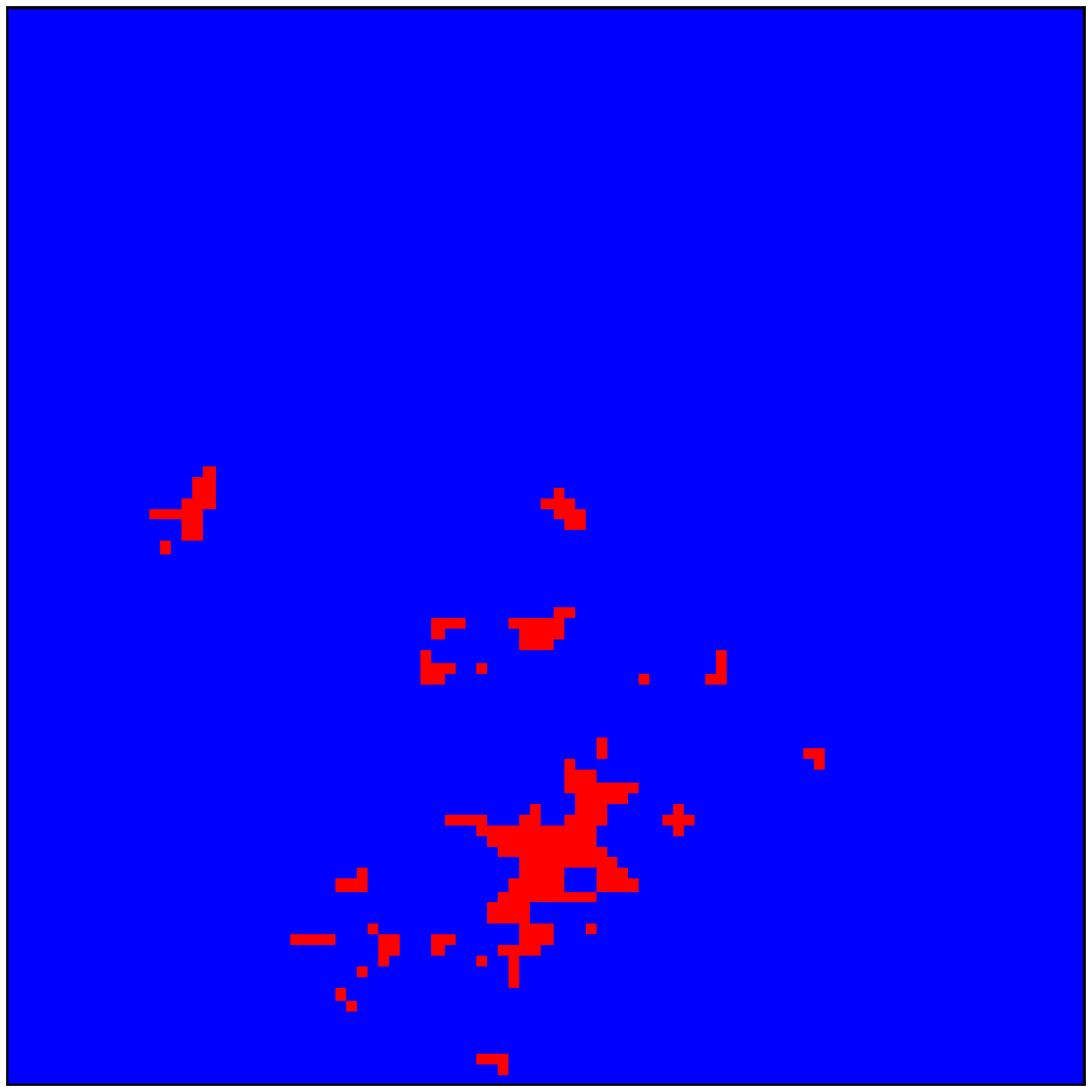}&\includegraphics[width=0.23\textwidth]{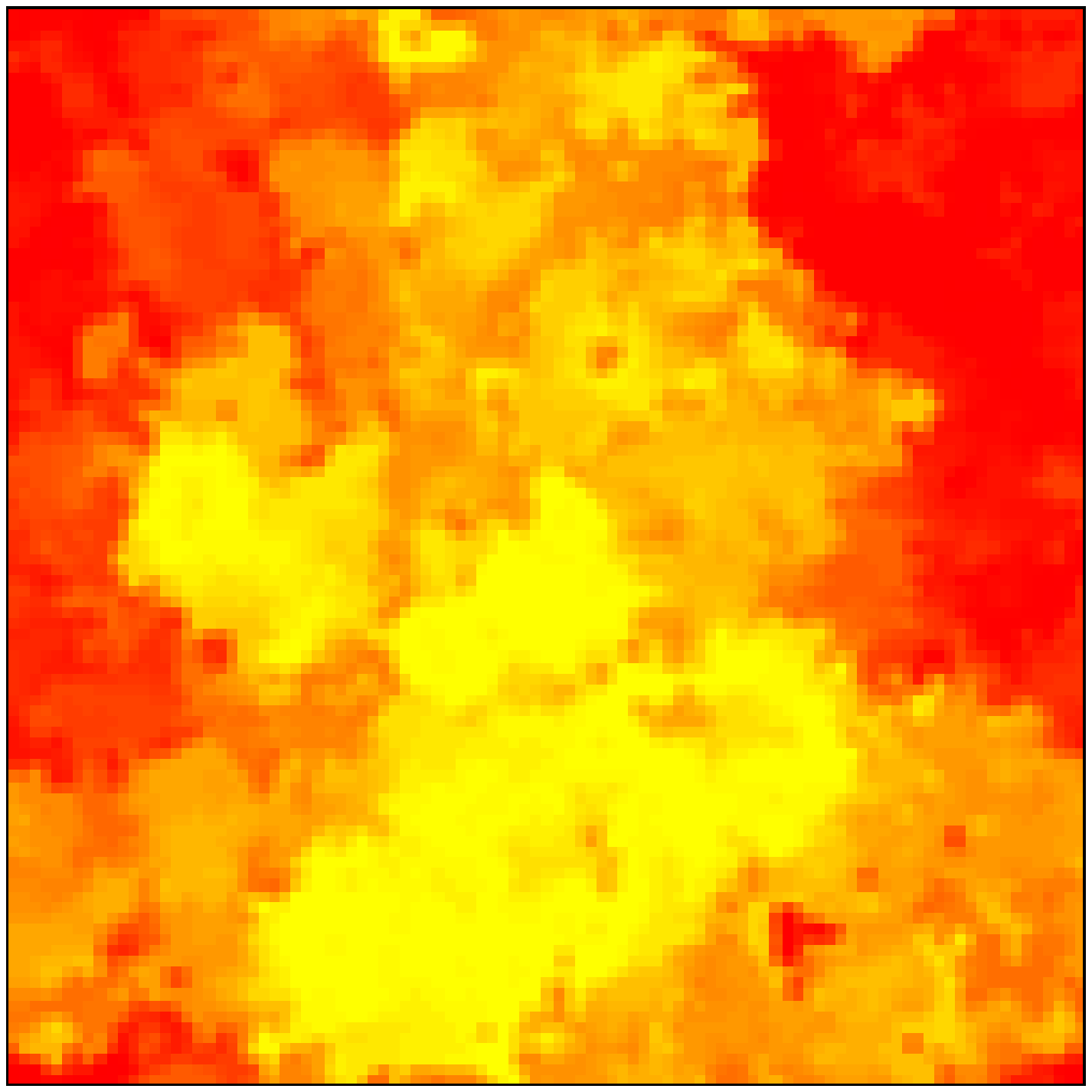}&\includegraphics[width=0.23\textwidth]{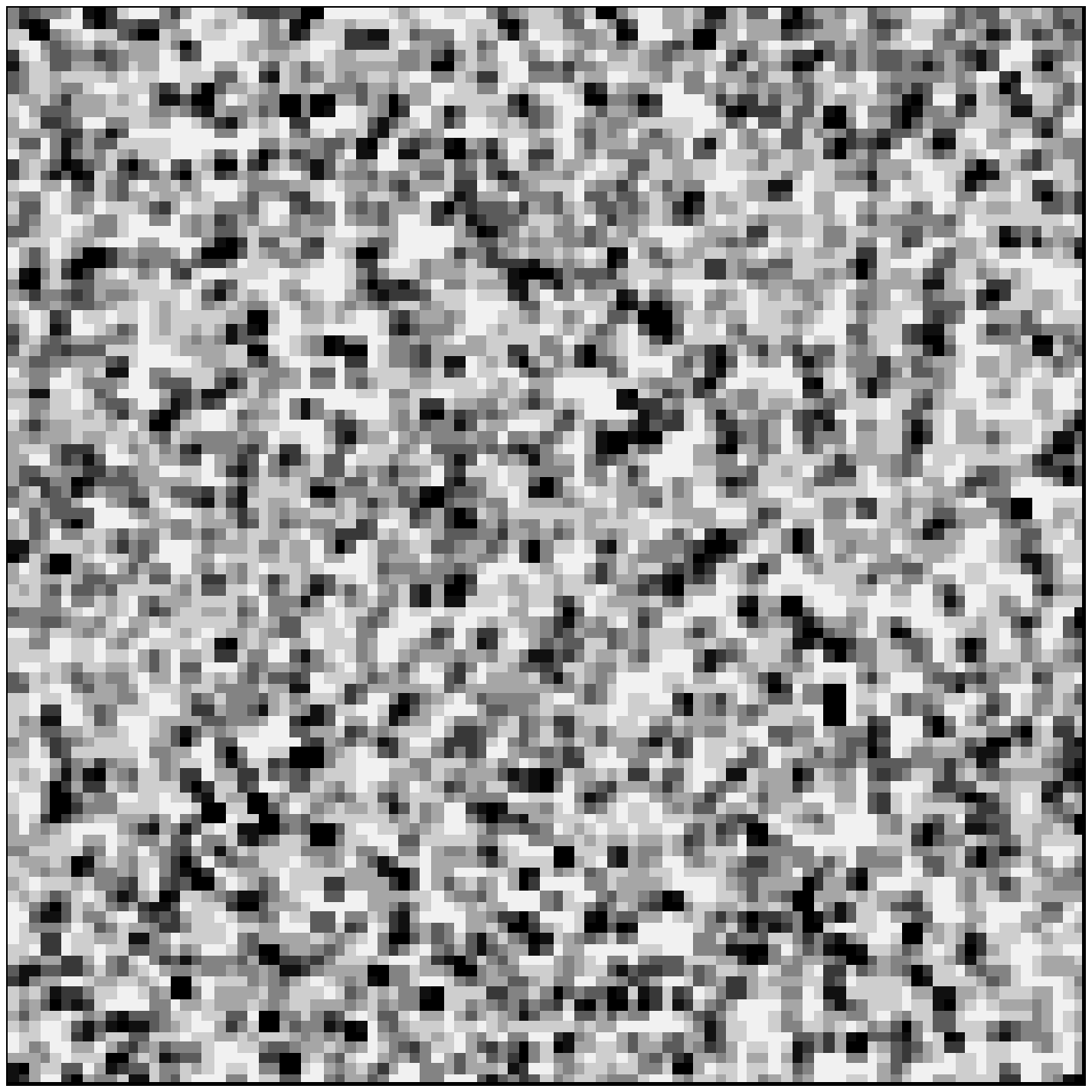}\\
\hline
\end{tabular}
\caption{Evolution of the voter model on a square lattice of $100\times100$ nodes with random asynchronous update (\emph{RAU}). The first column of images shows the states of the nodes in blue and red, the second one shows the time since the last change of state of each node, with red being a long time and yellow a small time. The third column shows the time since the last update, being dark gray for a long time and light gray for a small time. The updates of the nodes follow a Poisson process with a characteristic time of one Monte Carlo step. It is to third column shows three equivalent snapshots (spatial white-noise), because of the lack of memory of the system. The growth of domains proceeds via interfacial noise dynamics (first column). Nodes change state quite frequently, except when the system is approaching consensus (see middle column). }
\label{tab:rau}
\end{table}

\begin{table}
\centering
\begin{tabular}{|cccc|}
\hline
Time&Configuration&Time since state change&Time since update\\
\hline \hline
$t=1$&\includegraphics[width=0.23\textwidth]{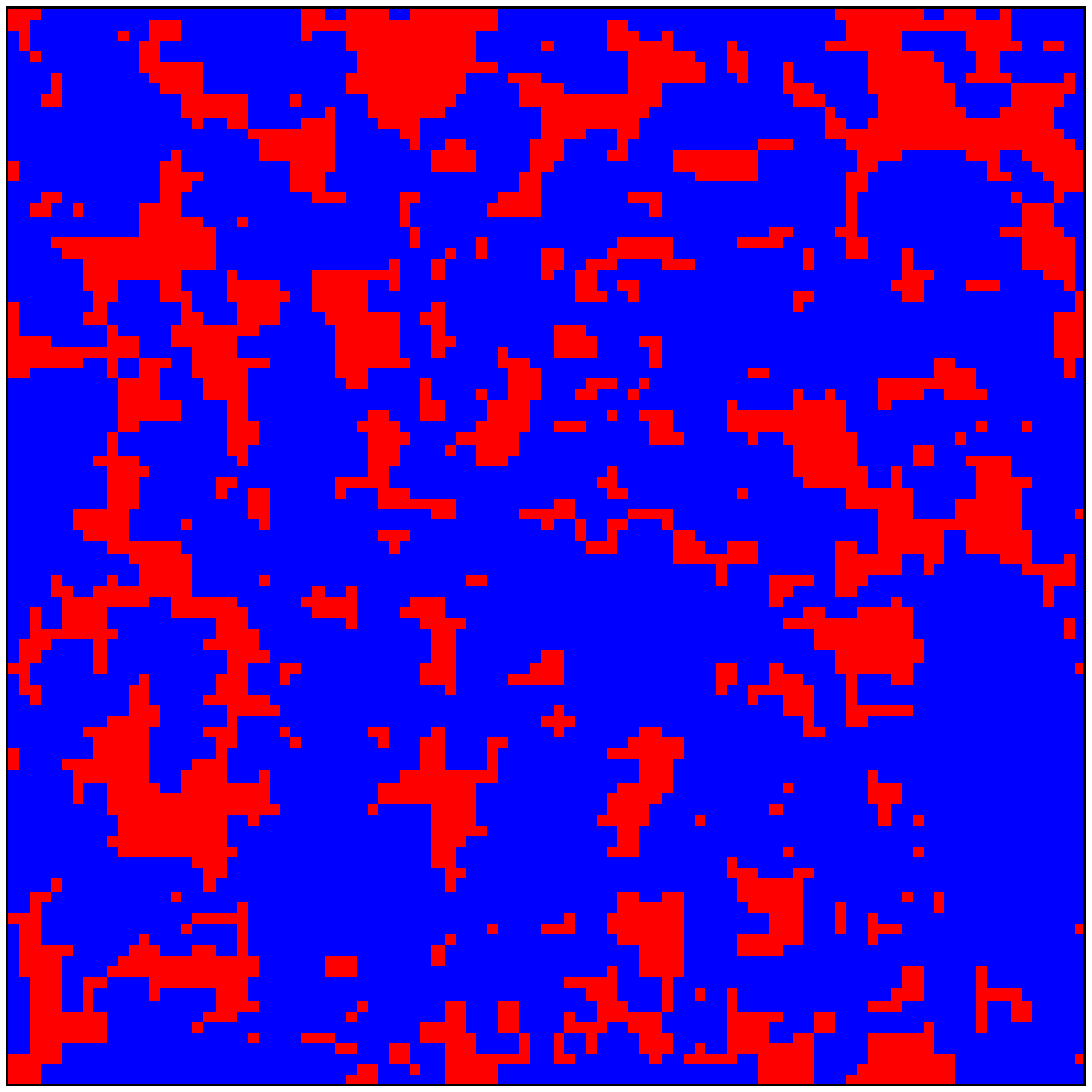}&\includegraphics[width=0.23\textwidth]{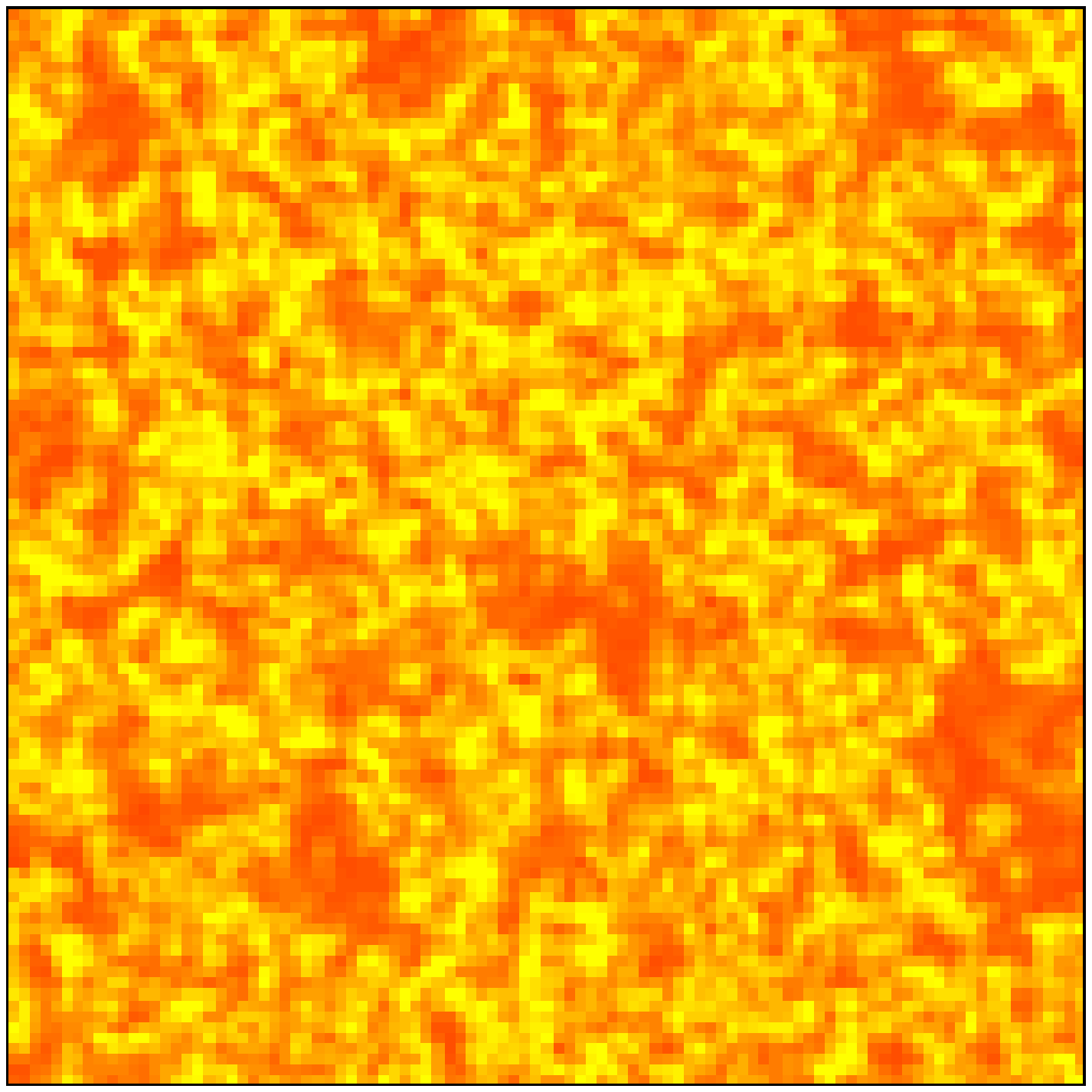}&\includegraphics[width=0.23\textwidth]{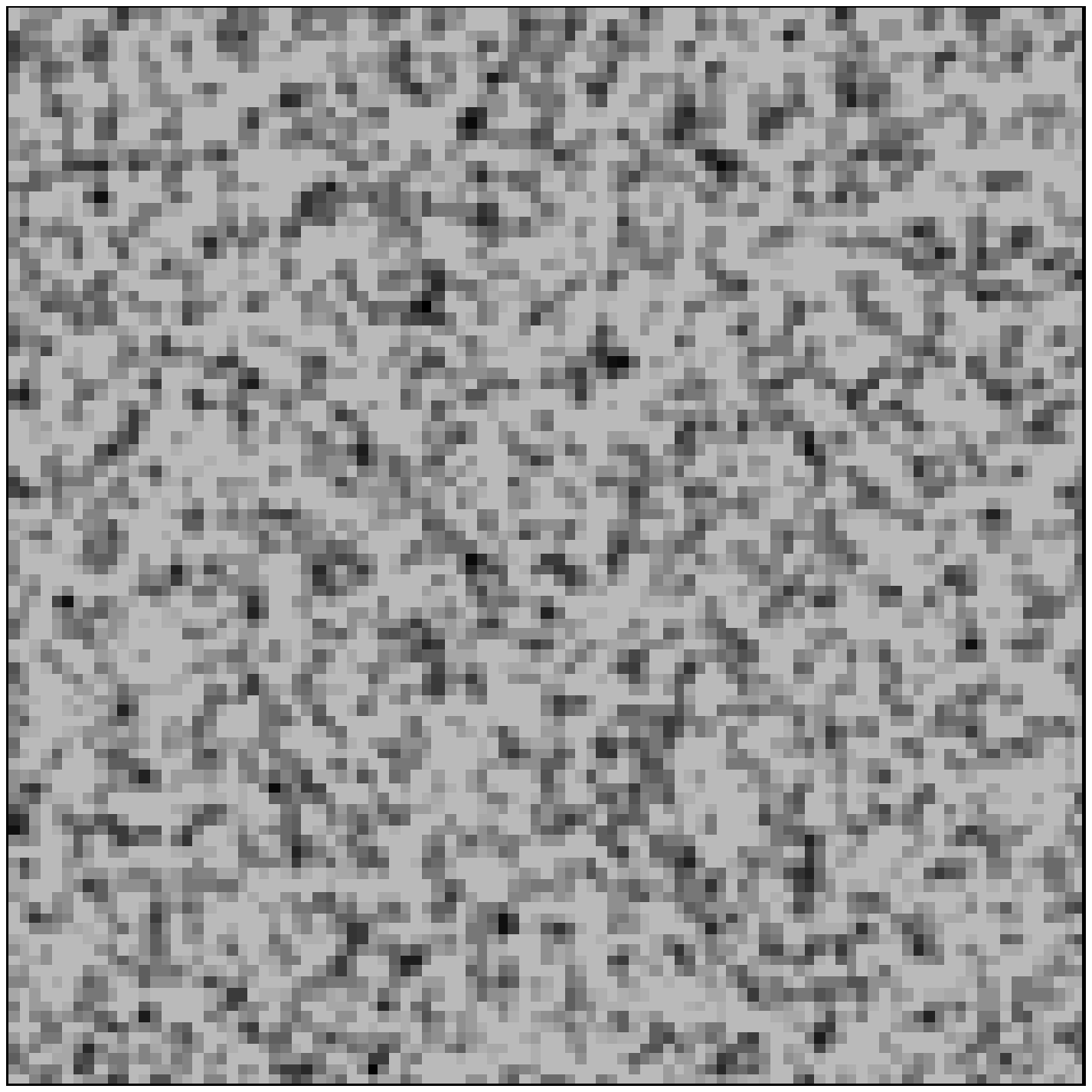}\\
\hline
$t=10$&\includegraphics[width=0.23\textwidth]{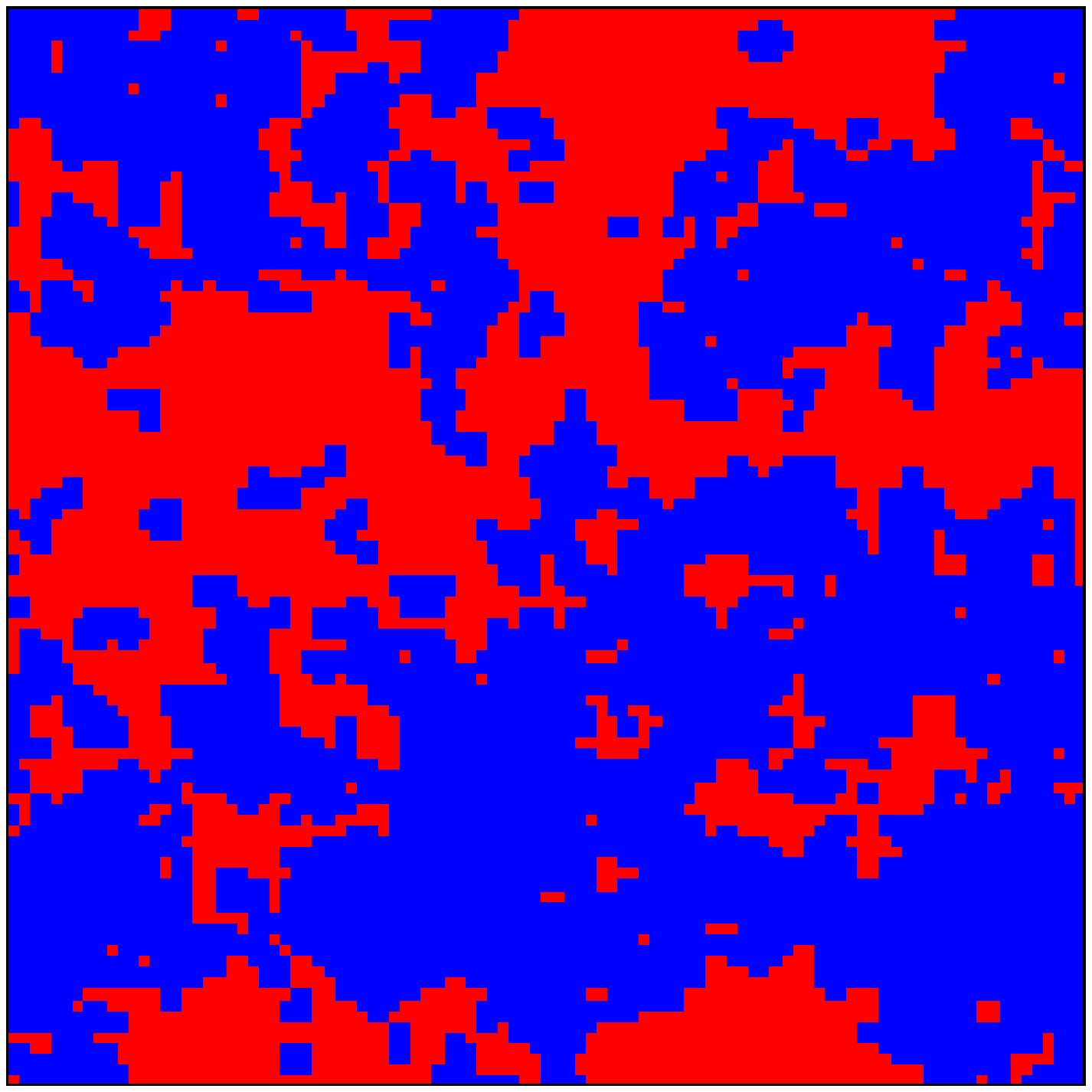}&\includegraphics[width=0.23\textwidth]{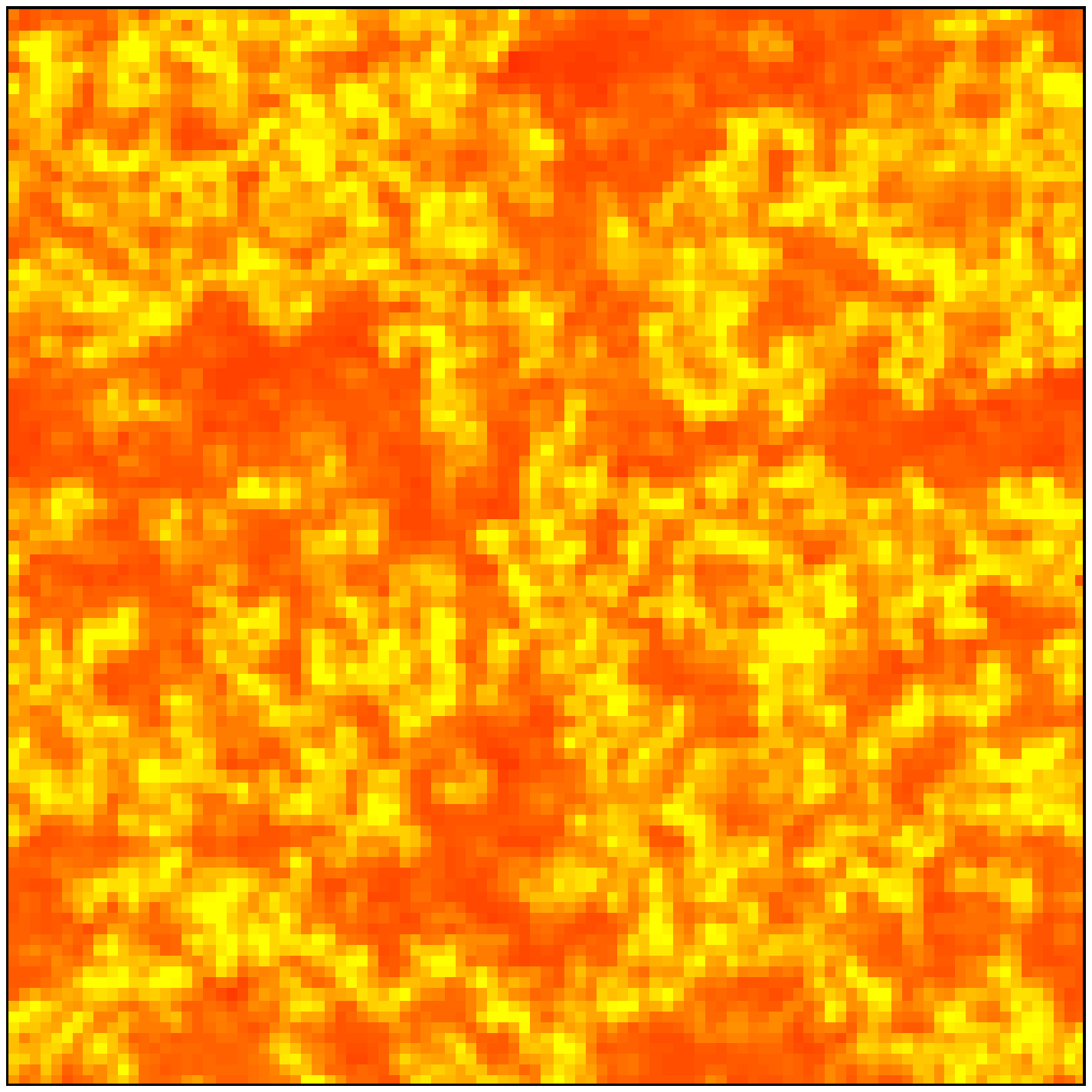}&\includegraphics[width=0.23\textwidth]{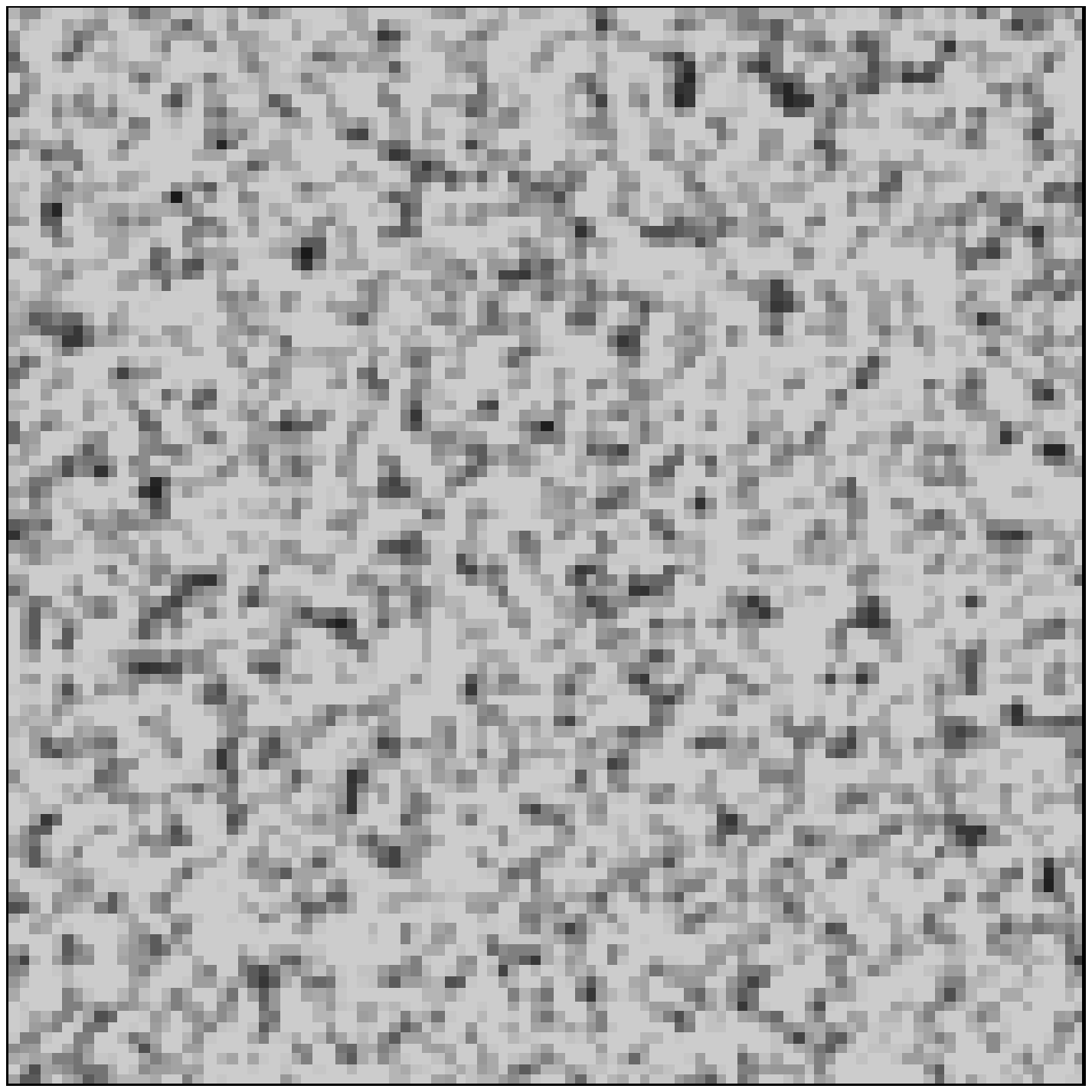}\\
\hline
$t=100$&\includegraphics[width=0.23\textwidth]{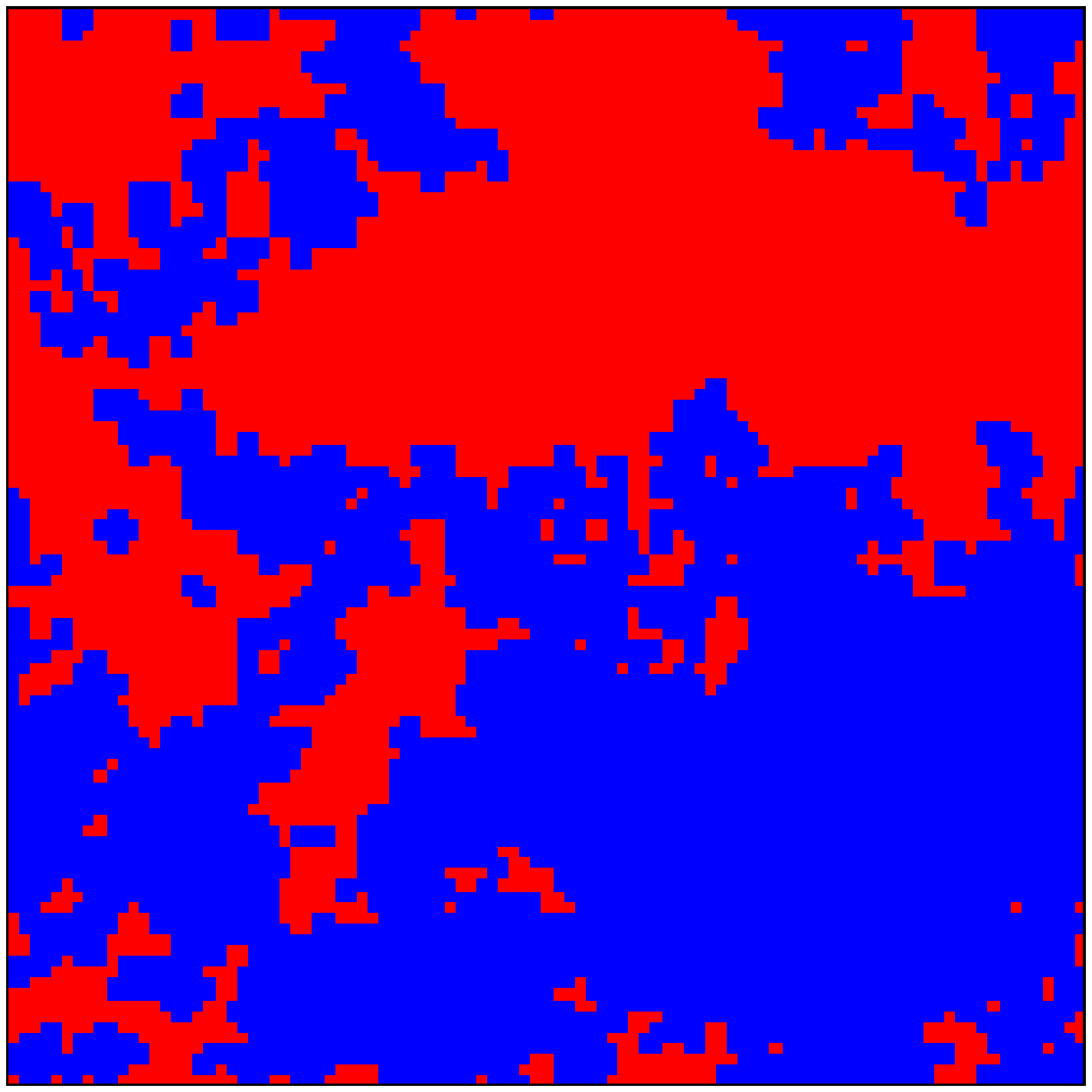}&\includegraphics[width=0.23\textwidth]{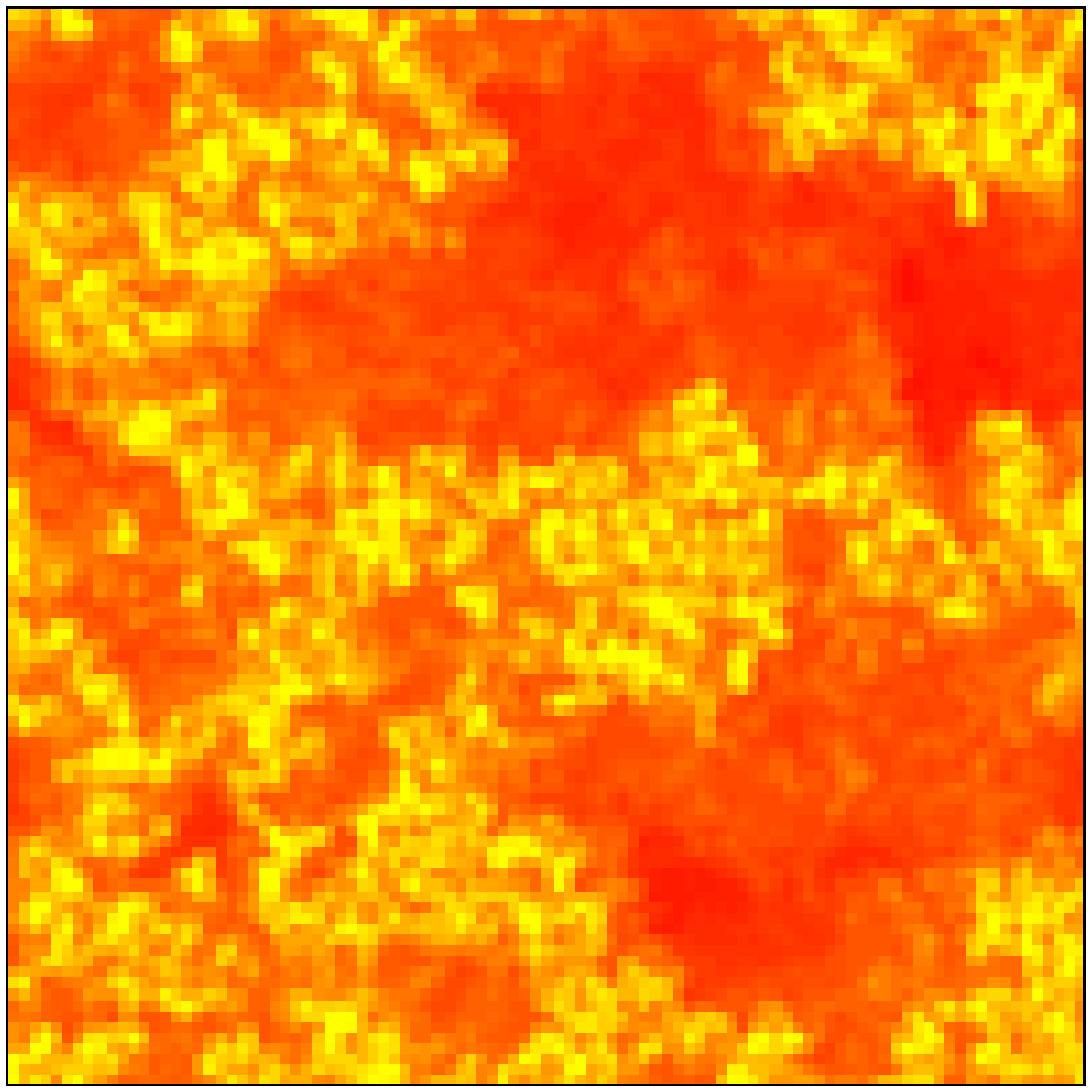}&\includegraphics[width=0.23\textwidth]{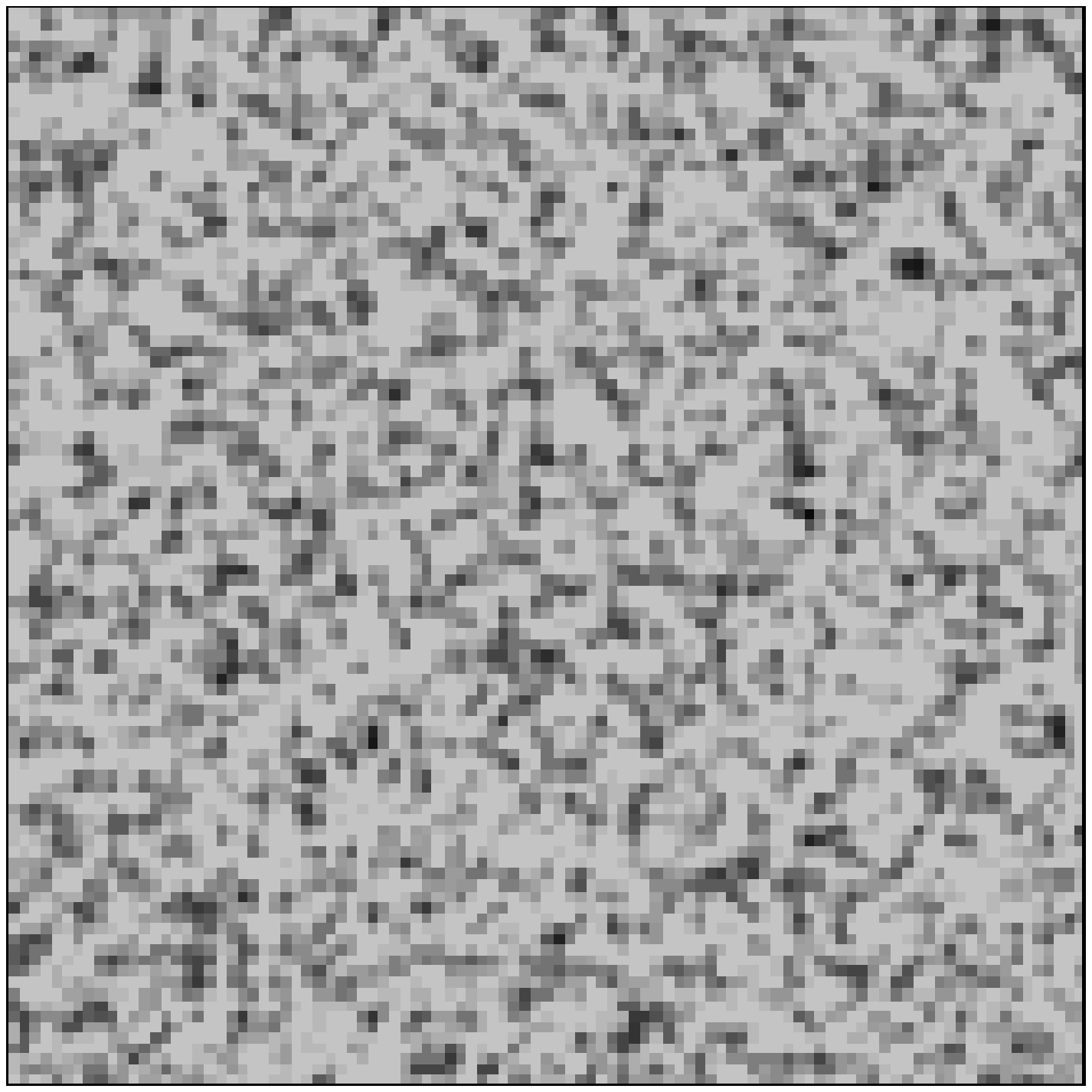}\\
\hline
\end{tabular}
\caption{Evolution of the voter model on a square lattice of $100\times100$ nodes with \emph{exogenous update}. Color codes as in Table \ref{tab:rau}. We observe the same coarsening  process (growth of domains) as with RAU (first column). Nodes also change state quite frequently (second column), with nodes that have kept their state for a longer time only inside of domains of the same state. Nevertheless,  times since the last update (third column) do not show any specific pattern: some form  of 1/f spatial noise with  nodes updated in the same way.}
\label{tab:exo}
\end{table}

\begin{table}
\centering
\begin{tabular}{|cccc|}
\hline
Time&Configuration&Time since state change&Time since update\\
\hline \hline
$t=1$&\includegraphics[width=0.23\textwidth]{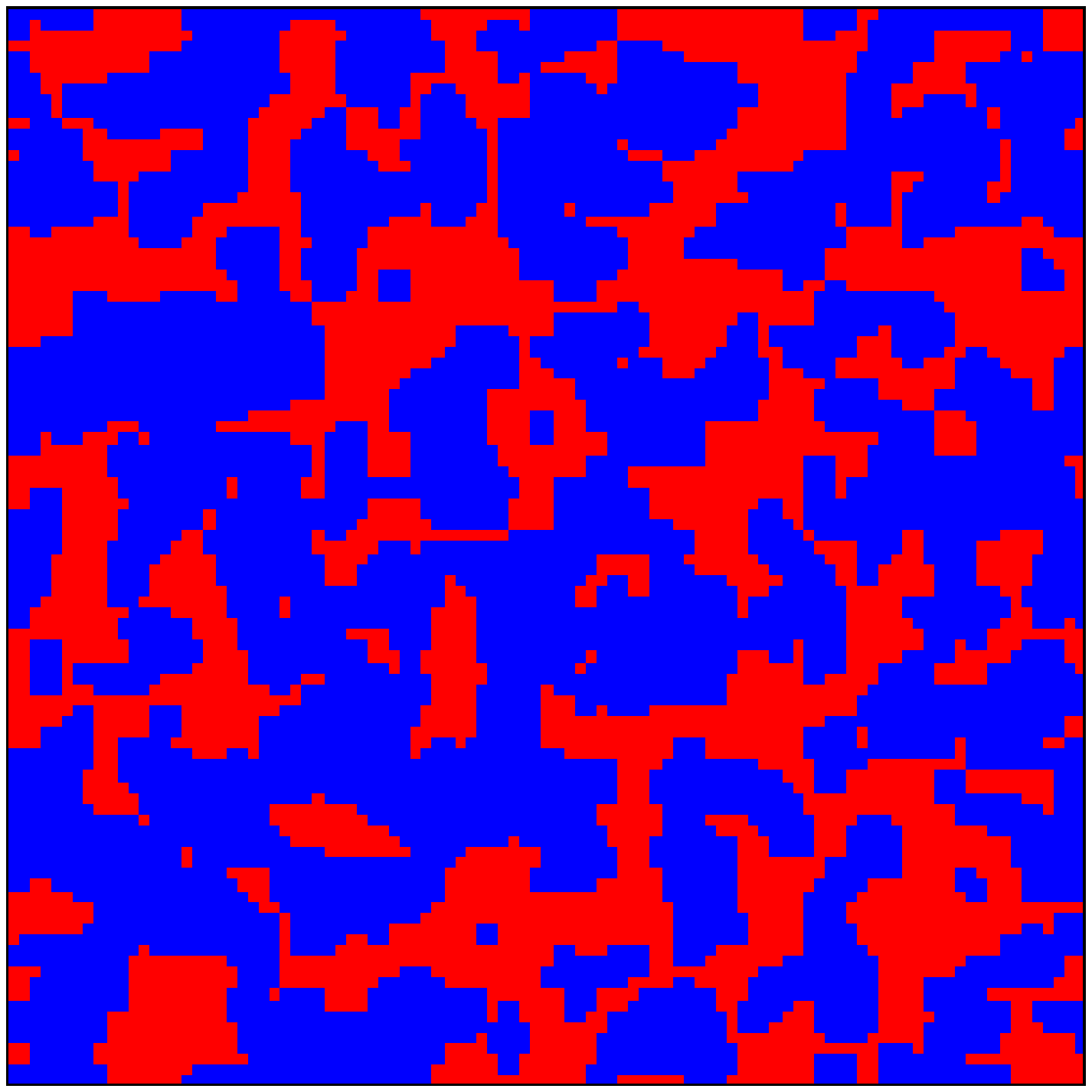}&\includegraphics[width=0.23\textwidth]{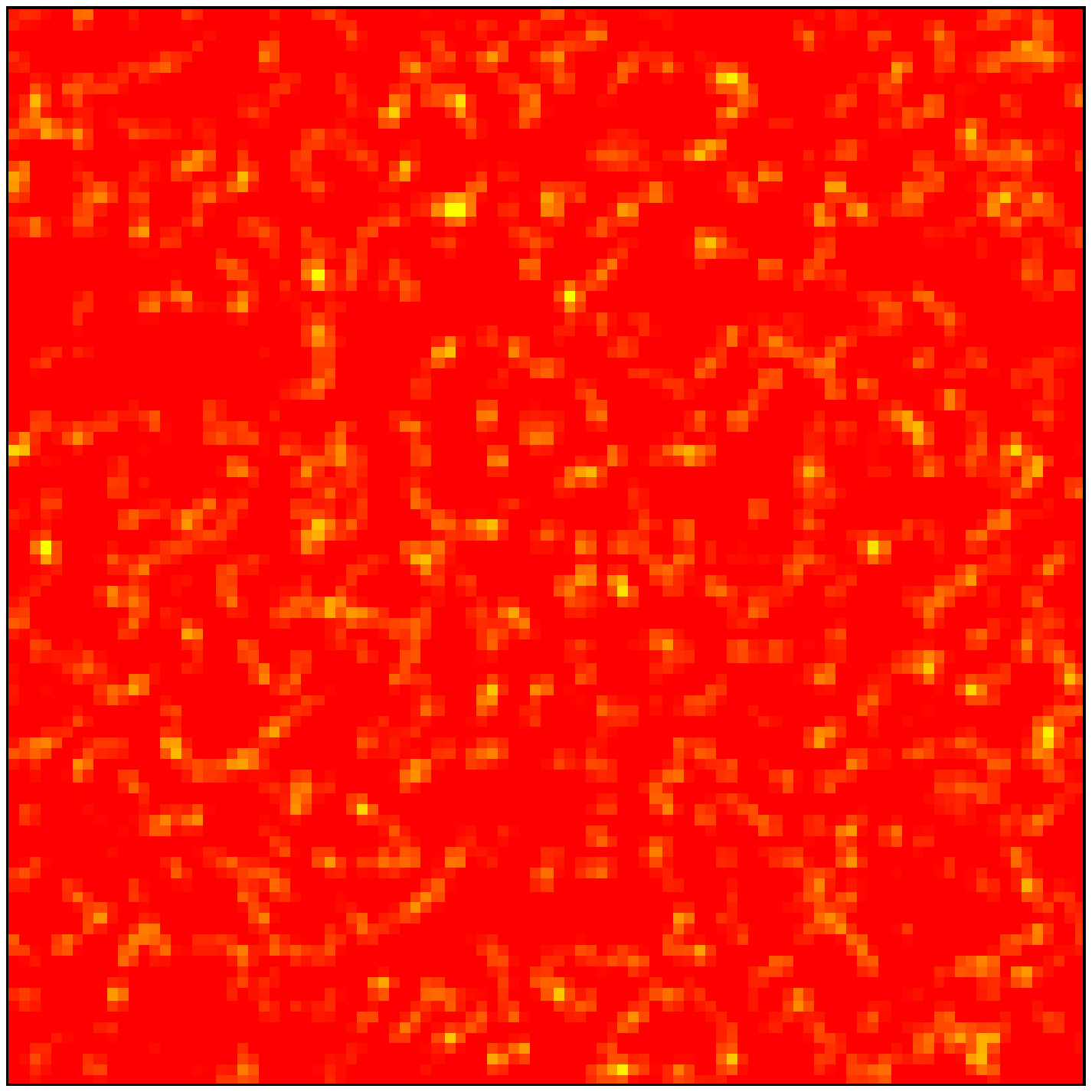}&\includegraphics[width=0.23\textwidth]{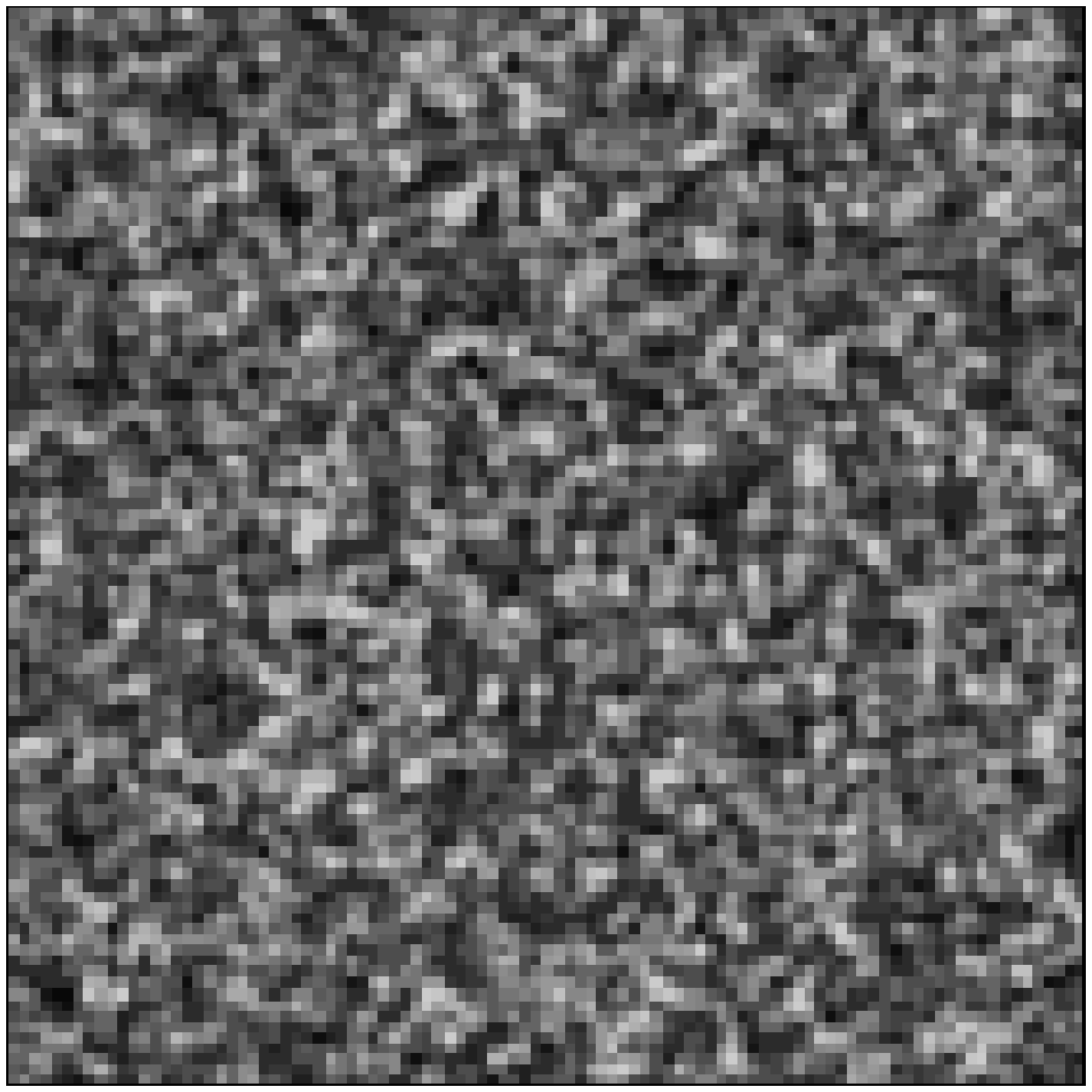}\\
\hline
$t=10$&\includegraphics[width=0.23\textwidth]{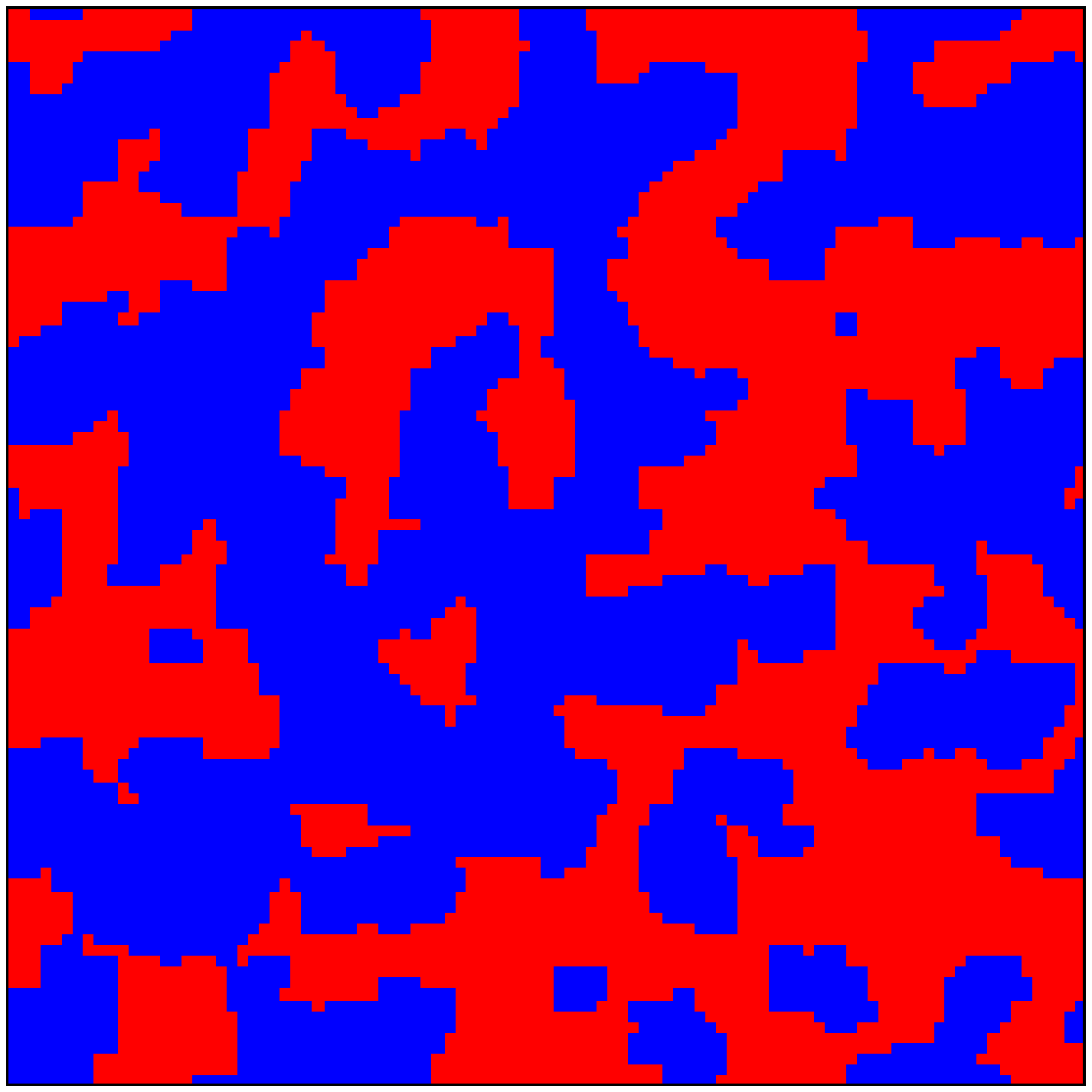}&\includegraphics[width=0.23\textwidth]{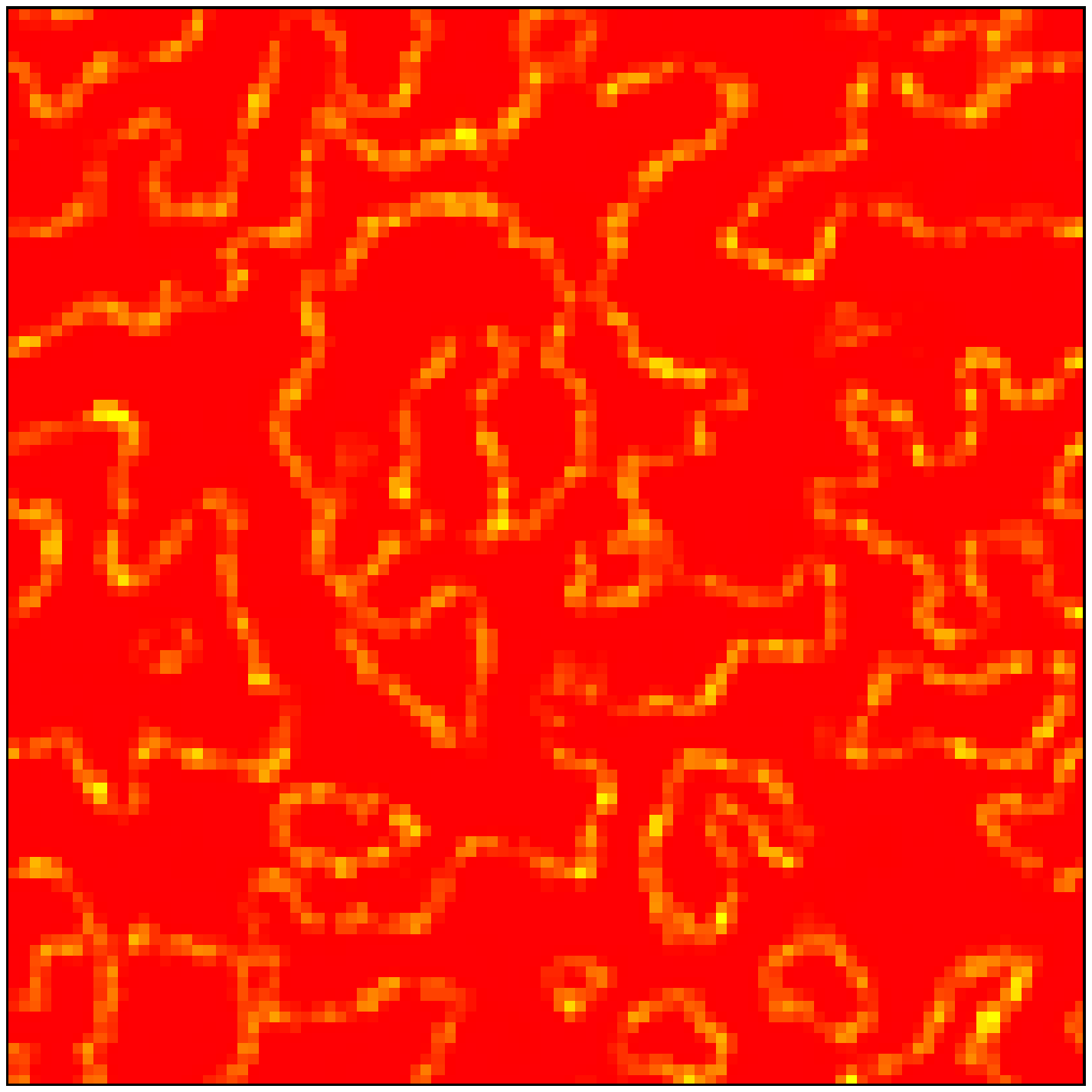}&\includegraphics[width=0.23\textwidth]{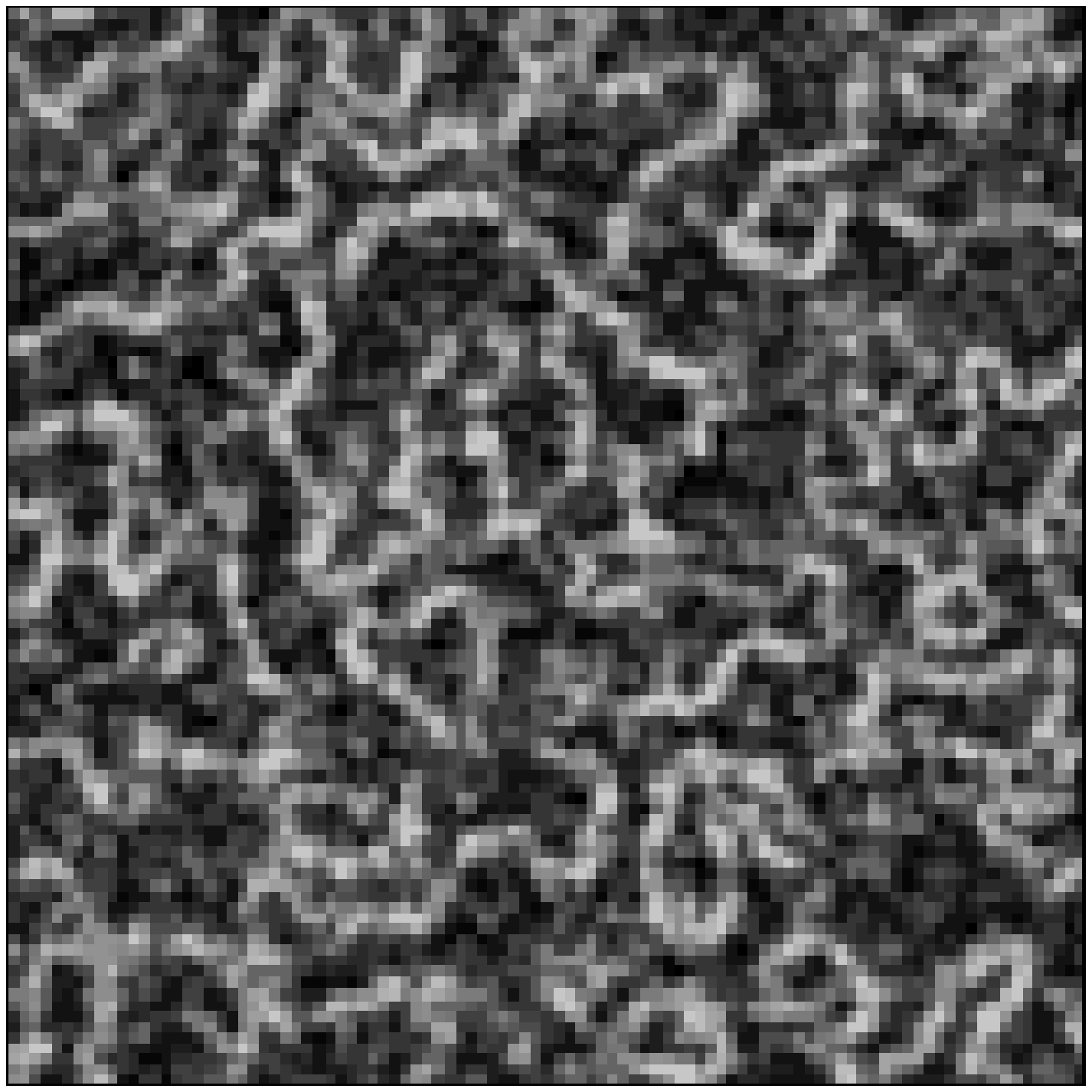}\\
\hline
$t=100$&\includegraphics[width=0.23\textwidth]{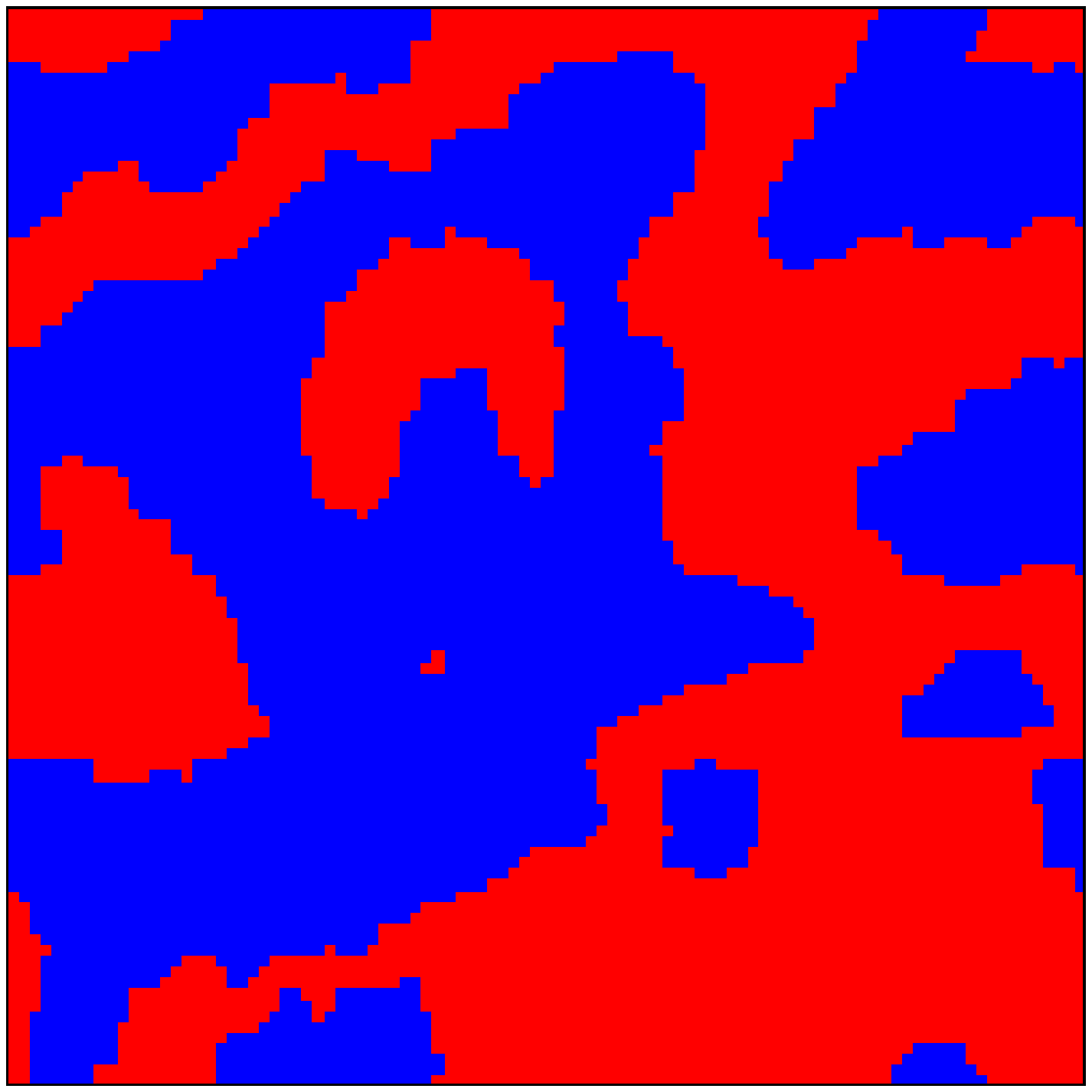}&\includegraphics[width=0.23\textwidth]{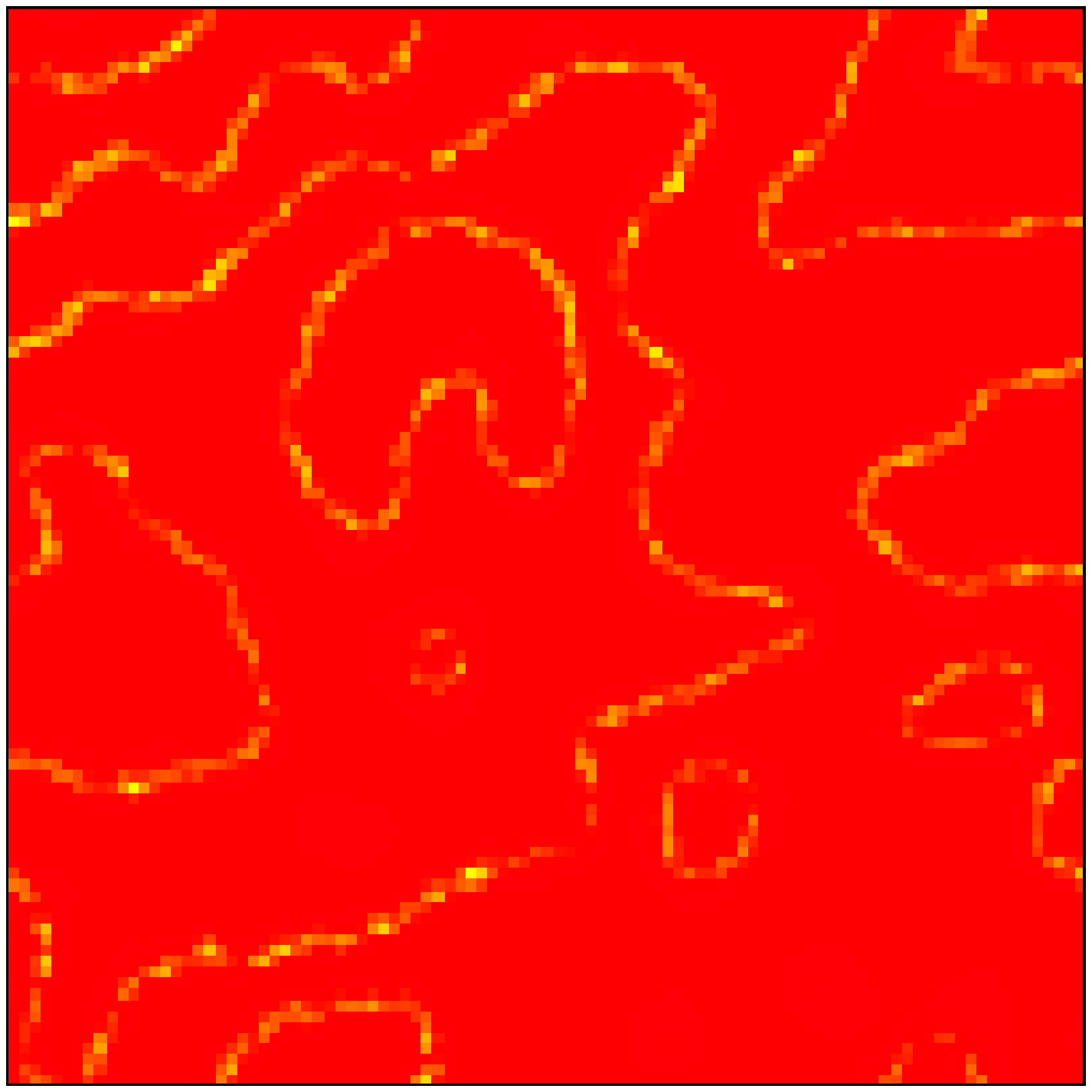}&\includegraphics[width=0.23\textwidth]{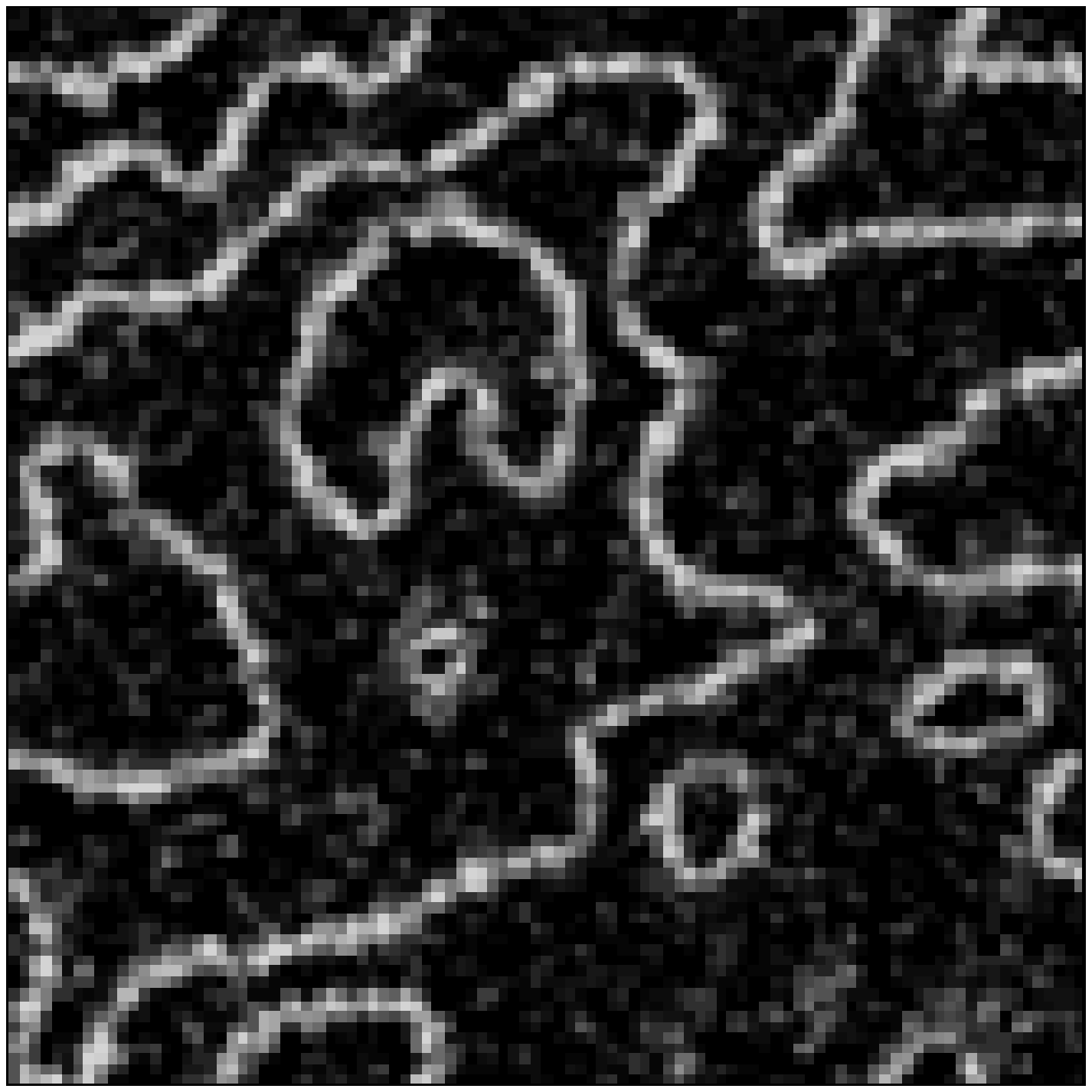}\\
\hline
\end{tabular}
\caption{Evolution of the voter model on a square lattice of $100\times100$ nodes with \emph{endogenous update}. Color codes as in Table \ref{tab:rau}. The effect of this update on the dynamics is striking and the same patterns are observed in the three columns. First, endogenous update introduces surfaces tension in the dynamics, so that the coarsening process (growth of domains) is now driven by curvature reduction (first column). In the second column we observe that the time since the last change of state is only small in the boundaries separating nodes with different states. Given that this time is now coupled to the update process, the same patterns are observed in the third column: the nodes at the interface (the ones which have changed less time ago) are updated much more frequently than the nodes in the bulk of a cluster of each state.}
\label{tab:endo}
\end{table}

\subsubsection{Voter model with exogenous update on complex networks}
If instead of the standard update rules discussed in section \ref{voter_standard} we now use the exogenous version of the new update, the agents will not be characterized only by its state $x_i$, but also by their internal time $\tau_i$, \textit{i.e.} the time since their last update event.

\begin{figure}
\centering
\begin{tabular}{ccc}
\includegraphics[height=0.31\textwidth,angle=-90]{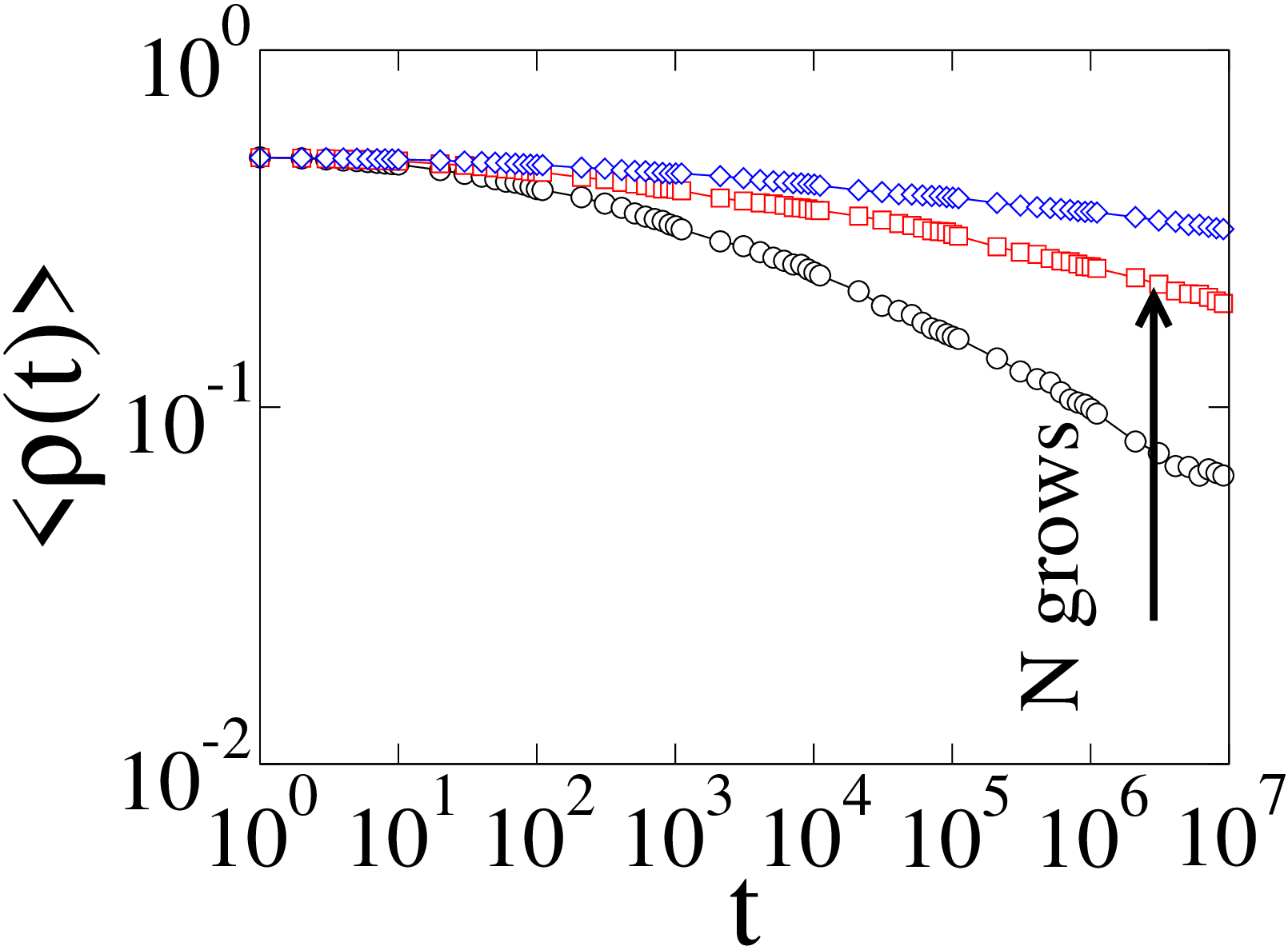}&\includegraphics[height=0.31\textwidth,angle=-90]{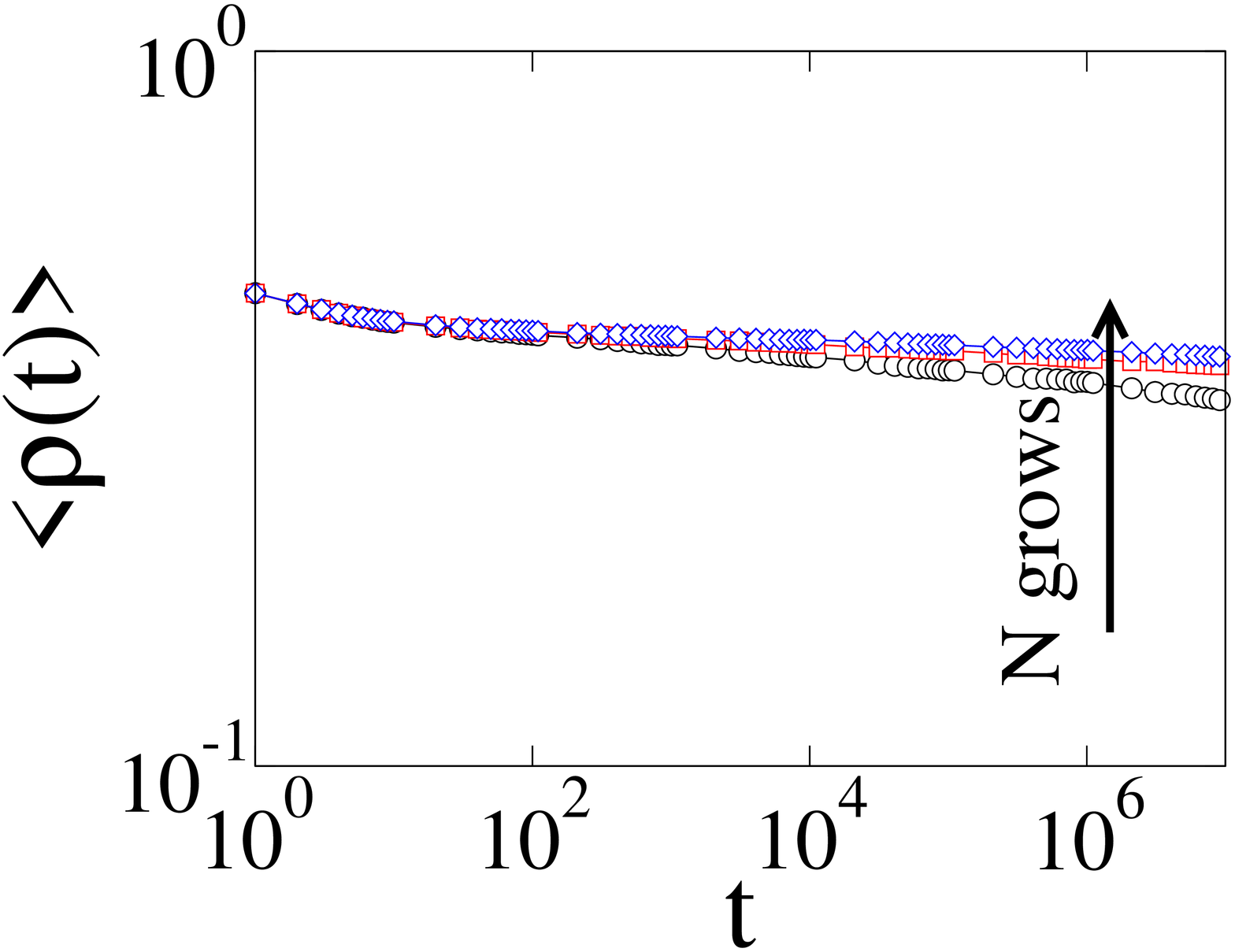}&\includegraphics[height=0.31\textwidth,angle=-90]{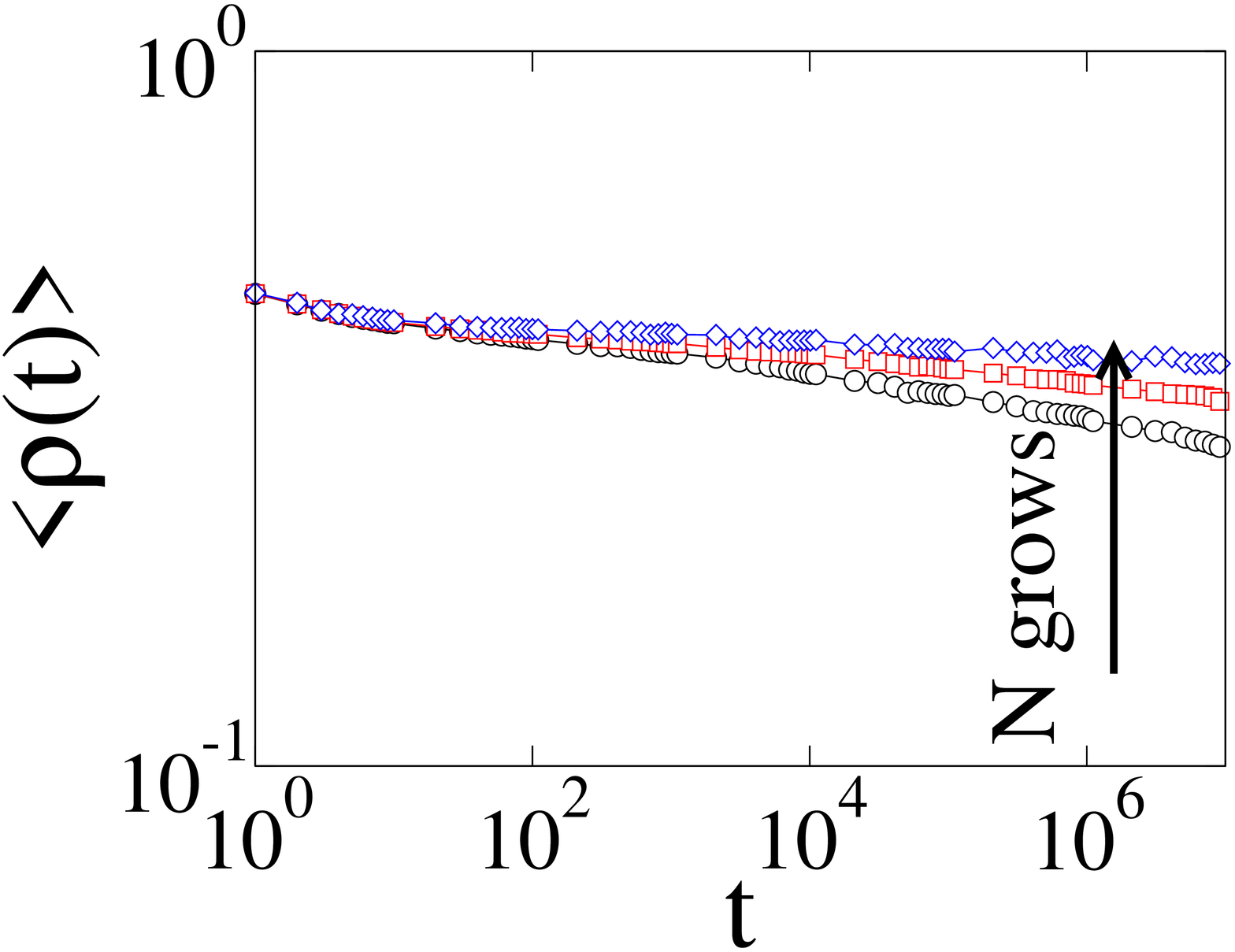}\\
\includegraphics[height=0.31\textwidth,angle=-90]{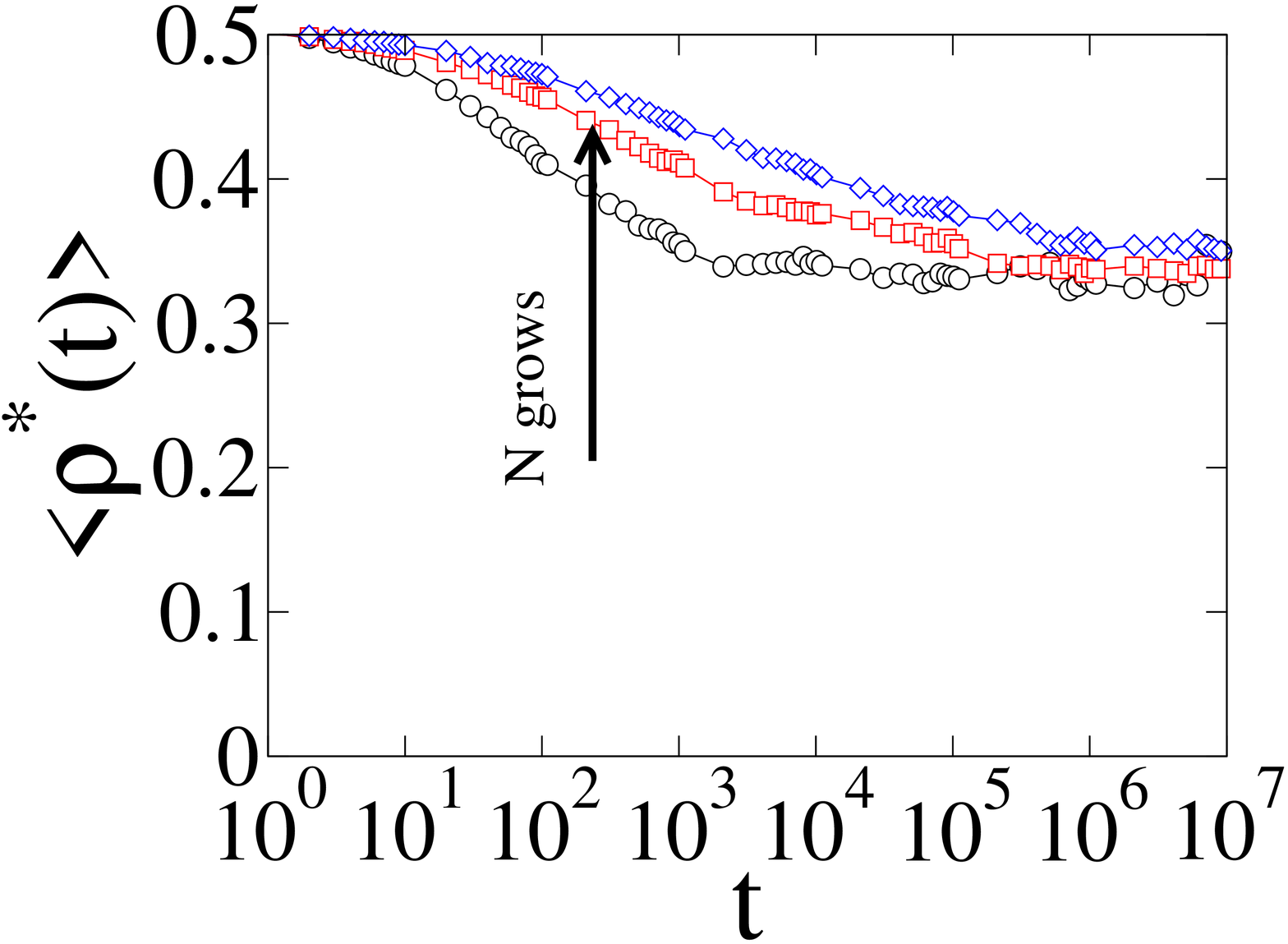}&\includegraphics[height=0.31\textwidth,angle=-90]{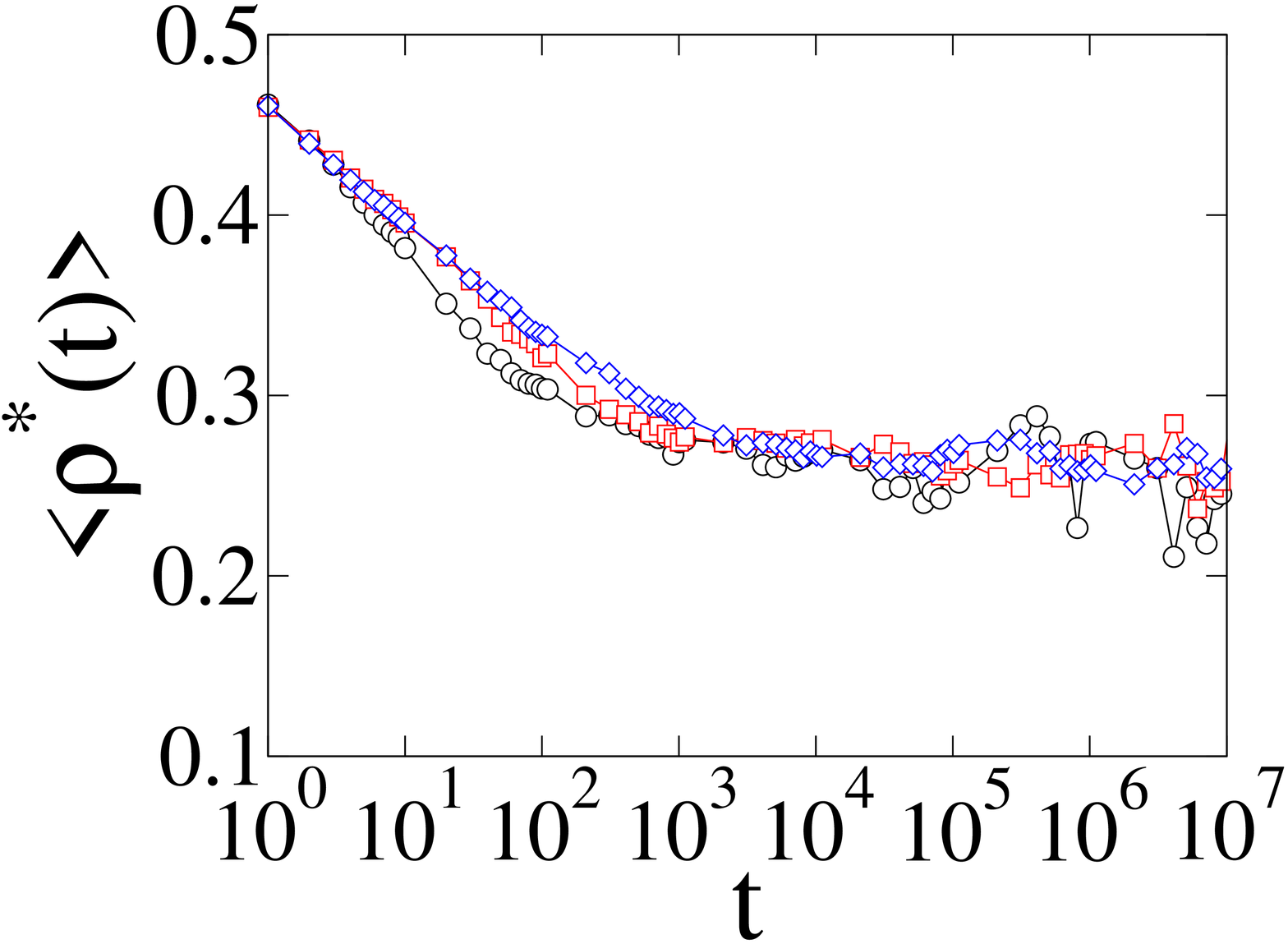}&\includegraphics[height=0.31\textwidth,angle=-90]{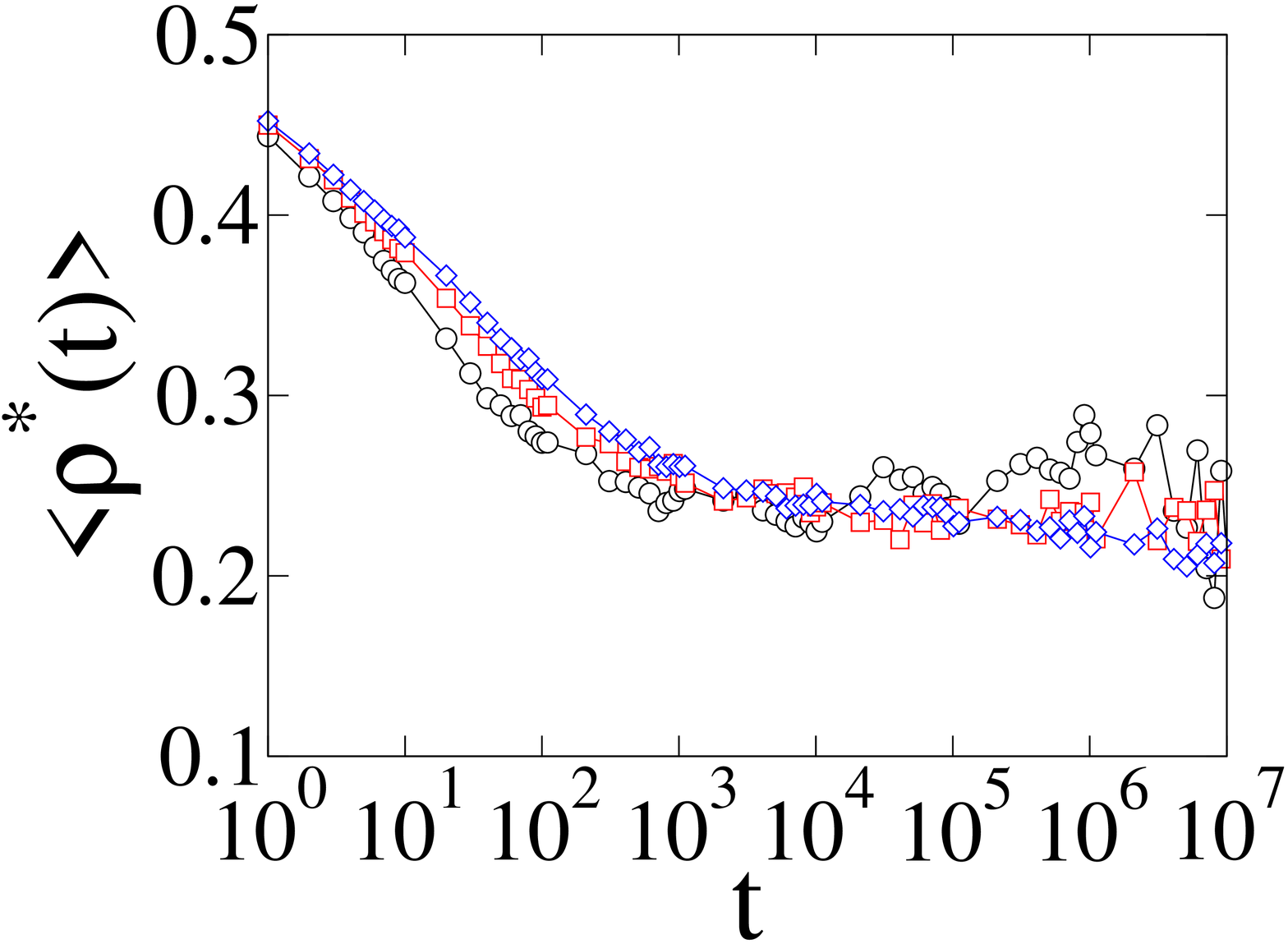}\\
\includegraphics[height=0.31\textwidth,angle=-90]{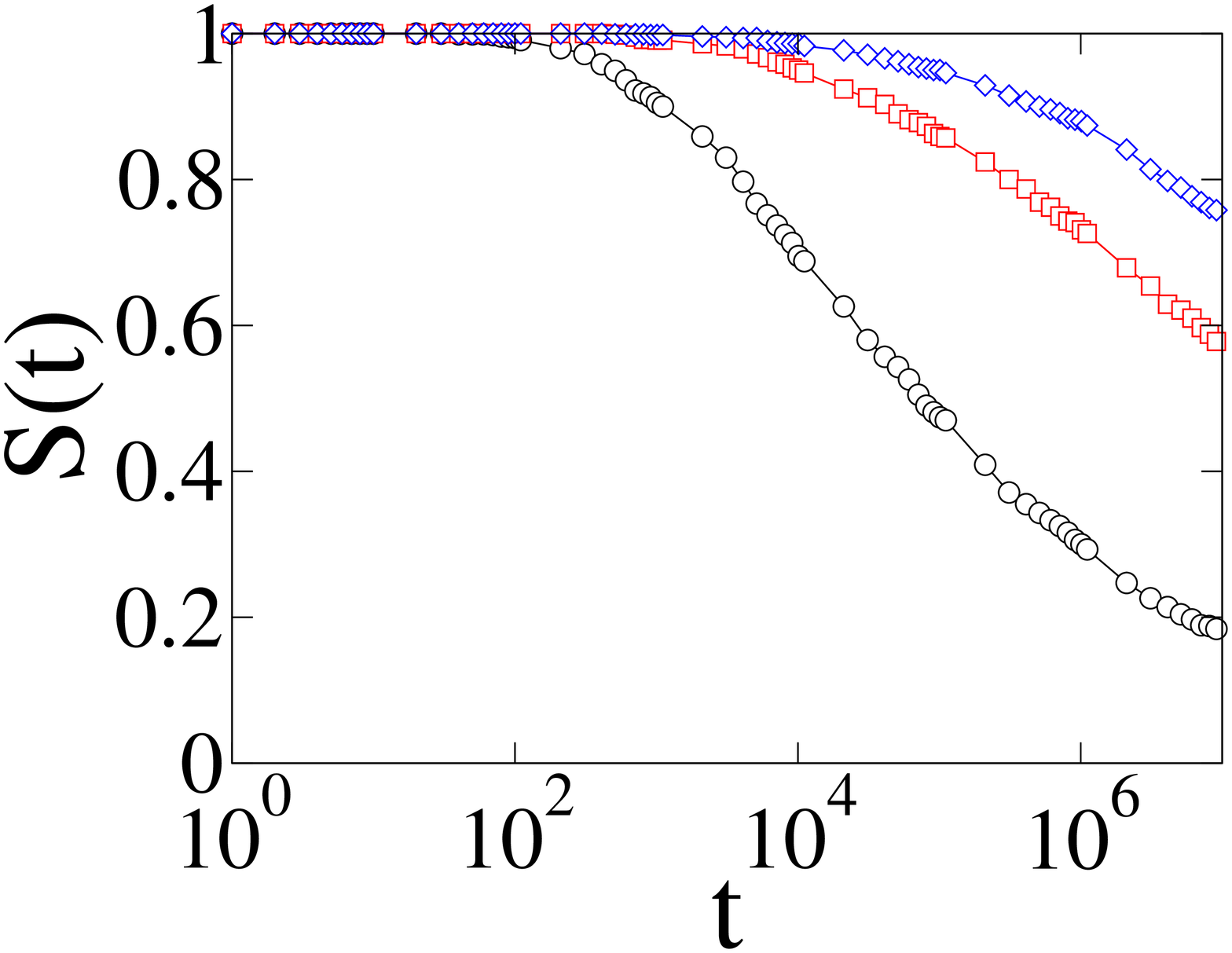}&\includegraphics[height=0.31\textwidth,angle=-90]{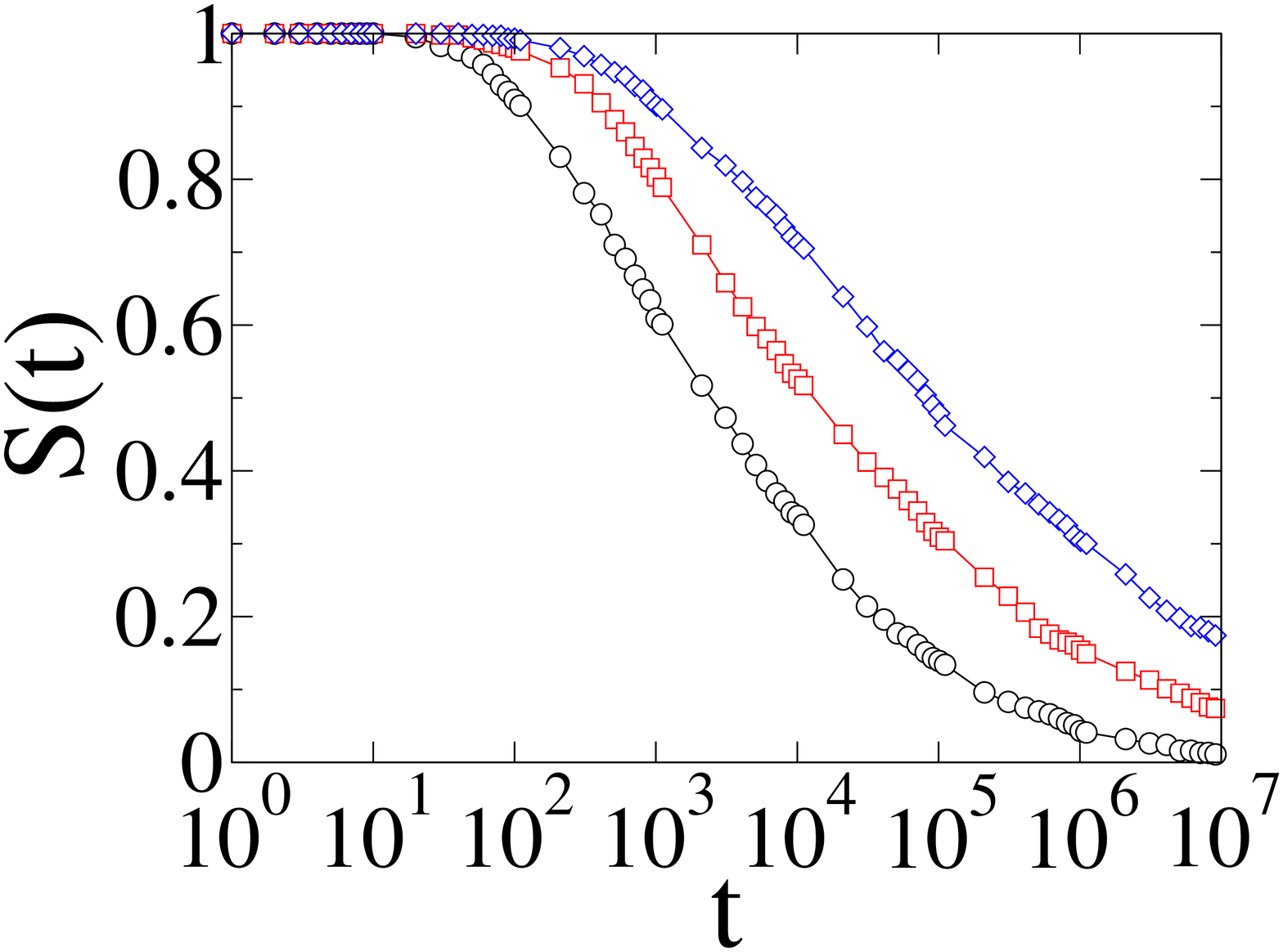}&\includegraphics[height=0.31\textwidth,angle=-90]{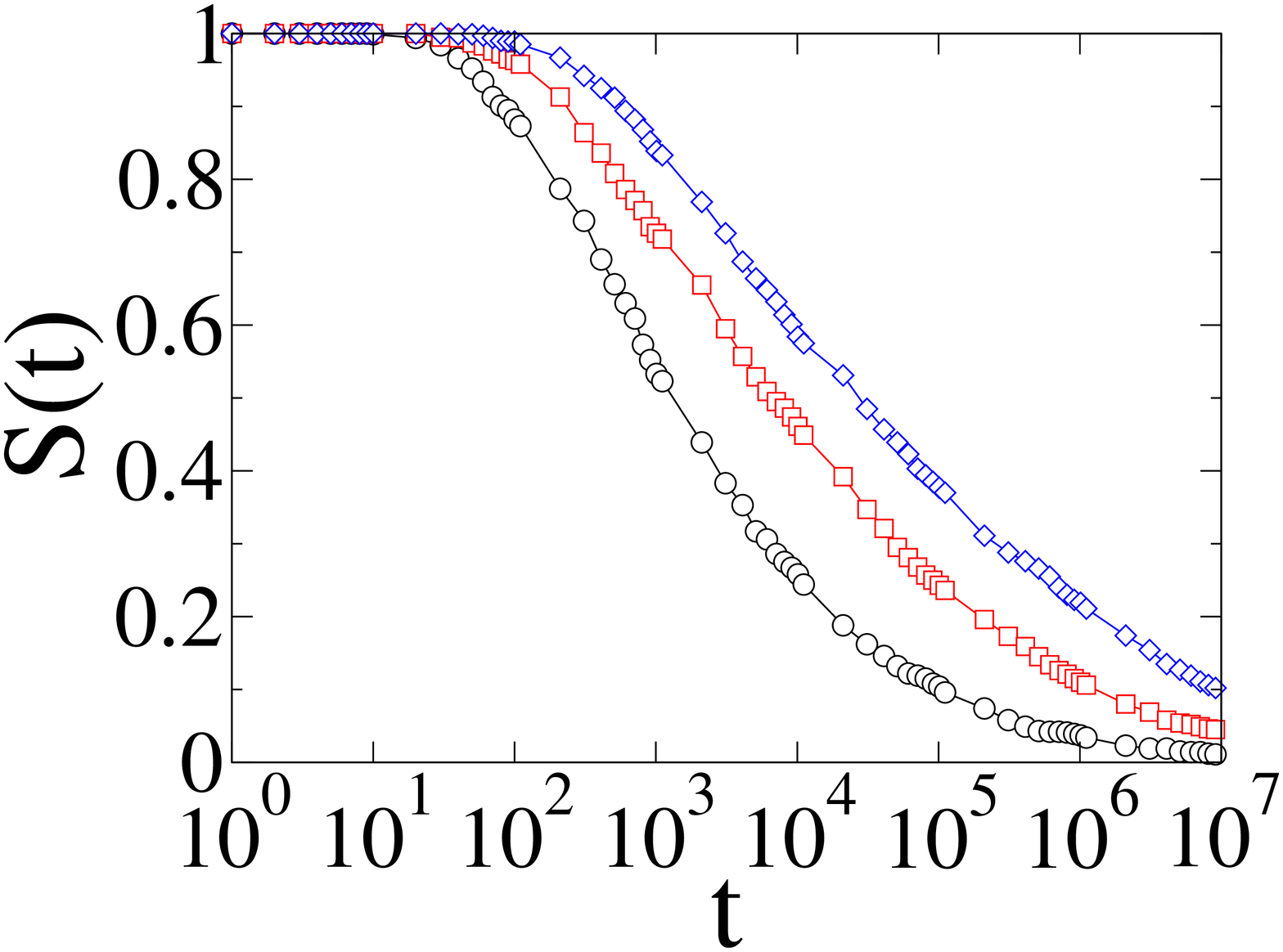}\\
\includegraphics[height=0.31\textwidth,angle=-90]{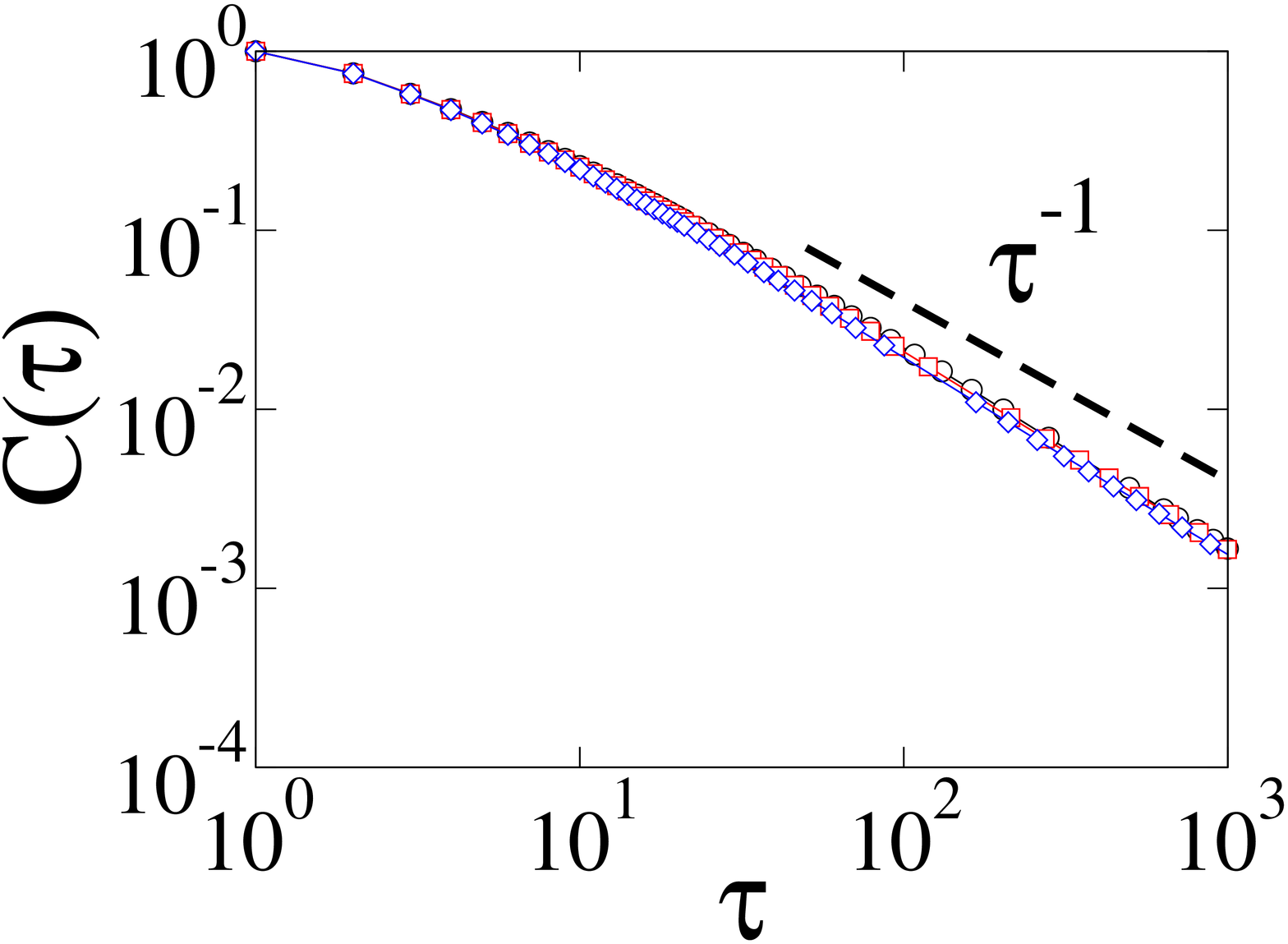}&\includegraphics[height=0.31\textwidth,angle=-90]{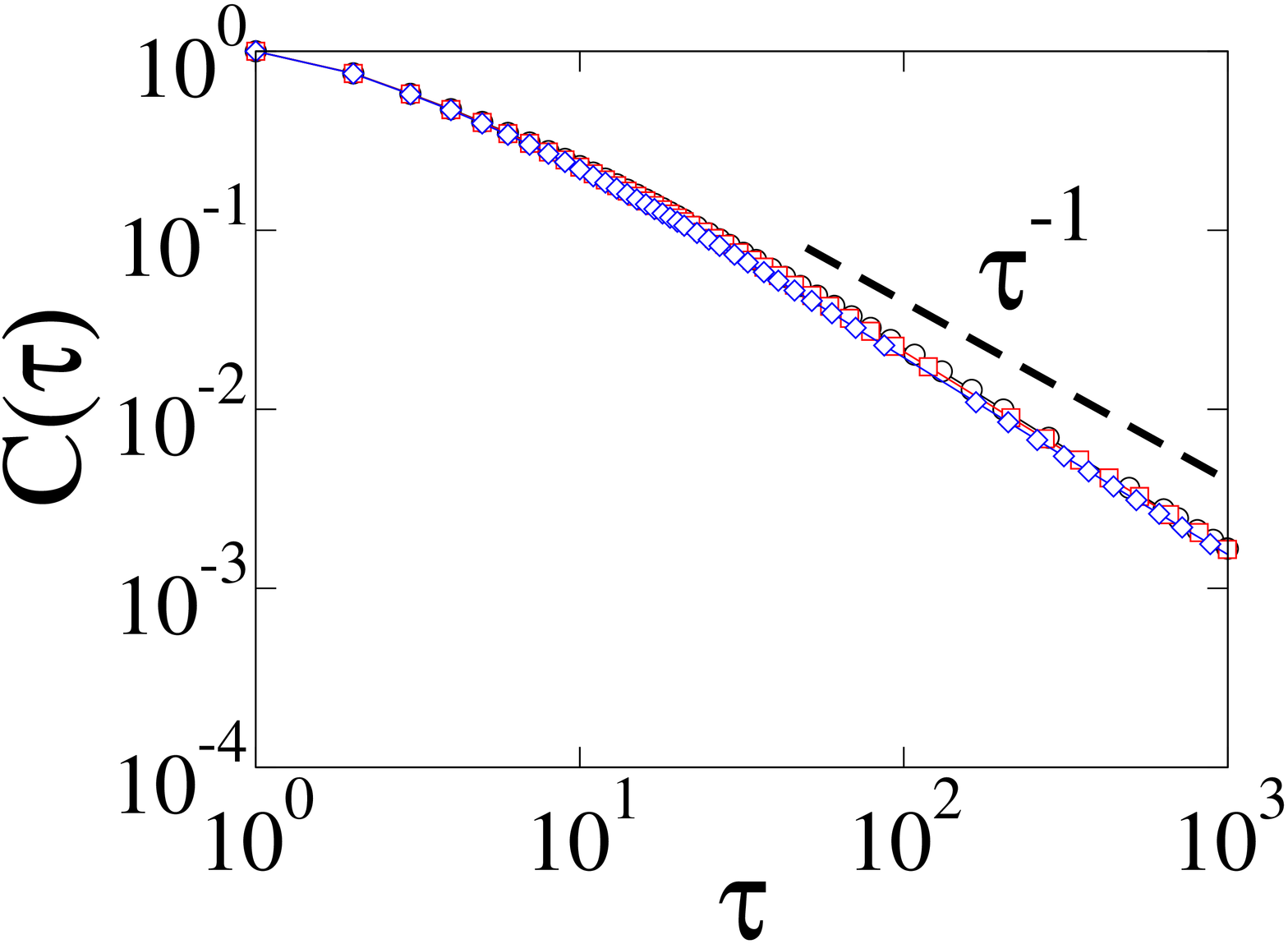}&\includegraphics[height=0.31\textwidth,angle=-90]{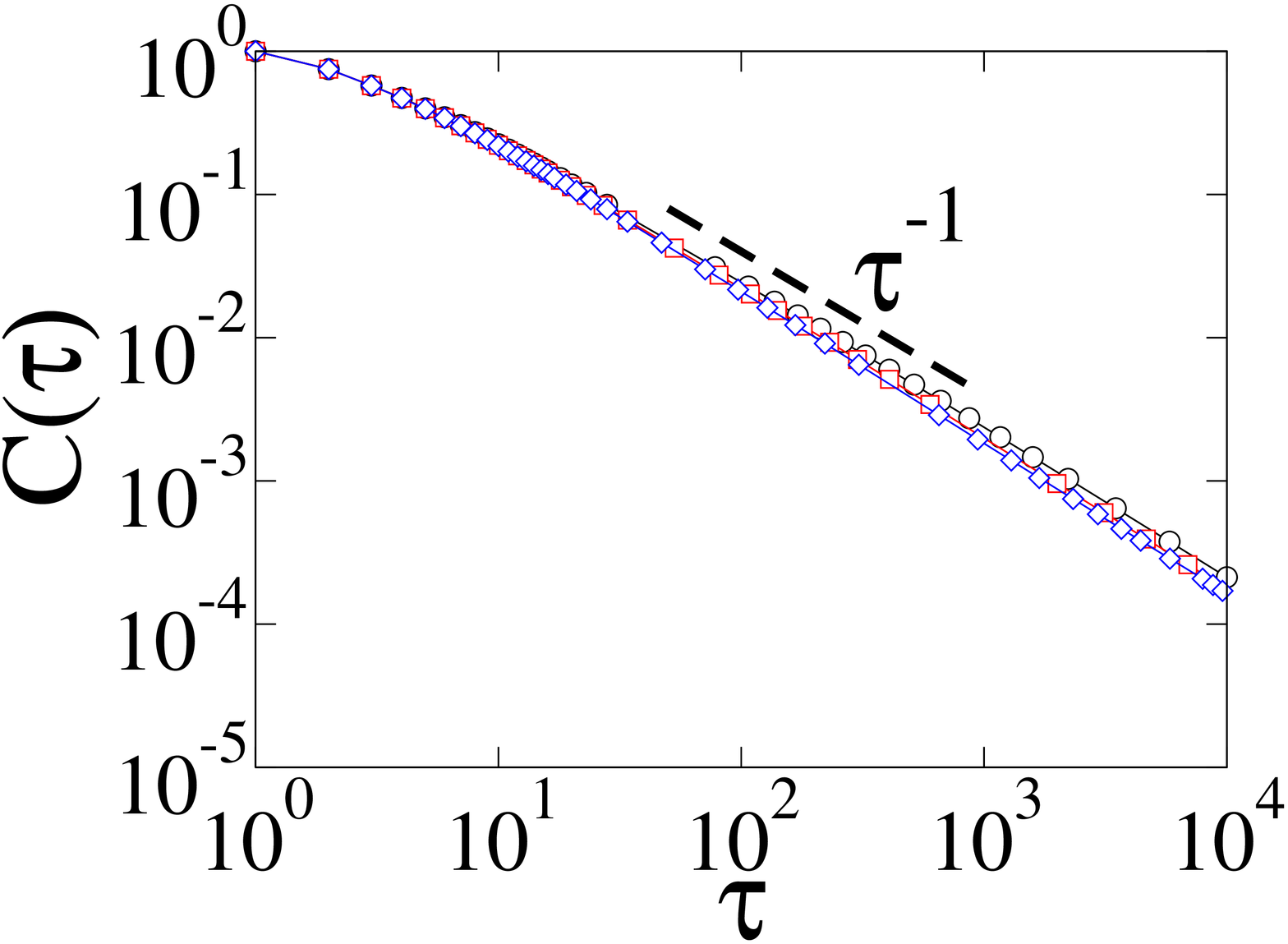}\\
\end{tabular}
\caption{Characteristics of the voter model with \emph{exogenous update} for several networks. Left column is for complete graphs of sizes $300$ in black,$1000$ in red and $4000$ in blue. Middle column is for random graphs with average degree $\langle k\rangle=6$ and sizes $1000$ in black,$2000$ in red and $4000$ in blue. Right column is for scale-free graphs with average degree $\langle k\rangle=6$ and sizes $1000$ in black,$2000$ in red and $4000$ in blue. Top row shows plots of the average density of interfaces $\langle\rho\rangle$, second row shows the density of interfaces averaged over surviving runs $\langle\rho^*\rangle$, middle column shows the survival probability $S(t)$ and right column shows the cumulative IET distribution $C(\tau)$. The averages where done over $1000$ realizations.}
\label{tab:exonets}
\end{figure}

The simulation steps for this modified voter model are as follows:
\begin{enumerate}
 \item With probability $p(\tau_i)$ every agent $i$ is given the opportunity of updating her state by interacting with a neighbor.
 \item If the agent interacts, one of its neighbors $j$ is chosen at random and agent $i$ copies $j$'s state. $x_i(t+1)=x_j(t)$. Agent $i$ resets $\tau_i=0$.
 \item The time is increased by a unit and we return to 1 to keep on with the dynamics.
\end{enumerate}
For an activation probability $p(\tau)=1/\tau$, \textit{i.e.} $\beta=1$ we ran simulations on a complete graph, on random graphs of different average degrees, and on a Barab\'asi-Albert scale-free graph of average degree $\langle k\rangle=6$ and for different system sizes (see Fig.\ref{tab:exonets}).

Our results can be summarized as follows:
\begin{description}
 \item{\textit{Density of active links $\langle\rho(t)\rangle$ and $\langle\rho^*(t)\rangle$: }}{When averaged over all runs, $\langle\rho(t)\rangle$ decays with different rates depending on the interaction networks and system sizes. For bigger system sizes the decay slows down, reaching a plateau in the thermodynamic limit (left column of Fig.~\ref{tab:exonets}). When averaged over active runs $\langle\rho^*(t)\rangle$ reaches a plateau (Inset of left column of Fig.~\ref{tab:exonets}), which is independent of the system size, showing that living runs stay, on average, on a dynamical disordered state, as happens with standard update rules.}
 \item{\textit{Survival probability $S(t)$: }}{No realizations order in some time, until the survival probability decays in a nontrivial way. It is not a purely exponential decay, but decays faster than any power law. Therefore no normalization problems are expected.}
 \item{\textit{Cumulative IET distribution $C(\tau)$: }}{Develops a power law tail consistent with the exponent $\beta=b$, which in this case is set to $1$, as we could expect if the approximation of Eq.\ref{cdetau} holds.}
\end{description}
\emph{The dynamics does not order the system with the exogenous update.}
\subsubsection{Voter model with endogenous update on complex networks}
We now use the endogenous update for the voter model. This is just the same as the exogenous update rule, but in this case the internal time of each agent $i$, $\tau_i$, is the time since her last change of state. In this way the update rule is coupled to the states of the agents.

The simulation steps for the modified voter model that we study are as follows:
\begin{enumerate}
 \item With probability $p(\tau_i)$ every agent $i$ is given the opportunity of updating her state by interacting with a neighbor.
 \item If the agent interacts, one of its neighbors $j$ is chosen at random and agent $i$ copies $j$'s state. $x_i(t+\delta)=x_j(t)$.
 \item If the update produces a change of state of node $i$, then $\tau_i$ is set to zero.
 \item The time is updated to a unit more and we return to 1 to keep on with the dynamics.
\end{enumerate}
The question now is if this modification will lead to qualitative changes in the outcome of the dynamics of the voter model.

\begin{figure}
\centering
\begin{tabular}{ccc}
\includegraphics[height=0.31\textwidth,angle=-90]{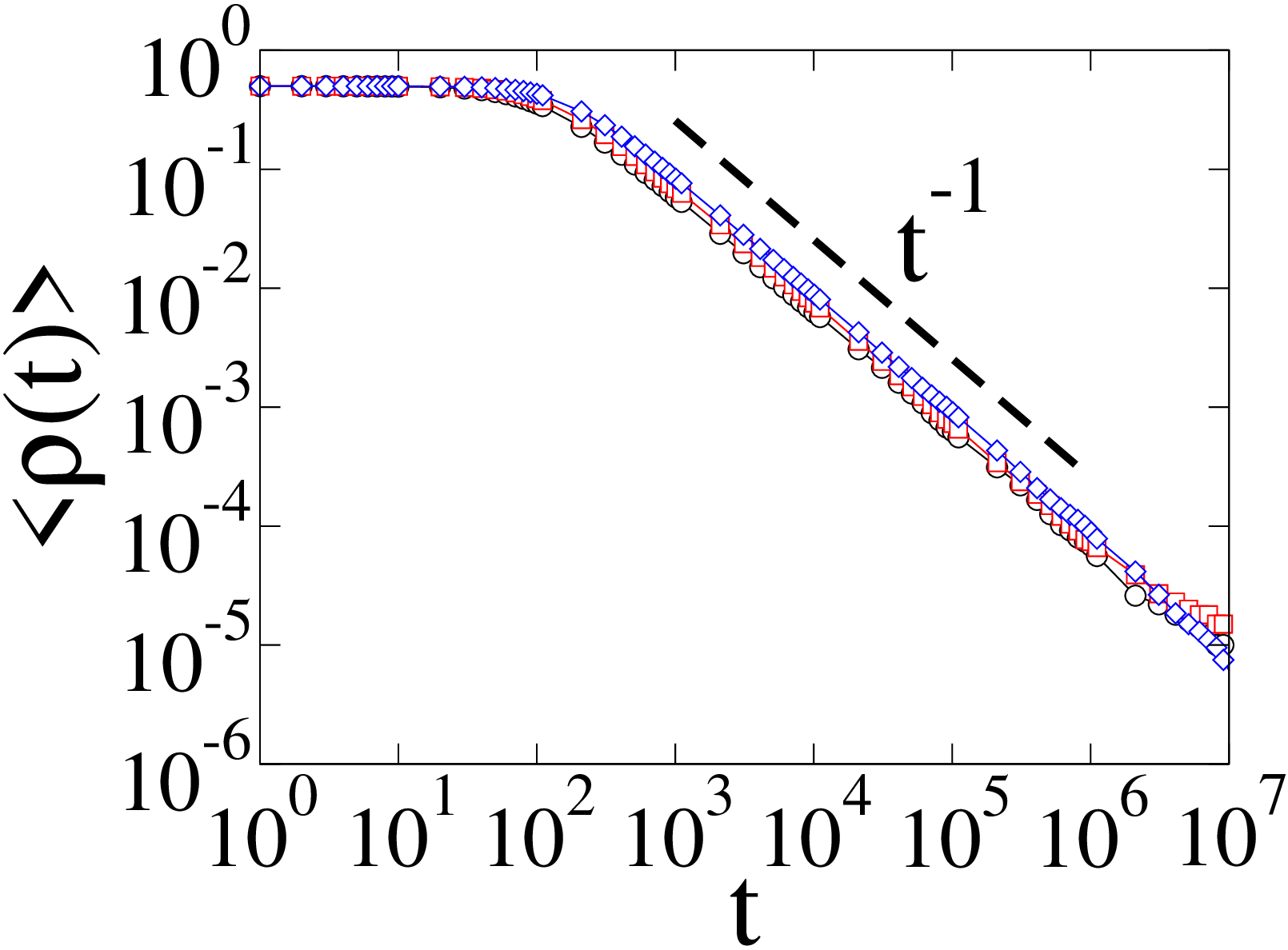}&\includegraphics[height=0.31\textwidth,angle=-90]{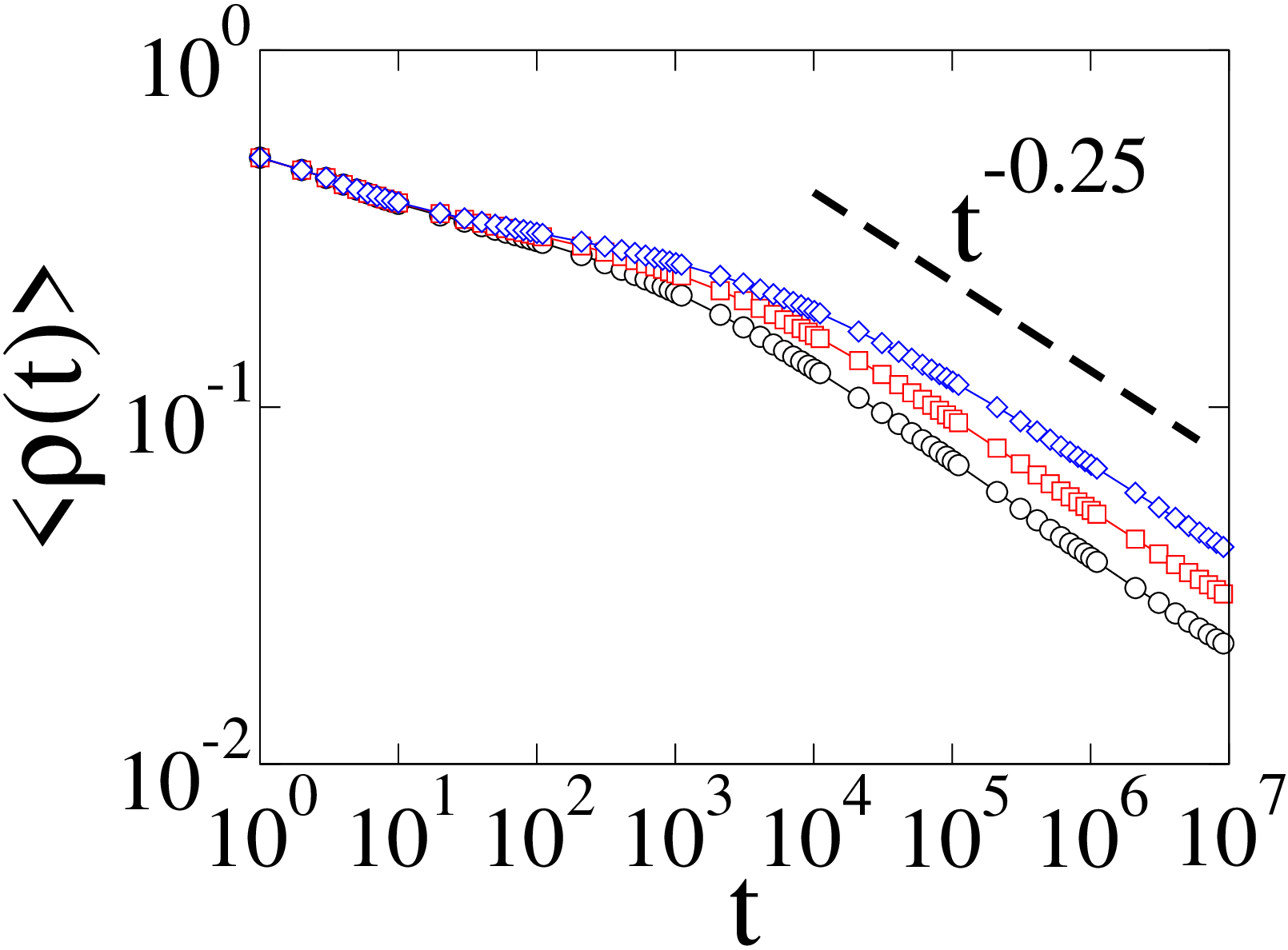}&\includegraphics[height=0.31\textwidth,angle=-90]{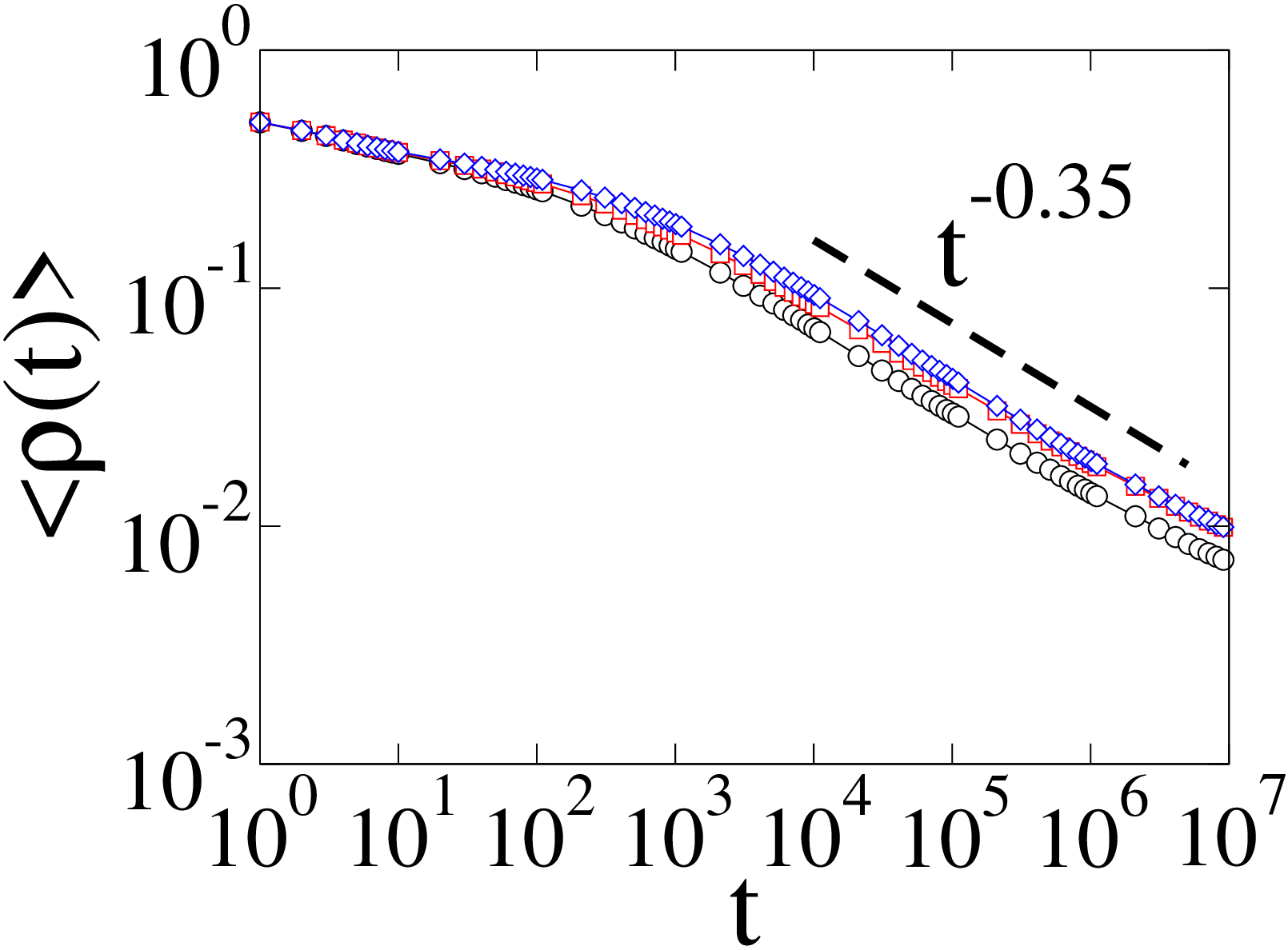}\\
\includegraphics[height=0.31\textwidth,angle=-90]{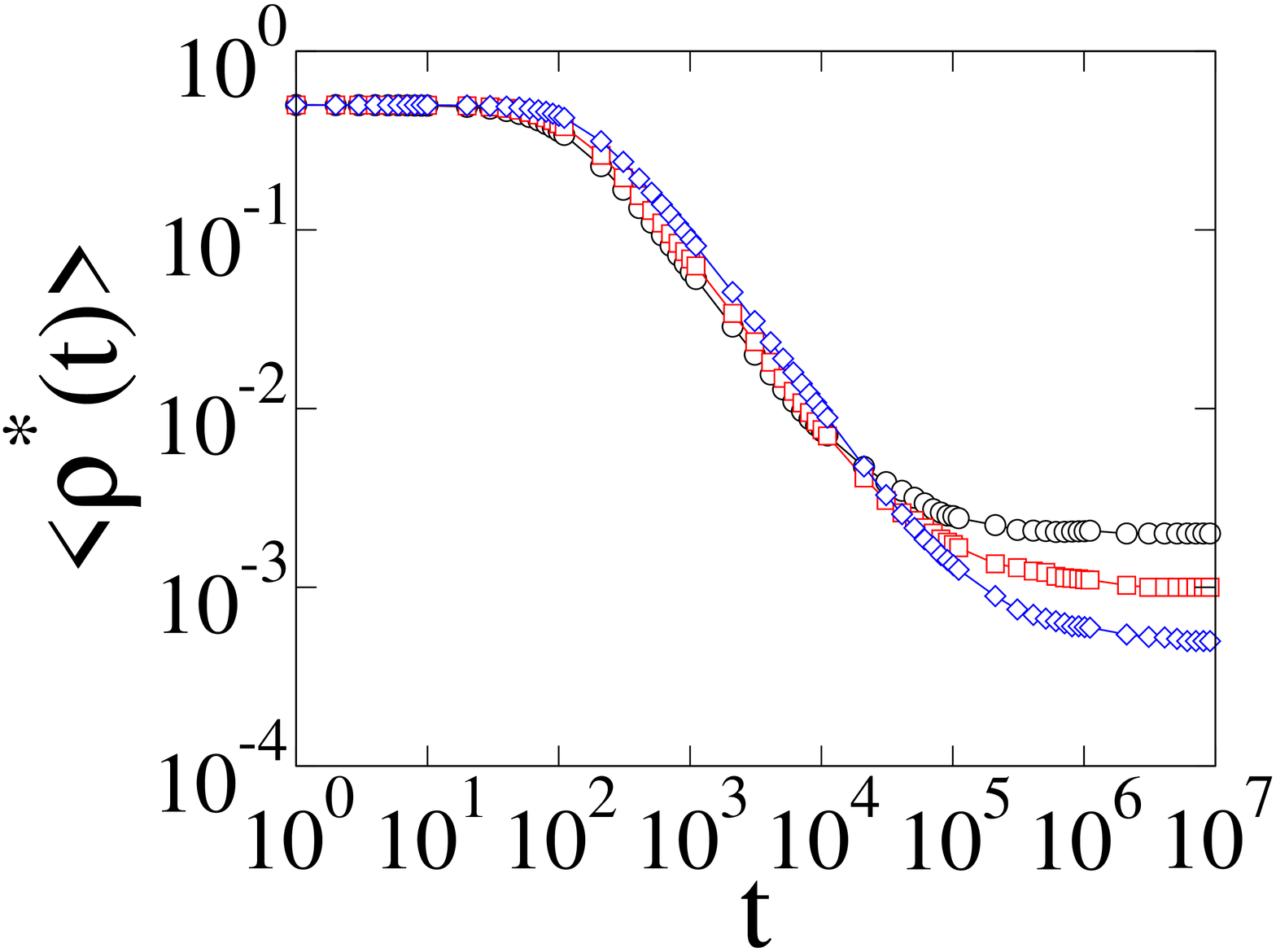}&\includegraphics[height=0.31\textwidth,angle=-90]{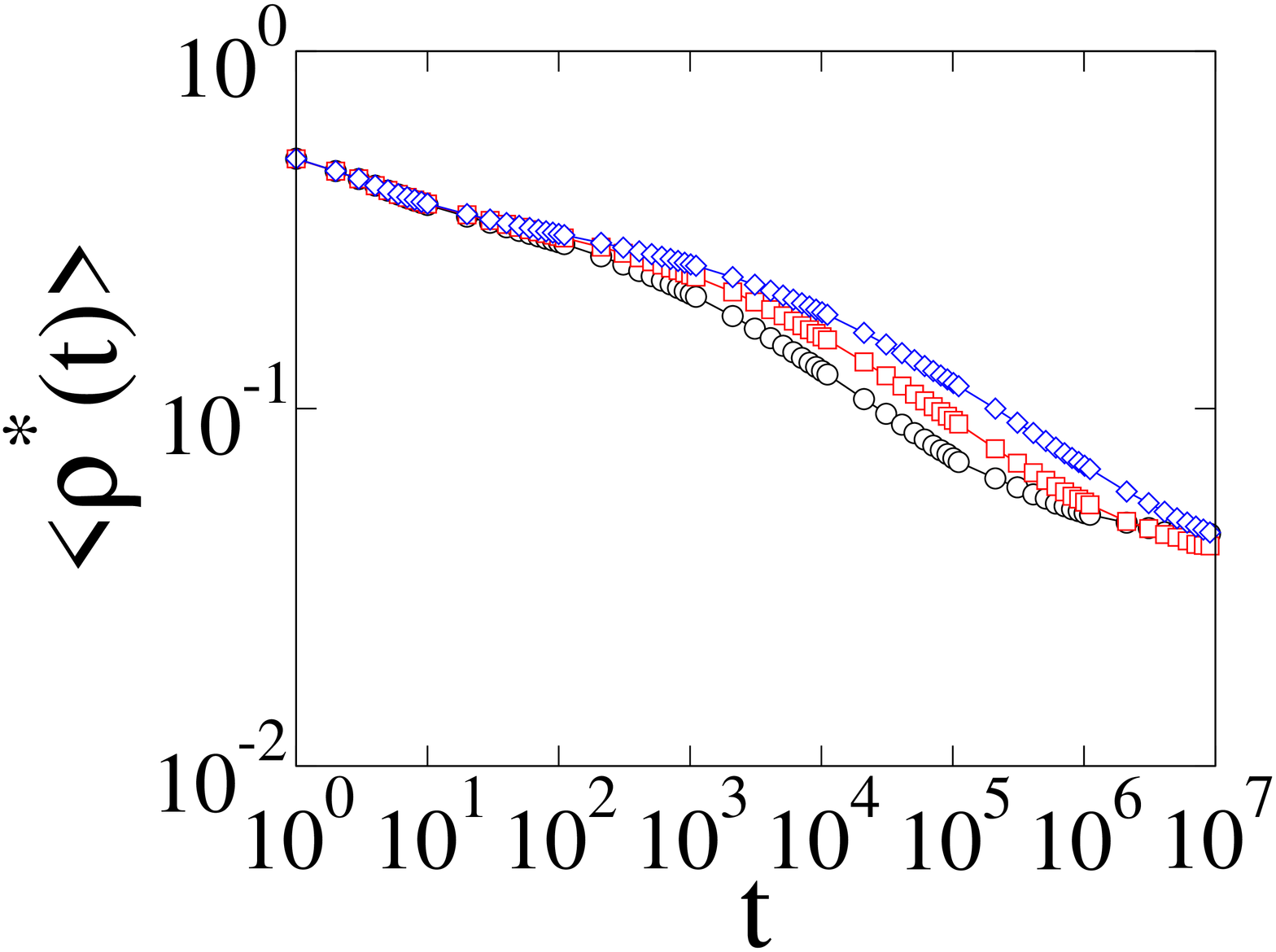}&\includegraphics[height=0.31\textwidth,angle=-90]{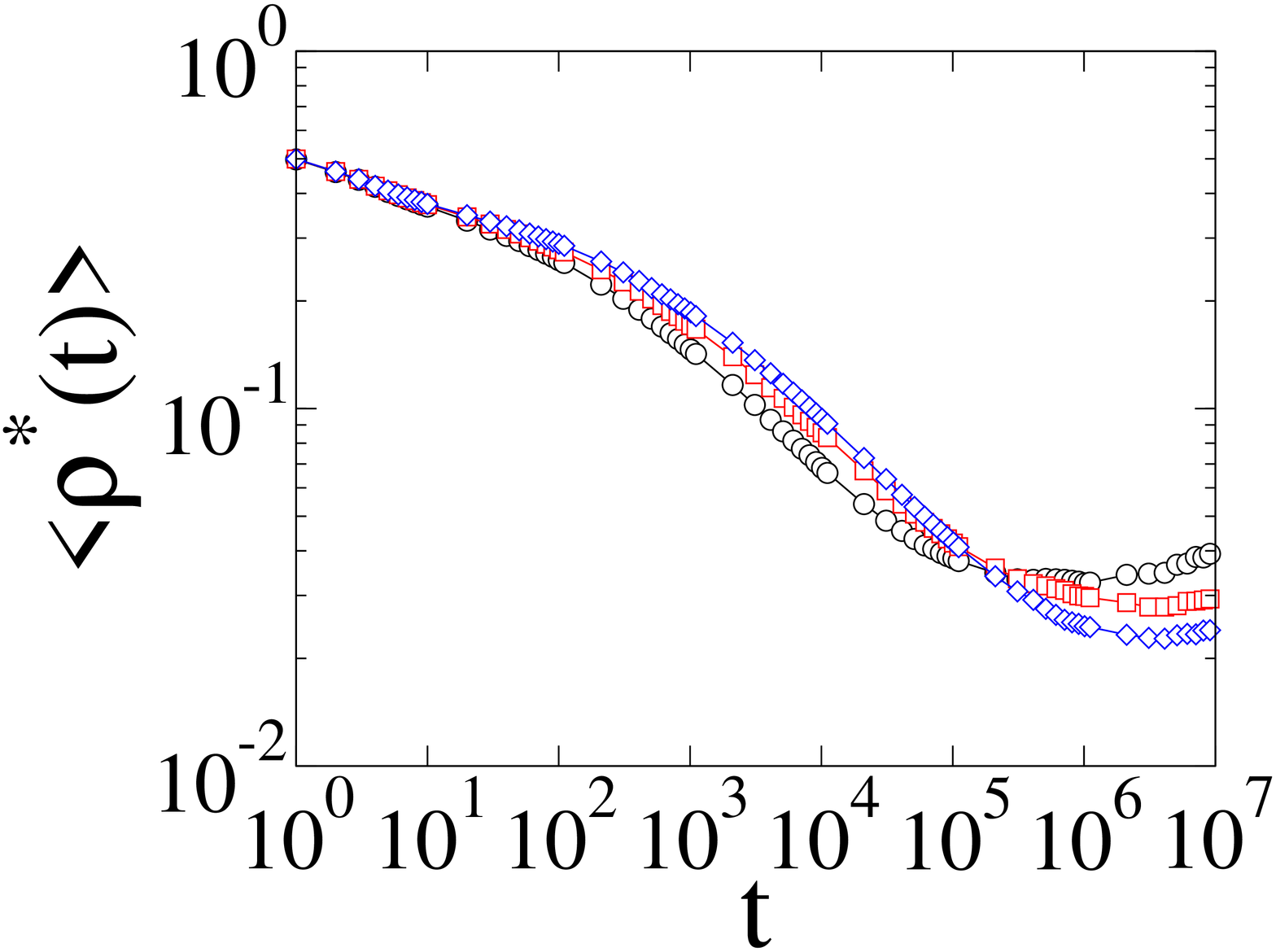}\\
\includegraphics[height=0.31\textwidth,angle=-90]{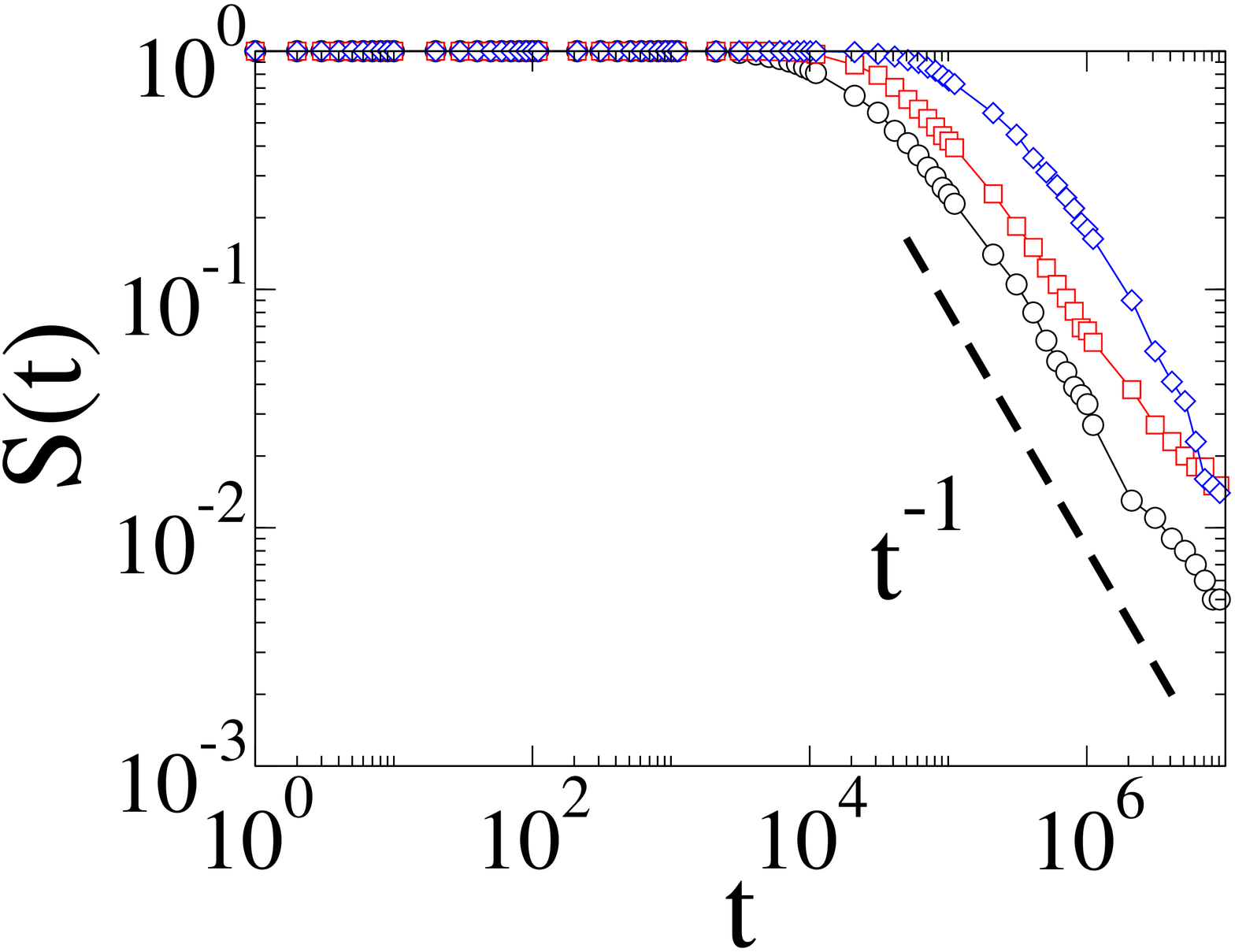}&\includegraphics[height=0.31\textwidth,angle=-90]{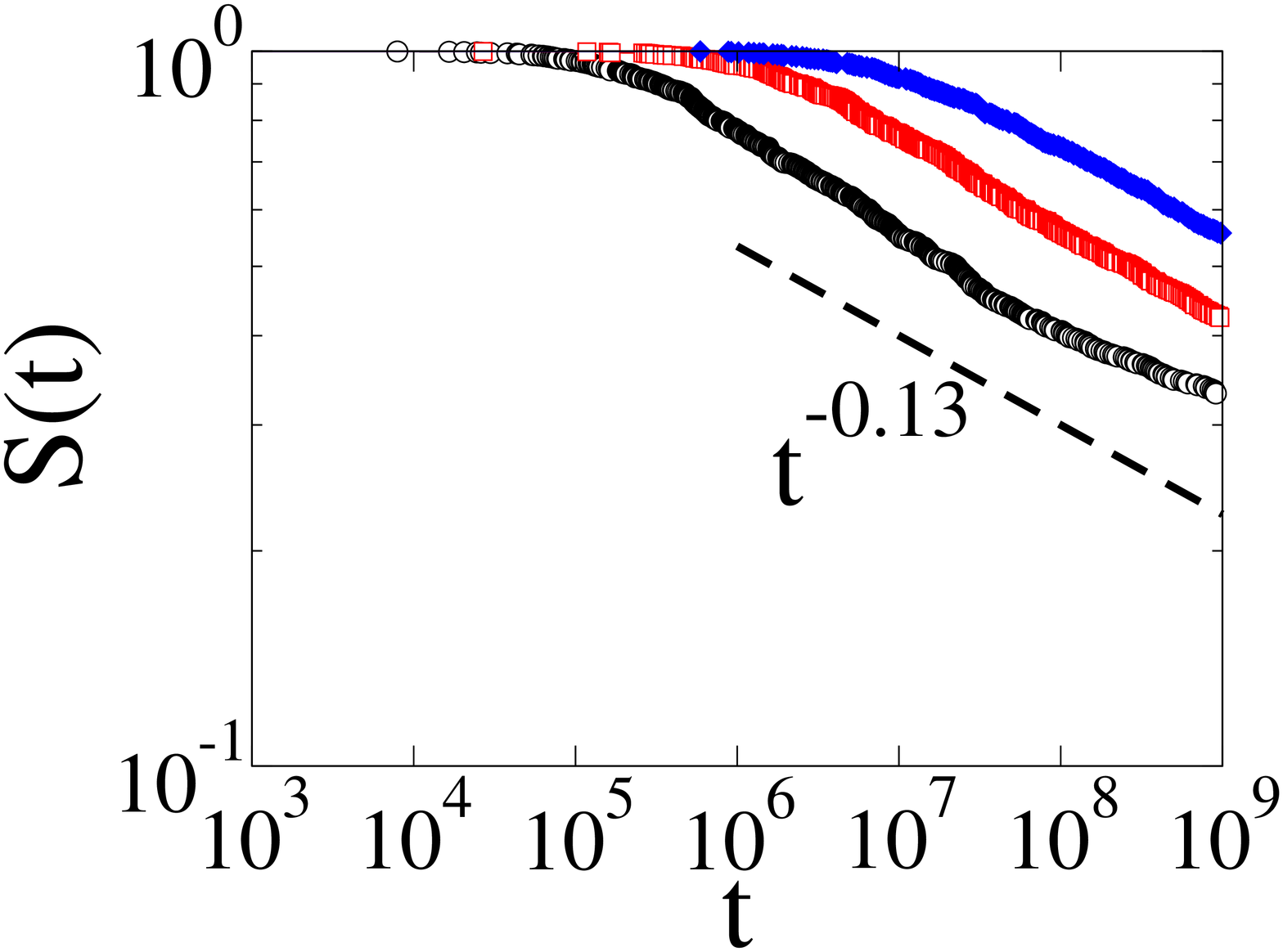}&\includegraphics[height=0.31\textwidth,angle=-90]{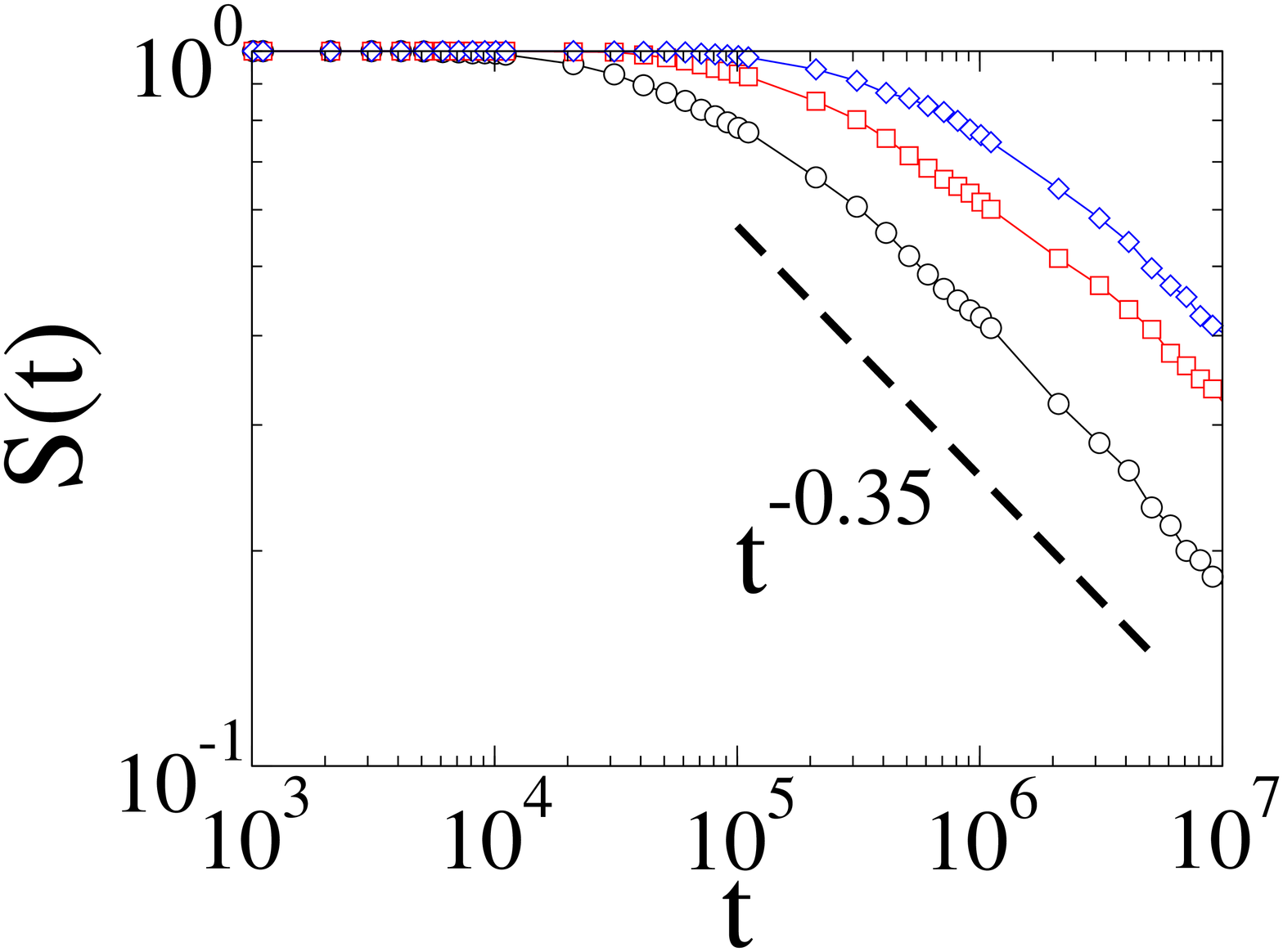}\\
\includegraphics[height=0.31\textwidth,angle=-90]{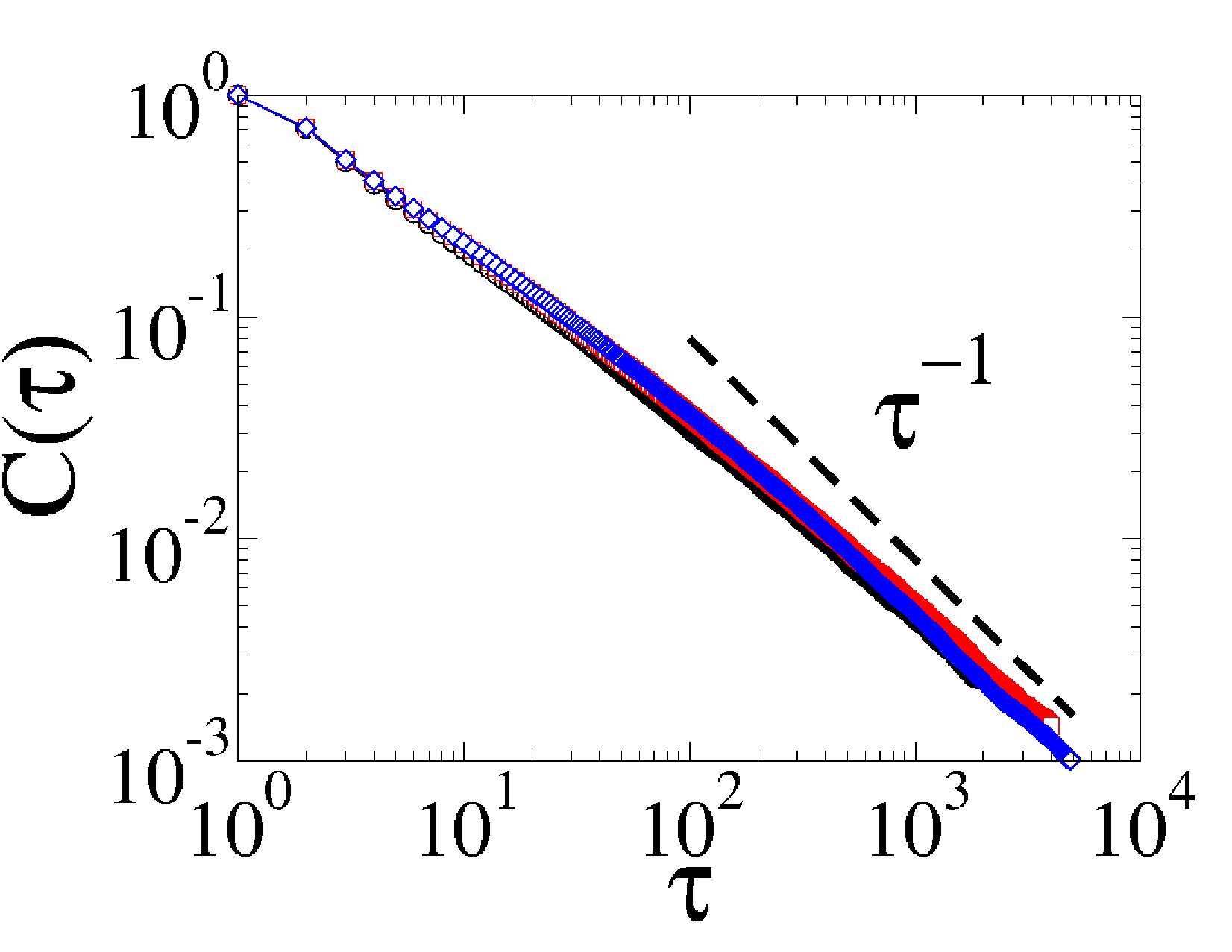}&\includegraphics[height=0.31\textwidth,angle=-90]{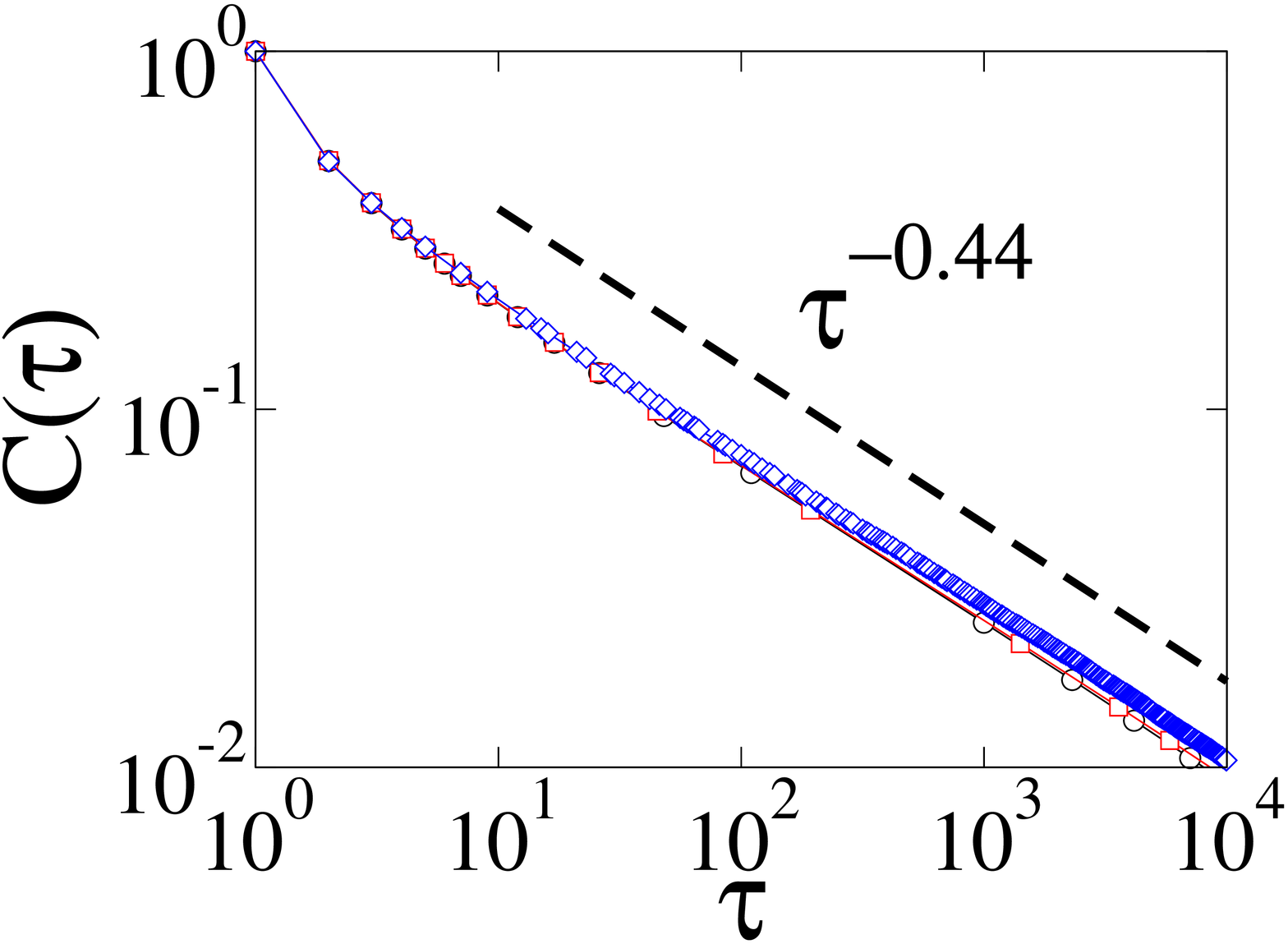}&\includegraphics[height=0.31\textwidth,angle=-90]{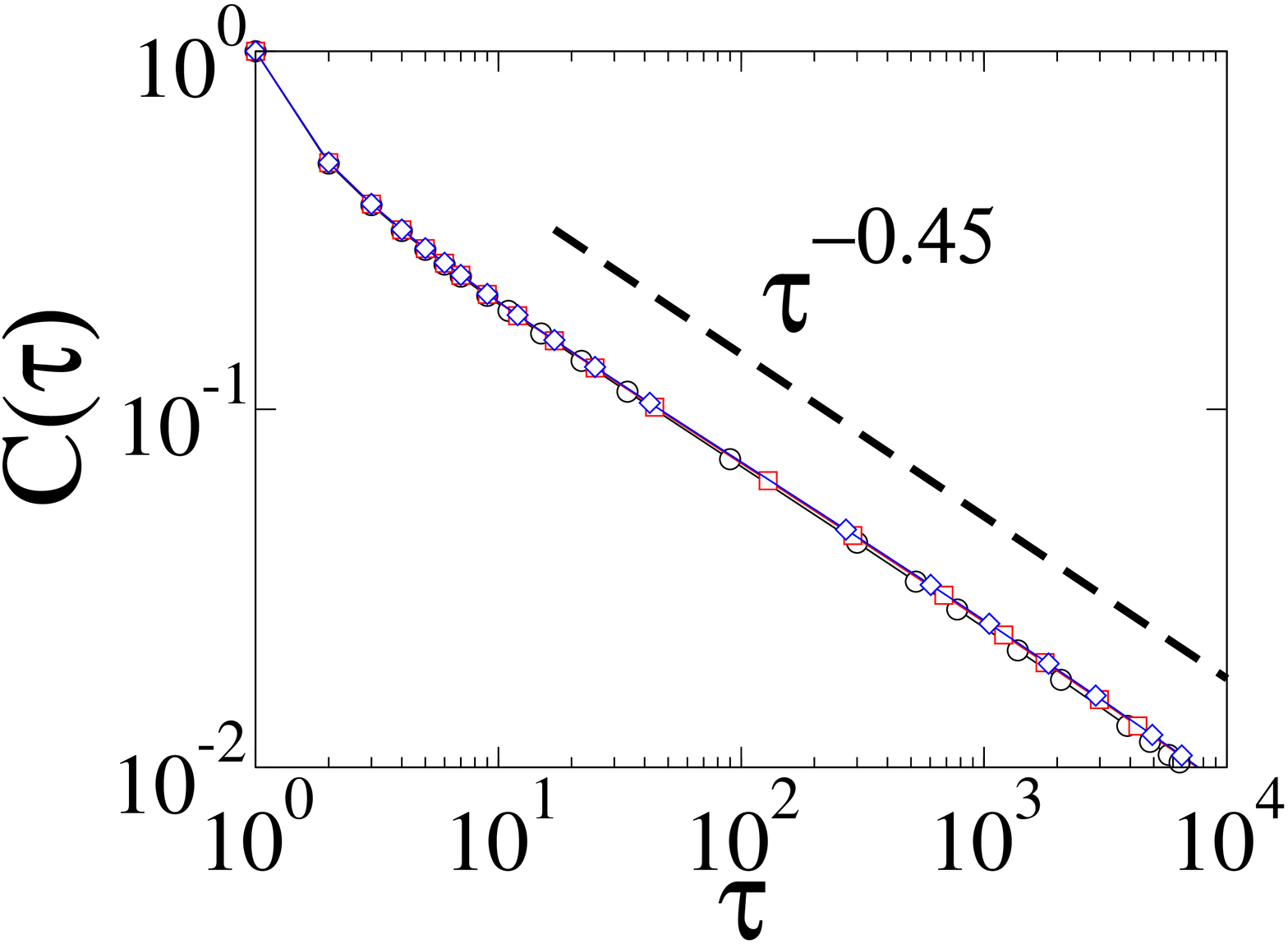}\\
\end{tabular}
\caption{Characteristics of the voter model with \emph{endogenous update} for several networks. Left column is for complete graphs of sizes $300$ in black,$1000$ in red and $4000$ in blue. Middle column is for random graphs with average degree $\langle k\rangle=6$ and sizes $1000$ in black,$2000$ in red and $4000$ in blue. Right column is for scale-free graphs with average degree $\langle k\rangle=6$ and sizes $1000$ in black,$2000$ in red and $4000$ in blue. Top row shows plots of the average density of interfaces $\langle\rho\rangle$, second row shows the density of interfaces averaged over surviving runs $\langle\rho^*\rangle$, third row shows the survival probability $S(t)$ and bottom row shows the cumulative IET distribution $C(\tau)$. The averages where done over $1000$ realizations.}
\label{tab:endonets}
\end{figure}

 For an activation probability $p(\tau)=1/\tau$, \textit{i.e.} $\beta=1$ we ran simulations on a complete graph, on random graphs of different average degree, and on a Barab\'asi-Albert scale-free graph of average degree $\langle k\rangle=6$ and for different system sizes (see Fig.\ref{tab:endonets}).

The exponents in the power laws of the quantities plotted in Fig.\ref{tab:endonets} are summarized in Table~(\ref{table_exp}) for the cases of complete, random and scale-free graph with mean degree $\langle k \rangle=6$.
\begin{table}
\label{table_exponents}
\begin{center}
\begin{tabular}{c|c|c|c|}
  & $\langle\rho(t)\rangle\propto t^{-\gamma}$ & $S(t)\propto t^{-\delta}$ & $C(\tau)\propto t^{-\beta}$ \\
\hline
 Complete graph & $\gamma=0.985(5)$ & $\delta=0.95(2)$ & $\beta=0.99(3)$ \\
 Random graph $\langle k \rangle=20$ & $\gamma=0.99(1)$ & $\delta=0.82(1)$ & $\beta=0.94(4)$ \\
 Random graph $\langle k \rangle=6$ & $\gamma=0.249(4)$& $\delta=0.13(1)$ & $\beta=0.45(1)$ \\
 Scale-free graph $\langle k \rangle=6$ & $\gamma=0.324(7)$ & $\delta=0.32(1)$ & $\beta=0.46(1)$ \\
\hline
\end{tabular}
\end{center}
\caption{Exponents for the power law decaying quantities $\rho(t)$, $S(t)$ and $C(\tau)$ for the voter model with the endogenous update rule.}
\label{table_exp}
\end{table}
We can see from the Table that, increasing the average degree of the random networks we get results that get closer to the ones on a complete graph.\\
Our results can be summarized as follows:
\begin{description}
 \item{\textit{Density of active links $\langle\rho(t)\rangle$ and $\langle\rho^*(t)\rangle$: }}{When averaged over all runs, $\langle\rho(t)\rangle$ decays as a power law with different exponents depending on the interaction network. When averaged over active runs $\langle\rho^*(t)\rangle$ we see that it decays as a power law until it reaches a plateau whose height depends on the system size and is smaller for bigger system sizes, which tells that \emph{the system is heading towards consensus}, contrary to what happens with the standard update rules. This is one of the main results of the present work.}
 \item{\textit{Survival probability $S(t)$: }}{It is one until it decays, also like a power law. The exponents are in all cases smaller or around $1$, so that \emph{the average time to reach consensus diverges for all system sizes}. Remember that the mean time to reach consensus is $\langle T\rangle=\int_0^{\infty}S(t)dt$. So, a proper average consensus time is not defined.}
 \item{\textit{Cumulative IET distribution $C(\tau)$: }}{Develops a power law tail. For a complete graph and a random network with high degree we recover an exponent $\beta$ in the tail of the interevent times cumulative distribution $C(\tau)$ that matches the one we wanted it to follow given our calculations and our choice $p(\tau)=1/\tau$. For the other two networks, random and scale-free with $\langle k\rangle=6$ we recover that the tail behaves approximately as $1/\sqrt{\tau}$.}
\end{description}
\emph{The dynamics does not order the system with the endogenous update through a coarsening process that leads to the divergence of the mean time to reach consensus for all system sizes.}
As a summary, the complete graph case gives us already the qualitative behavior: for the voter model with exogenous update the timescales are much larger than in the voter model with RAU, but it has the same qualitative behavior: the system doesn't order in the thermodynamic limit, but stays in a disordered dynamical configuration with asymptotic coexistence of both states. This contrasts with what happens with the endogenous update, where the timescales are also perturbed, but with the difference that a coarsening process occurs, slowly ordering the system. We have checked that the ensemble average of the magnetization $\langle m(t) \rangle=\frac{1}{N} \sum_{i=1}^N \langle s_i(t)\rangle$ is conserved for the exogenous update, whereas for the endogenous update this conservation law breaks down, as previously discussed in Ref.~\cite{tessone}. The non-conservation of the magnetization leads to an ordering process. The conservation law is broken due to the different average values of the persistence time in both populations of agents (+1 and -1) leading to different average activation probabilities.

\subsubsection{Varying the exponents of the cumulative IET distribution $C(\tau)$}
\begin{figure}[H]
  \centering
  \subfloat{\label{cemeanf}\includegraphics[draft=false,height=0.31\textwidth,angle=-90]{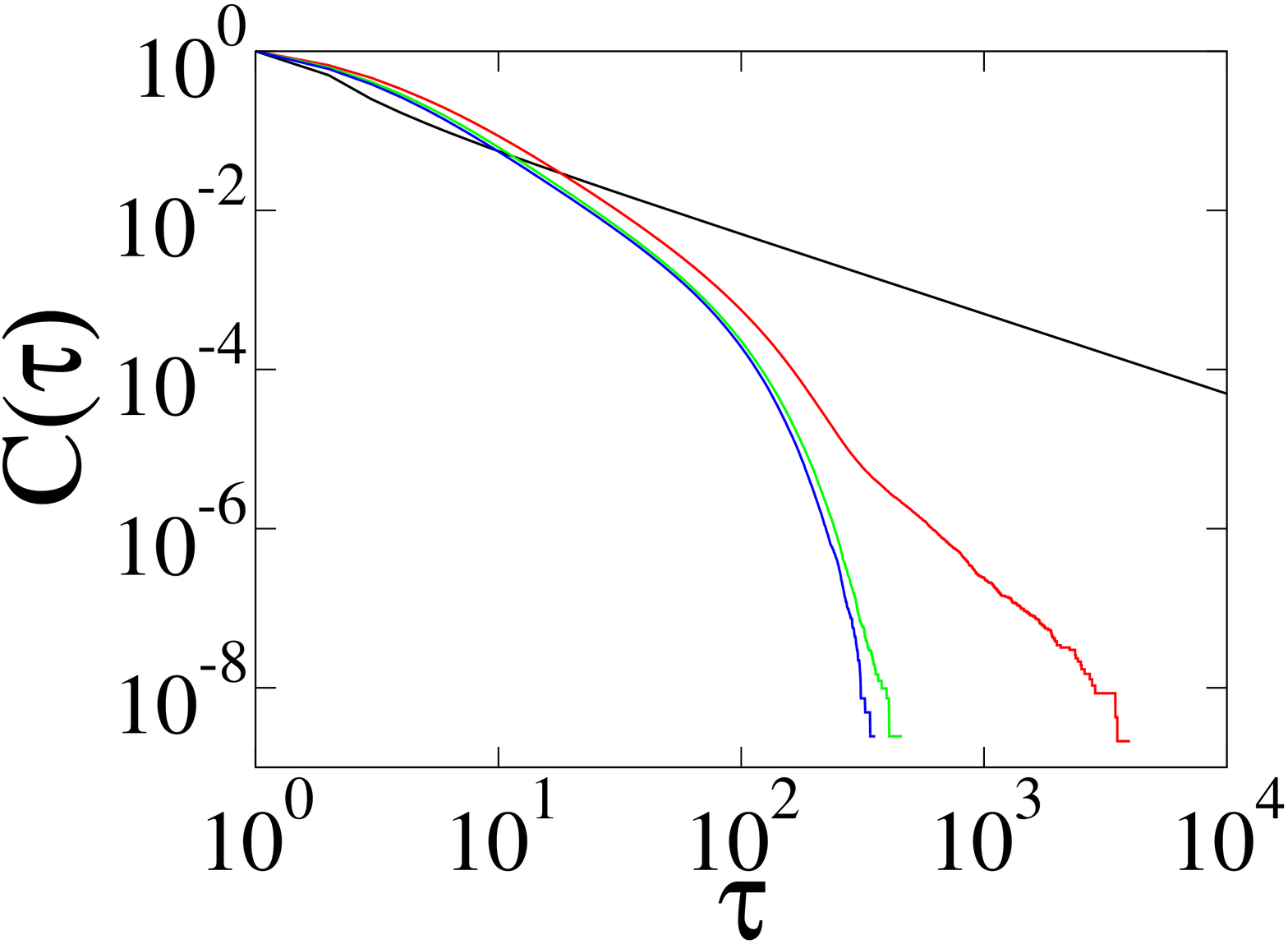}}\quad
  \subfloat{\label{cerand}\includegraphics[draft=false,height=0.31\textwidth,angle=-90]{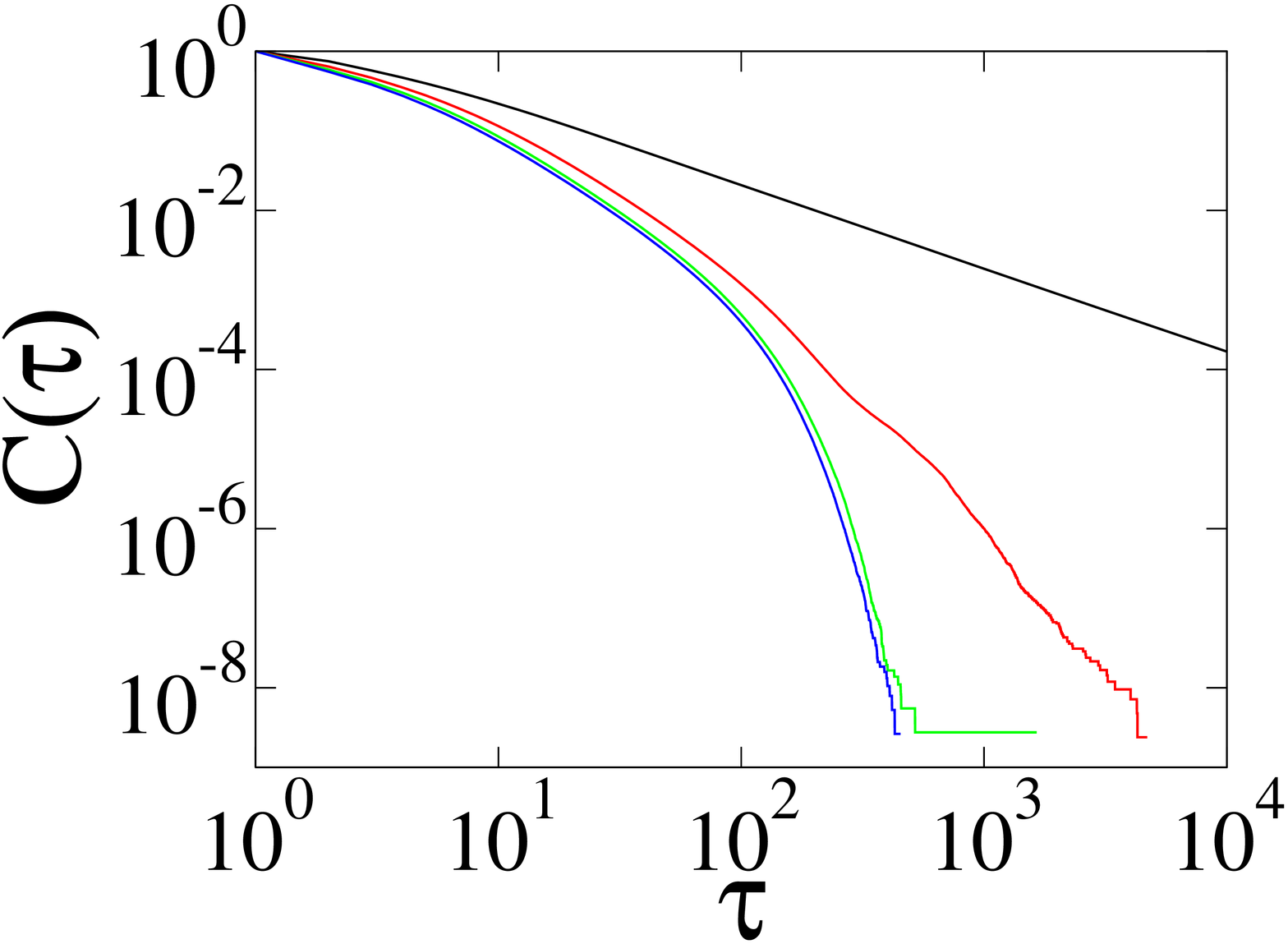}}\quad
  \subfloat{\label{ceBA}\includegraphics[draft=false,height=0.31\textwidth,angle=-90]{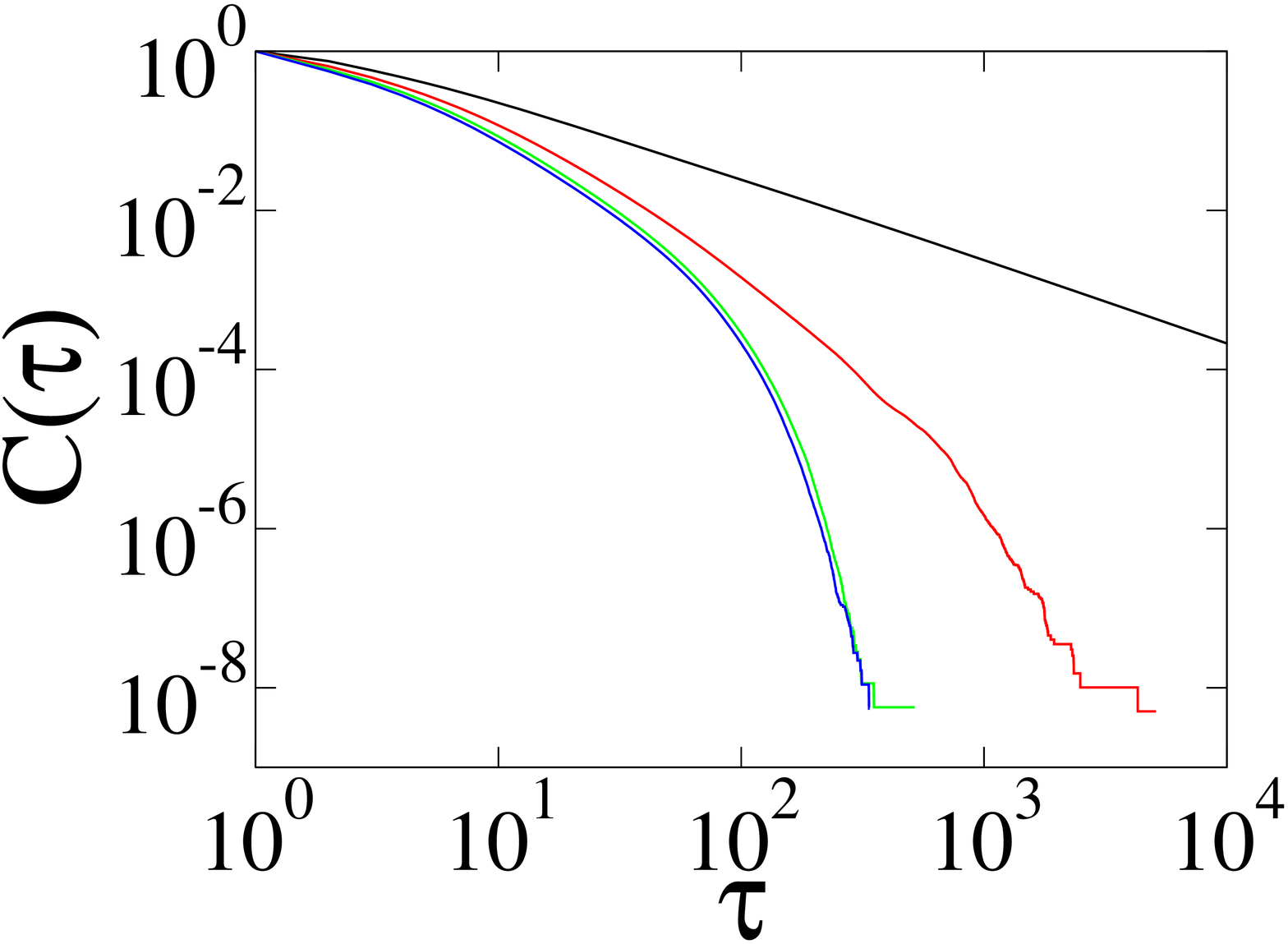}}
  \caption{\emph{Exogenous update:} cumulative IET distribution $C(\tau)$ for different values of the parameter $b$ (grows from right to left) appearing in the activation probability $p(\tau)$ for complete graph, random graph with $\langle k\rangle=6$ and Barab\'asi-Albert scale-free network with $\langle k\rangle=6$ and for system size $N=1000$.}
  \label{cumul_exo_varbeta}
\end{figure}

As was shown in section \ref{coev_update_rule} the exponent in the cumulative IET distribution $C(\tau)\propto \tau^{-\beta}$ should be related to the parameter $b$ appearing in the activation probability $p(\tau)=b/\tau$. If every time we let an agent be updated, this one changes state, this relation is such that $\beta=b$. When introducing the dynamics, this relation is not so clear and depends also on the kind of network where the dynamics are taking place.
In Fig.~\ref{cumul_exo_varbeta} we can see the interevent times cumulative distributions for different values of $b$ for the exogenous update.
We can see that for $b=1$ the power law tail is recovered with an exponent that matches $\beta=b$. For higher values of $b$ the form of the tail is rapidly lost and we have cumulative IET distribution $C(\tau)$ are similar to those with standard update rules, \textit{i.e.}, do not display heavy tails.

\begin{figure}[H]
  \centering
  \subfloat{\label{cemeanf}\includegraphics[draft=false,height=0.31\textwidth,angle=-90]{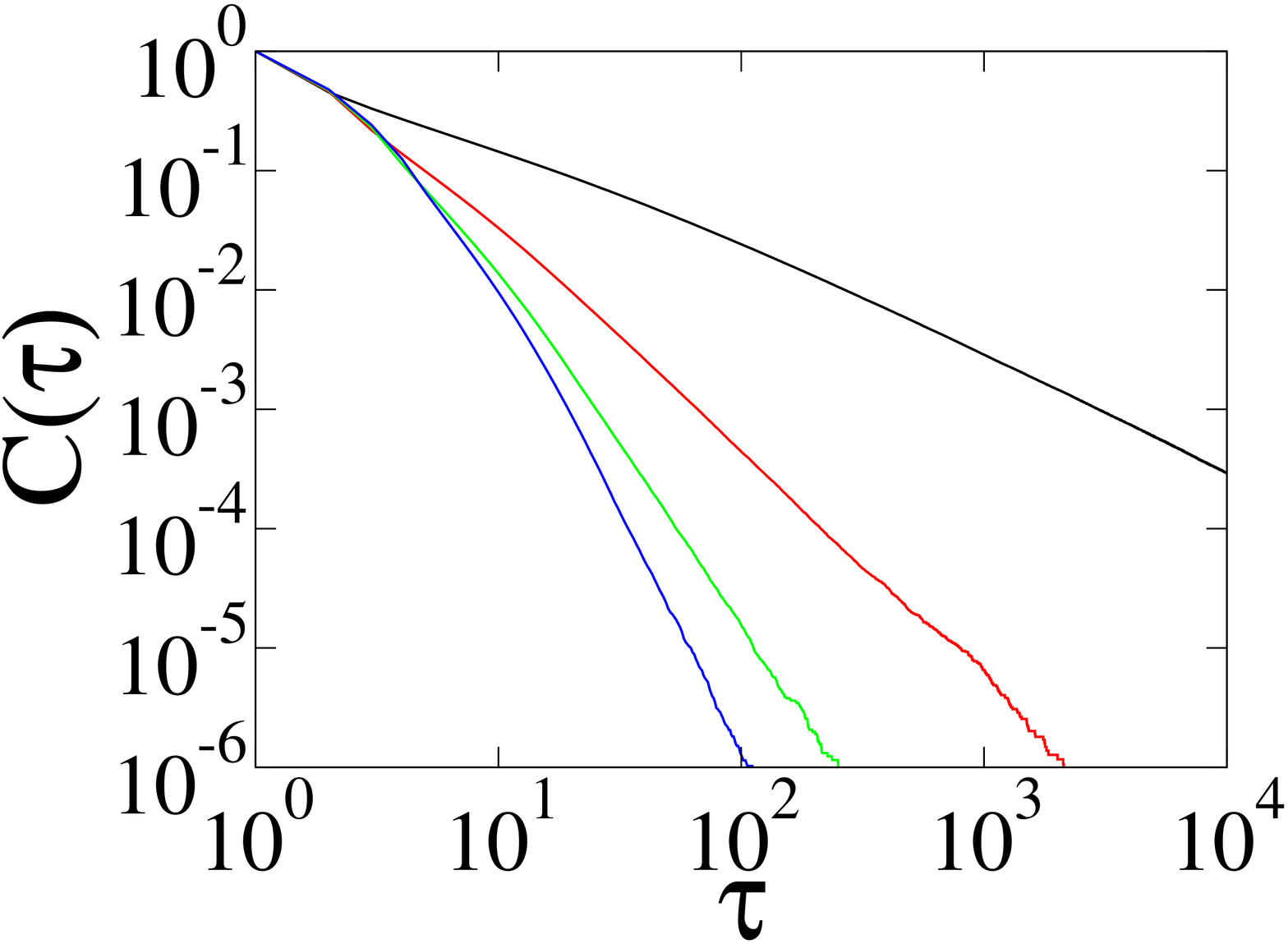}}\quad
  \subfloat{\label{cerand}\includegraphics[draft=false,height=0.31\textwidth,angle=-90]{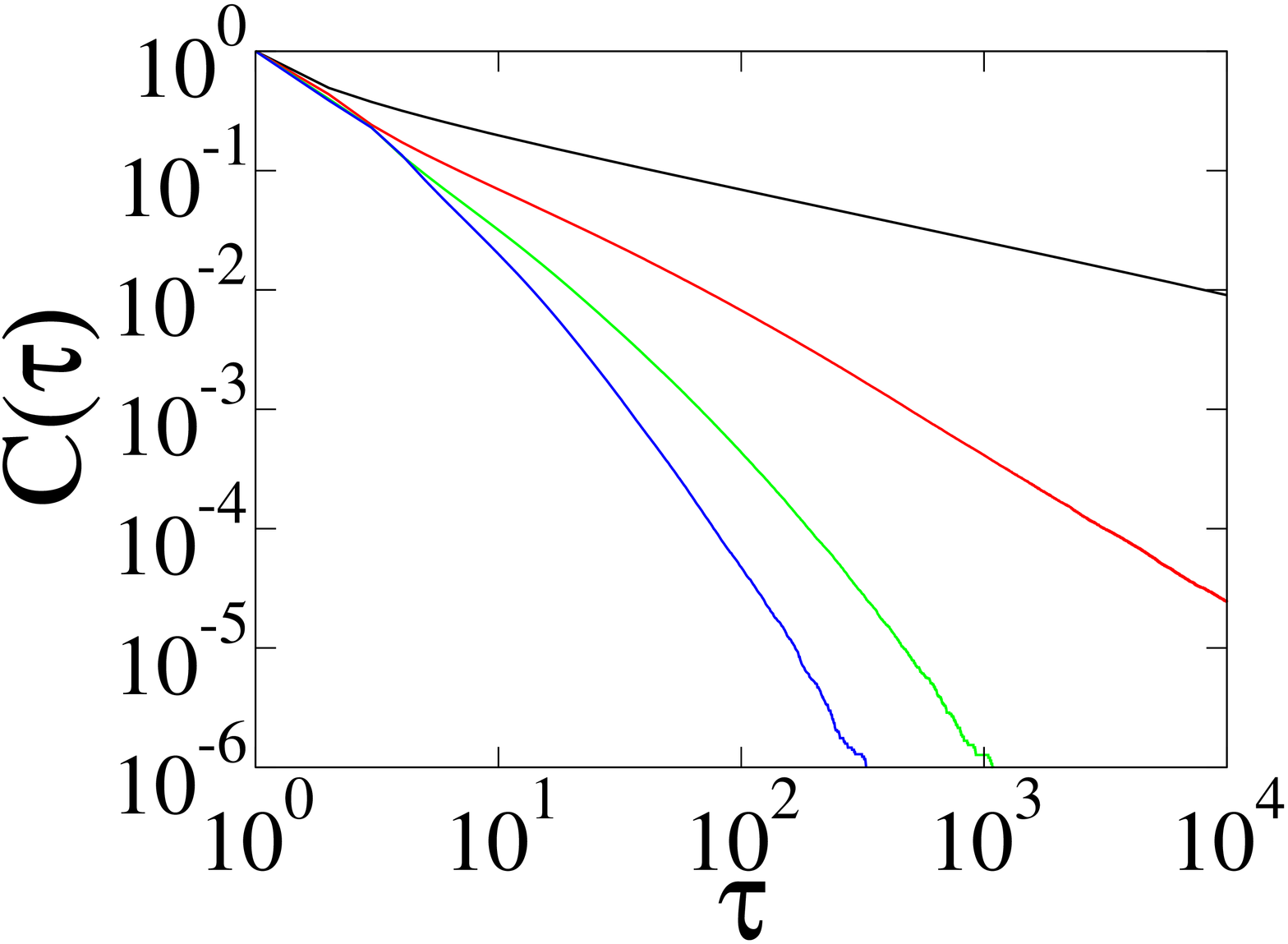}}\quad
  \subfloat{\label{ceBA}\includegraphics[draft=false,height=0.31\textwidth,angle=-90]{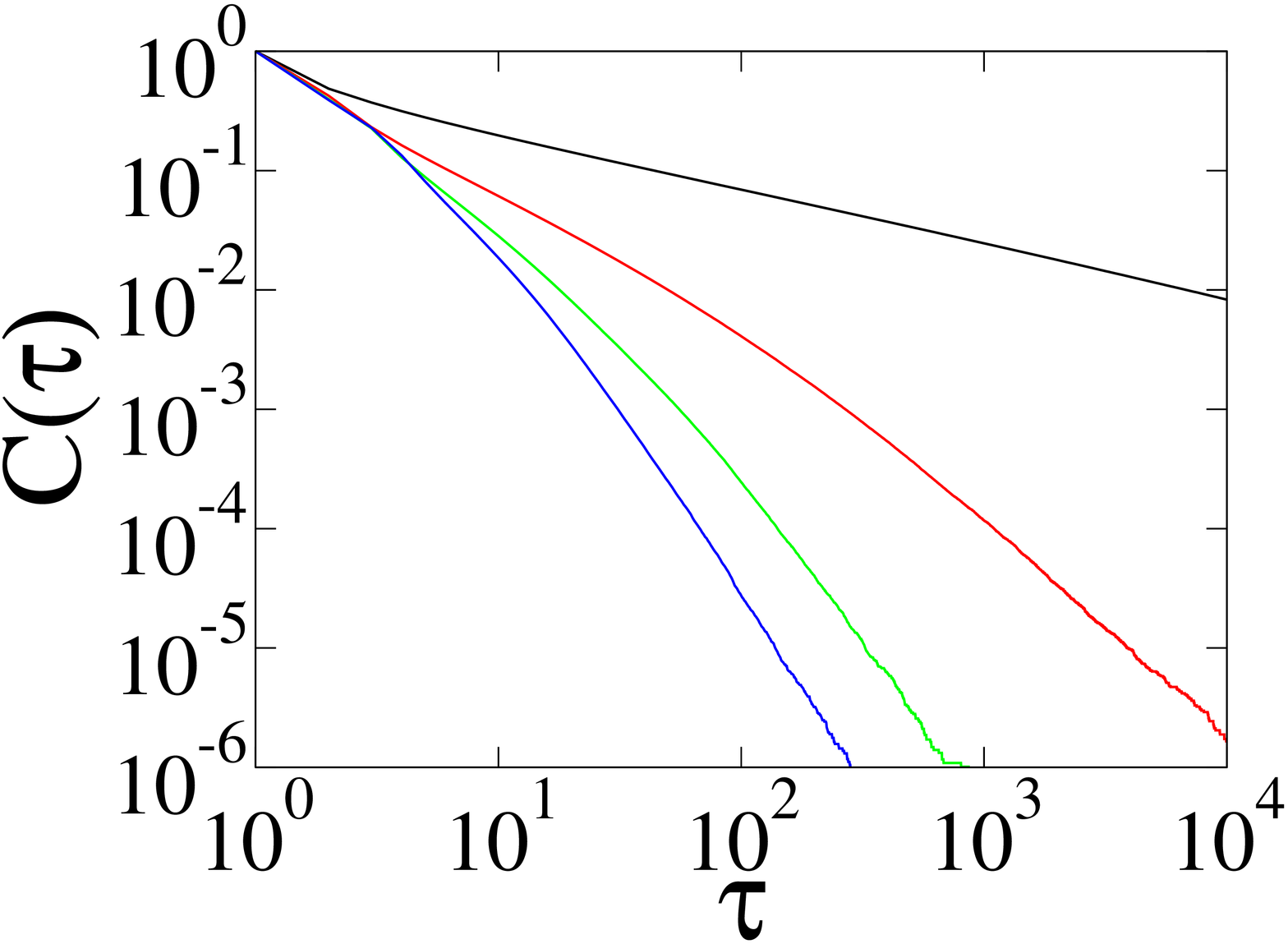}}
  \caption{\emph{Endogenous update:} cumulative IET distribution $C(\tau)$ for different values of the parameter $b$ (grows from right to left) appearing in the activation probability $p(\tau)$ for complete graph, random graph with $\langle k\rangle=6$ and Barab\'asi-Albert scale-free network with $\langle k\rangle=6$ and for system size $N=1000$.}
  \label{cumul_varbeta}
\end{figure}

\begin{figure}
 \centering
 \includegraphics[height=0.95\textwidth,angle=-90,draft=false]{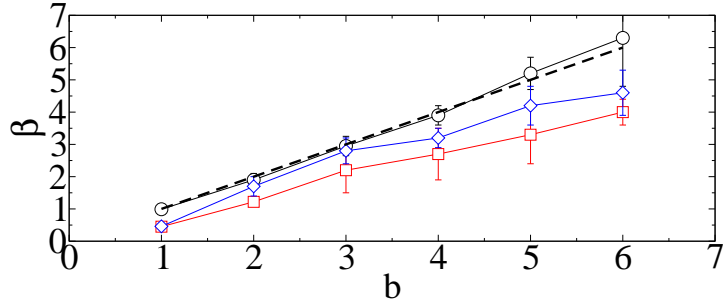}
 \caption{\label{rel_betas_topologies}\emph{Endogenous update.} Relation of $\beta$, the exponent of the cumulative IET distribution $C(t)\sim t^{-\beta}$, and $b$, the parameter in the function $p(\tau)=b/\tau$ for three different topologies; fully connected (circles), random with $\langle k\rangle=6$ (squares) and scale free  with $\langle k\rangle=6$ (diamonds) networks. As a guide to the eye we plot the curve $\beta=b$ with a dashed line. The bars stand for the associated standard errors of the measures.}
\end{figure}

In Fig.~\ref{cumul_varbeta} we can see the interevent times cumulative distributions for different values of $b$ for the endogenous update.
The endogenous update rule has a wider range of $b$-values for which the heavy tail is recovered. We measured the exponents of the tails for different values $b$ in the different topologies. The results can be seen in Fig.~\ref{rel_betas_topologies}

Surprisingly, for the case of the complete graph, we recover the relation predicted, \textit{i.e.} a linear relation between $\beta$ in the cumulative distribution function and $b$, the parameter in the probability $p(\tau)$.

In the case of other topologies we find that the relation $b-\beta$ is not the one predicted in the case of no interactions, but it displays a reminiscent behavior of the one observed for a complete graph: the exponent $\beta$ found in the cumulative interevent time distribution increases monotonically with the parameter $b$ in the activation probability.

\subsubsection{Effective events}

\begin{figure}[H]
\centering
\begin{tabular}{cc}
\includegraphics[height=0.5\textwidth,angle=-90]{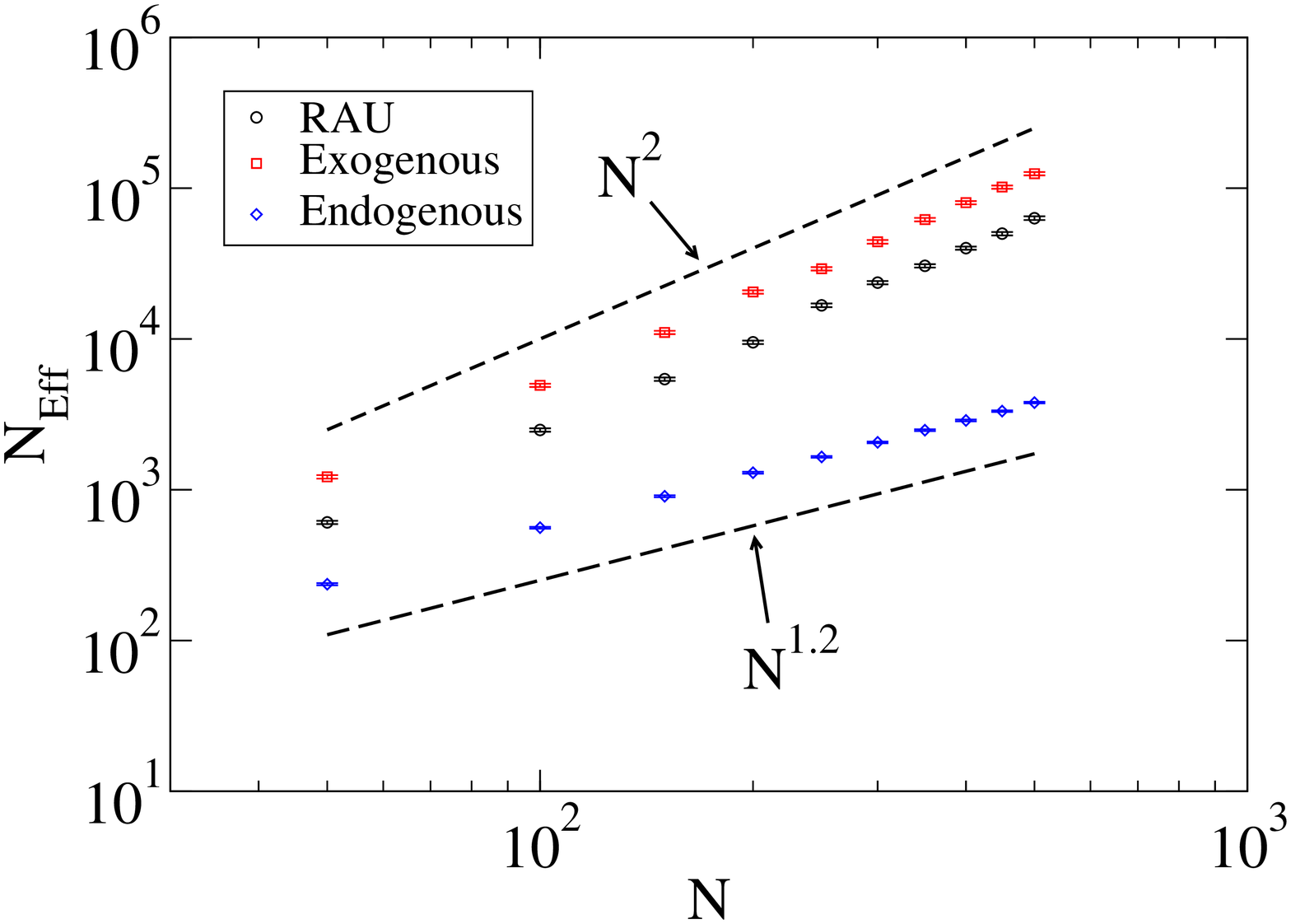}&\includegraphics[height=0.5\textwidth,angle=-90]{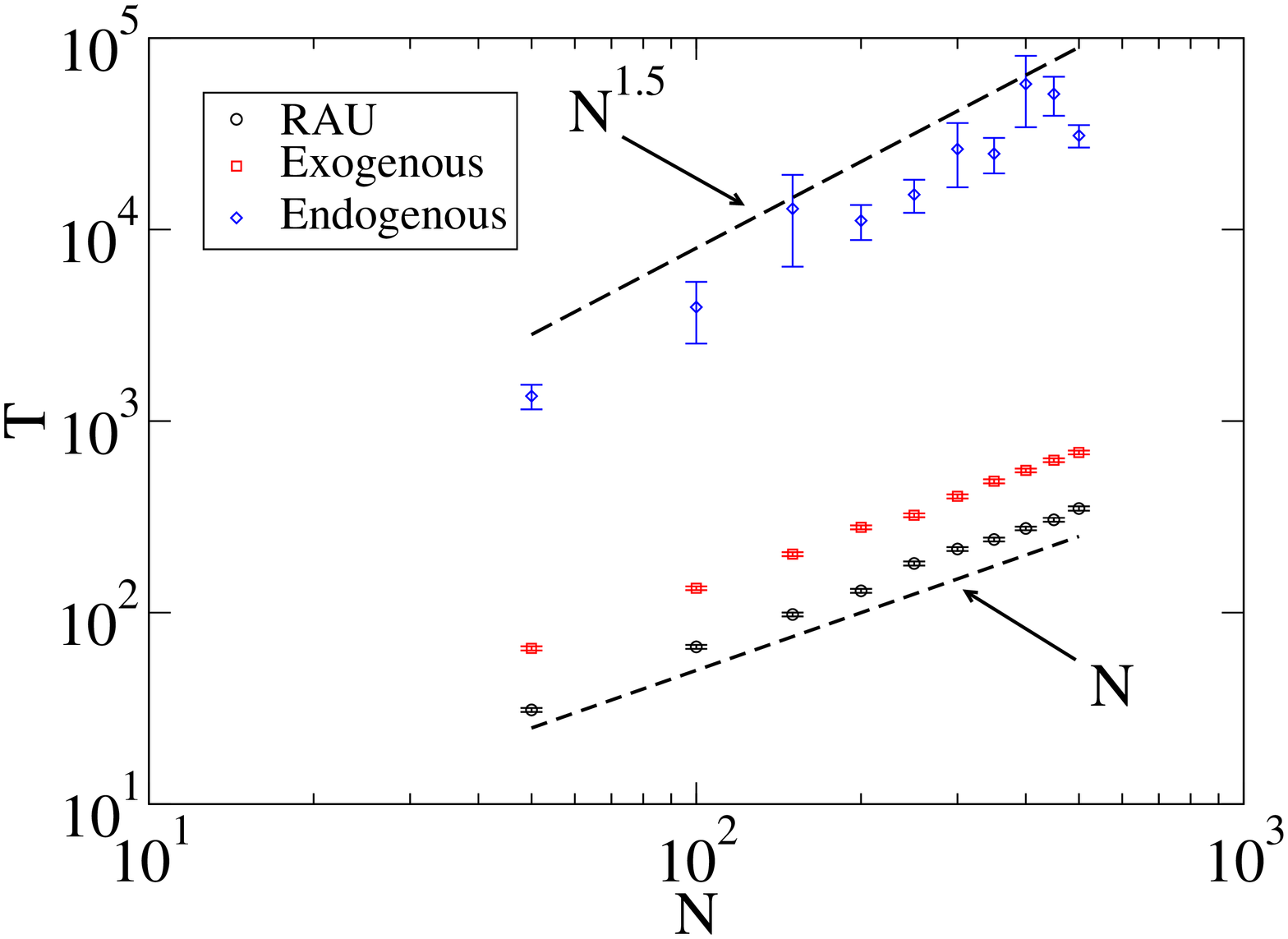}\\
\end{tabular}
\caption{On the left we can see the scaling of the number of effective events with system size for a complete graph and three different update rules, RAU, exogenous and endogenous. On the right we can see the scaling of the consensus time with system size for a complete graph and three different update rules, RAU, exogenous and endogenous. }
\label{tab:effevents}
\end{figure}

An interesting feature is the number of effective events, \textit{i.e.} updates that result in a change of state, are needed to get to consensus. It happens that for the usual update rules and the exogenous update, the scaling with system size is the same, while the endogenous update follows a different scaling (\textit{cf.} left plot in Fig.\ref{tab:effevents} for the case of complete graph), signaling the difference due to the coarsening process that appears for the endogenous update. Furthermore the number of effective events needed with the endogenous update to order the system is much less than with the other update rules. This efficiency in ordering is due to the coarsening process that occurs with the endogenous update. Even though, in terms of time steps, the exogenous update is much slower, such that the time to reach consensus diverges. In the right plot of Fig.\ref{tab:effevents} we see a time for reaching consensus for the endogenous update, but this time will diverge if the sample of realizations taken for the average is big enough.

\section{Discussion}

The take home message of this work is to beware of social simulations of interacting individuals based on a constant activity rate: Human activity patterns need to be implemented as an essential part of social simulation. We have shown that heterogeneous interevent time distributions can produce a qualitative change in the voter model of social consensus, leading from dynamical coexistence of equivalent states to ordering dynamics. More specifically, we have shown that for standard update rules (SAU, RAU, SU) of the voter model dynamics in networks of high dimensionality (Fully connected, random, scale free) the system remains in long lived disordered dynamical states of coexistence of the two states, and activity patterns are homogeneous with a well defined characteristic interevent time. A power law tail for the cumulative interevent time distribution is obtained with two forms of the update rule accounting for heterogeneous activity patterns. For an exogenous update rule the dynamics is still qualitatively the same than for standard update rules: the system does not order, remaining trapped in long lived dynamical states.  However, when the update rule is coupled to the states of the agents (endogenous update) it becomes part of the dynamical model, modifying in an essential way the dynamical process: there is coarsening of domains of nodes in the same state, so that the system orders approaching a consensus state. Also the times to reach consensus in the endogenous version of the update rule are such that a mean time to reach consensus is not well defined. In fact the scaling of effective events needed for consensus is able to give a signature of which of the updates is ordering the system. In summary, when drawing conclusions from microscopic models of human activity, it is necessary to take into account that the macroscopic outcome depends on the timing and sequences of the interactions. Even if recovering heterogeneous interevent time distributions the type of update (exogenous vs. endogenous rule) can modify the ordering dynamics.

Recent research on human dynamics has revealed the ``small but slow" paradigm \cite{small_but_slow,vazquez_non_poiss}, that is, the spreading of an infection can be slow despite the underlying small-world property of the underlying network of interaction. Here, with the help of a general updating algorithm accounting for realistic interevent time distributions, we have shown that the competition of two states can lead to slow ordering not only in small-world networks but also in the mean field case. Our results provide a theoretical framework that bridges the empirical efforts devoted to uncover the properties of human dynamics with modeling efforts in opinion dynamics.

Works closely related to our research are those in Refs. \cite{tessone,baxter,masuda}. Stark \textit{et al.} \cite{tessone} introduced an update rule similar to our \textit{endogenous update} and focused on consensus times. However they did not explore the activity patterns followed a heavy tail distribution for the interevent intervals. They found that by slowing the dynamics, introducing a probability to interact that decays with the time since the last change of state, consensus formation could be actually accelerated. Baxter \cite{baxter} introduced a time dependence in the flip rates of the voter model. He explored the case when the flip rates vary periodically obtaining that consensus times depended non-trivially on the period of the flip-rate oscillations, having larger consensus times for larger periods, until it saturates. Finally, Takaguchi and Masuda \cite{masuda} investigated some variations of the voter model, where the intervals between interactions of the agents were given by different distributions. The models they used are similar to our \textit{exogenous update}. They found that the times to consensus in the case of a power law distributed interevent interval distribution were enlarged, in agreement with our results.

Possible future avenues of research following the ideas of this work are to study other dynamics and  topologies. An example is the possibility that fat-tailed IET distributions appear as a consequence of topological traps in the network of interaction under majority rule dynamics. These traps can lead to anomalous scaling of consensus times for a majority rule dynamics \cite{epl_anomalous,pre_anomalous}. A consensus time is a global property of the system, but it remains unclear if this is also reflected in the microscopic dynamics, giving rise to broad IET distributions.

\section{Acknowledgements}

We acknowledge financial support from the MINECO (Spain) and FEDER (EU) through projects FISICOS (FIS2007-60327) and MODASS (FIS2011-24785). JFG acknowledges support from the Government of the Balearic Islands through the Conselleria d'Educació, Cultura i Universitats and the ESF.

\bibliographystyle{spphys}
\bibliography{references}

\end{document}